\documentclass[letterpaper,11pt]{article}
\pdfoutput=1

\usepackage{jheppub}
\usepackage{multirow}
\usepackage{diagbox}

\usepackage{subcaption}
\usepackage[countmax]{subfloat}
\usepackage{amssymb}
\usepackage{amsmath}
\usepackage{color}
\usepackage{graphicx}
\usepackage{verbatim}
\usepackage{amsthm}
\usepackage{slashed}
\usepackage{hyperref}
\usepackage{makecell}

\usepackage{side}
\usepackage{ulem}

\newcommand{\tmop}[1]{\ensuremath{\operatorname{#1}}}

\preprint{ANL-179489}

\title{
Simultaneous CTEQ-TEA extraction of PDFs and SMEFT parameters from jet and $t{\bar t}$ data
}

\author[a,b]{Jun Gao,}
\author[a,b]{MeiSen Gao,}
\author[c]{T.~J.~Hobbs,}
\author[a,b]{DianYu Liu,}
\author[a,b]{XiaoMin Shen}

\affiliation[a]{INPAC, Shanghai Key Laboratory for Particle Physics and Cosmology, \\
School of Physics and Astronomy, Shanghai Jiao-Tong University, Shanghai 200240, China}
\affiliation[b]{Key Laboratory for Particle Astrophysics and Cosmology (MOE), Shanghai 200240, China}
\affiliation[c]{High Energy Physics Division, Argonne National Laboratory, Lemont, IL 60439, USA}

\emailAdd{jung49@sjtu.edu.cn}
\emailAdd{gmason@sjtu.edu.cn}
\emailAdd{tim@anl.gov}
\emailAdd{dianyu.liu@sjtu.edu.cn}
\emailAdd{xmshen137@sjtu.edu.cn}

\abstract{
Recasting phenomenological Lagrangians in terms of SM effective field theory (SMEFT) provides a valuable means of connecting
potential BSM physics at momenta well above the electroweak scale to experimental signatures at lower energies.
In this work we jointly fit the Wilson coefficients of SMEFT operators as well as the PDFs
in an extension of the CT18 global analysis framework, obtaining self-consistent constraints
to possible BSM physics effects.
Global fits are boosted with machine-learning techniques in the form of neural networks
to ensure efficient scans of the full PDF+SMEFT parameter space.
We focus on several operators relevant for top-quark pair and
jet production at hadron colliders and obtain constraints on the Wilson coefficients
with Lagrange Multiplier scans.
We find mild correlations between the extracted Wilson coefficients, PDFs, and other
QCD parameters, and see indications that these correlations may become more prominent in future
analyses based on data of higher precision.
This work serves as a new platform for joint analyses of SM and BSM physics based on the CTEQ-TEA framework. 
}
  

\begin{document}

\maketitle
\section{Introduction}
The lack of unambiguous evidence at the LHC for new fundamental particles beyond the
Standard Model (BSM) suggests that the energy scale associated with possible nonstandard interactions may be beyond the direct reach of contemporary hadron colliders.
This possibility suggests a complementary need to search for possible {\it indirect} BSM signatures which might be realized in non-resonant deviations from SM predictions.
Assuming the typical BSM energy scale to be much larger than the electroweak scale,
deviations from the SM can be parametrized phenomenologically using
the framework of SM effective field theory (SMEFT) 
\cite{Weinberg:1978kz,Buchmuller:1985jz,Leung:1984ni},
in which the presence of novel interactions or degrees-of-freedom is encoded in operators of dimension greater than $4$; these operators are then associated with corresponding Wilson coefficients and are suppressed by the
UV scale characterizing the BSM physics.

With the end of Run 2, the Large Hadron Collider (LHC) has now accumulated data with an
integrated luminosity of about 150~${\rm fb}^{-1}$,
allowing for an extensive battery of precision tests of the SM.
Although this already represents a voluminous data set, the LHC is
expected to produce an $\sim$order-of-magnitude more data over the coming
decade.
Getting the most from these data requires that both SM and SMEFT theoretical predictions
be calculated with an accuracy and precision comparable to that of
the experimental data.
To that end, production rates at hadron colliders can be calculated via collinear factorization, whereby partonic matrix
elements, which may involve nonstandard interactions, are convoluted with the 
parton distribution functions (PDFs).
However, the PDFs used in the theoretical predictions are conventionally fitted
by assuming the absence of BSM. 
As a consequence, effects of BSM physics may inadvertently be absorbed into the fitted PDFs, such that using general-purpose PDFs may lead to various statistical and other biases in BSM searches.
In principle, one might hope to avoid such complications by limiting the energy of experimental data sets in
PDF fits used for BSM searches to below a threshold thus minimizing the contamination from possible BSM
with the price of removal of many PDF sensitive data~\cite{2112.11266}. 
To more systematically exploit the full range of high-energy data for SMEFT-based BSM hypothesis testing, it is necessary to combine experimental measurements and theoretical
predictions in a {\it consistent} analysis framework. Such an approach is intended to minimize potential bias while maximizing the sensitivity to
the BSM scenarios encoded in the SMEFT matrix elements.
Thus, one possible solution in this direction involves extending PDF analyses to joint fits of both PDFs
and BSM matrix elements as pioneered in an earlier study by the CTEQ collaboration~\cite{hep-ph/0303013}
and developed later in Refs.~\cite{1902.03048,1905.05215,2104.02723,2110.13204,2111.10431,2201.07240,2203.13923}. 
In addition to avoiding statistical biases due to the use of frozen PDFs in SMEFT-based BSM fits,
simultaneous SMEFT-PDF global analyses can shed light on the complicated correlations
that may potentially exist among the PDF parameters, SMEFT Wilson coefficients, and between members of
each of these sets.

One avenue to extracting information from a simultaneous fit of PDFs and BSM is the method
of Lagrange Multiplier (LM) scans~\cite{hep-ph/0008191,hep-ph/0101051}.
The uncertainties
of any input parameters or derived variables can be determined
from the behavior of the profiled log-likelihood function ($\chi^2$) as a function of the prescribed
variable, without any assumptions about the behavior of the $\chi^2$ in the neighborhood of the
global minimum.
However, the LM method is less used especially for fits with a large number of degrees of freedoms, since
it requires a detailed scan of the parameter space and is computationally expensive.
Fortunately, it was demonstrated that this drawback can be overcome using machine learning in the form
of neural networks (NNs) in Ref.~\cite{2201.06586}.
The profile of the $\chi^2$ in the PDF parameter space can be modeled by NNs, which ensures efficient
scans of the PDF parameter space with almost no time cost.
In this paper we further develop the framework to include key input parameters of the SM,
the mass of the top-quark, the strong coupling constant, and, importantly, the Wilson
coefficients in a specific realization of the SMEFT expansion, up to dimension-$6$, in addition to the usual PDF parameters.
We derive constraints on coefficients of a series of dimension-$6$ operators related
to top-quark pair and jets production with a similar setup to the CT18 global
analysis~\cite{1912.10053} of SM QCD but with extended data sets and theory predictions.
The paper is organized as follows.  
Given its central place to this analysis, we first present the essential theoretical details
and methodology of the simultaneous SMEFT-PDF calculation in Section~\ref{sec:the_cal}.
In Section~\ref{sec:nn}, we describe the architectures of the NNs used
in this work, followed by its validation.
In Section~\ref{sec:exp}, we list
the experimental data of top-quark pair production and jet production
that are used in this work.
Results of Lagrange multiplier scans for Wilson coefficients associated with the top-quark pair production and the jet
production are shown in Section~\ref{sec:tt_np} and Section~\ref{sec:jet_contact}
respectively.
We include discussions on the tolerance criteria and correlations of parameters
in Section~\ref{sec:disc}.
Finally, we conclude in Section \ref{sec:conc}.

\section{Theoretical calculations}\label{sec:the_cal}
The effects of new physics can be described as effective
interactions in the framework of standard model effective field
theory.
This section will firstly describe the SMEFT operators
used in our work.
Then we state the theoretical calculations of the
modified top-quark pair and jet production.
Theoretical calculations on the other DIS and Drell-Yan (DY) processes
are not affected by the effective interactions considered in this
work and are the same as in the CT18 global analyses of SM QCD.

\subsection{Top-quark pair and jet production within SMEFT}
\label{sec:operators}
In SMEFT the deviations with respect to the SM can be parametrized using
a basis of higher-dimensional operators constructed from the SM fields and gauge
symmetries~\cite{Weinberg:1978kz,Buchmuller:1985jz}. 
The full Lagrangian thus consists of the SM Lagrangian and additional terms
expanded in $\Lambda$,
\begin{equation}
\mathcal{L}_{\mathrm{SMEFT}}=\mathcal{L}_{\mathrm{SM}}+\sum_{i} \frac{C_{i} O_{i}^{(6)}}{\Lambda^{2}}+\dots,
\end{equation}
where $\Lambda$ is the matching scale usually chosen as the energy scale of
new physics, and is well above the electroweak scale.
$O_{i}^{(6)}$ are the dimension-6 operators, and $C_{i}$ are the
respective Wilson coefficients which contain information about the
ultraviolet (UV) theory. 
We do not consider operators of dimension-7 and higher of which contributions
are suppressed for the processes of interests.

Due to the large number of potential operators of higher dimension relative to the available
data, it is typically necessary to impose a number of symmetries and other constraints
to simplify the full SMEFT parameter space.
Following Refs.~\cite{1412.7166,1412.5594,1802.07237,1901.05965,1910.03606,2008.11743},
we impose a $U(2)_{Q} \times U(2)_{u} \times U(2)_{d}$
flavor symmetry among the left-handed quark doublets, right-handed up-type quarks singlets and right-handed down-type quarks singlets of the first and second generation.
For top-quark pair production, this flavor symmetry leads to 14 independent four-quark operators and 8 independent operators with two heavy quarks and bosons~\cite{1910.03606}.
In this work, we only focus on the following four typical operators.
\begin{eqnarray}
\label{eq:top_SMEFT}
  O_{t u}^{1}&=& \sum_{i=1}^2 \left(\bar{t} \gamma_{\mu} t\right)\left( \bar{u}_{i} \gamma^{\mu} u_{i}\right)\,, \\
  O_{t d}^{1}&=& \sum_{i=1}^3 \left(\bar{t} \gamma^{\mu} t\right)\left( \bar{d}_{i} \gamma_{\mu} d_{i}\right)\,, \nonumber\\
  O_{tG} &=& ig_s  (\bar{Q}_{L, 3} \tau^{\mu \nu} T^A t) 
  \tilde{\varphi} G_{\mu \nu}^A + \text{h.c.}\,, \nonumber\\
  O_{tq}^8 &=&\sum_{i=1}^2 (\bar{Q}_{i} \gamma_{\mu} T^A Q_{i})
               (\bar{t}  \gamma^{\mu} T^A t)\,, \nonumber
\end{eqnarray}
where $u_{i}, d_{i}$ are the right-handed quarks and $Q_{i}$ is the left-handed quark doublet
of the $i^\mathit{th}$ generation, and $t$ is the right-handed top quark. 
In addition, $T^A$ is the Gell-Mann matrix;
$\tau^{\mu \nu} = \frac{1}{2}  (\gamma^{\mu} \gamma^{\nu} - \gamma^{\nu} \gamma^{\mu})$;
$\varphi$ is the Higgs doublet;
$G^A_{\mu \nu}$ is the gluon field strength tensor; and $g_s$ is the
strong coupling. 
It is assumed that $C_{tu}^1=C_{td}^1$ and all Wilson coefficients are taken to be real
throughout this work.
In Sec.~\ref{sec:jet_contact}, we also study quark contact interactions in the chiral basis
relevant for jet production; in general, these are
\cite{Eichten:1983hw, Eichten:1984eu, Chiappetta:1990jd, 1201.6510, 1204.4773}
\begin{eqnarray}
  \label{eq:CI}
  O_1 & =2\pi\left(\sum_{i=1}^3 \bar{q}_{L i}
  \gamma_{\mu} q_{L i}\right)\left(  \sum_{j=1}^3 \bar{q}_{L j} \gamma^{\mu} q_{L j}\right) \,, \nonumber\\
  O_3 & =2\pi\left(\sum_{i=1}^3 \bar{q}_{L i}
  \gamma_{\mu} q_{L i}\right)\left(  \sum_{j=1}^3 \bar{q}_{R j} \gamma^{\mu} q_{R j}\right) \,,\\
  O_5 & =2\pi\left(\sum_{i=1}^3 \bar{q}_{R i}
        \gamma_{\mu} q_{R i}\right)\left(  \sum_{j=1}^3 \bar{q}_{R j} \gamma^{\mu} q_{R j}\right)  \,, \nonumber
\end{eqnarray}
where $i, j$ are generation indices and $q_{L(R)}$ denotes left(right)-handed quark field
of either up or down type.
Here, the factor of $2\pi$ is due to the convention used in studies of models of quark
compositeness.
Aside from quark-compositeness models, these interactions may arise from various kinds
of BSM scenarios, such as $Z'$ models.
The relative sizes of the corresponding Wilson coefficients depend on the details of the UV-complete models.
In this work, it is assumed that the quark contact interactions 
are purely left-handed, and hence, only $C_1\neq 0$, such that the Wilson coefficients for the
latter two operators of Eq.~(\ref{eq:CI}) are taken to be zero, $C_3=C_5=0$.
Also, all currents involving these operators are assumed to be diagonal and universal in flavor space
to suppress tree-level flavor-changing neutral currents (FCNC).

\subsection{Theoretical computations}
\label{sec:theory}

If the associated Wilson coefficients are nonzero, the SMEFT operators discussed in Sec.~\ref{sec:operators} above have the
potential to affect the total and differential cross sections computed in typical PDF analysis. 
Assuming these Wilson coefficients, $C_i$,
are input parameters, we can write their contribution to the cross sections for some arbitrary observable, $\hat{O}$, as
\begin{equation}\label{eq:xsec}
\frac{d\sigma}{d\hat{O}}=\frac{d\sigma_{\mathrm{SM}}}{d\hat{O}}+\sum_{i} \frac{d\tilde{\sigma}_{i}}{d\hat{O}}  \frac{C_{i}}{\Lambda^{2}}+\sum_{i, j} \frac{d\tilde{\sigma}_{i j}}{d\hat{O}} \frac{C_{i} C_{j}}{\Lambda^{4}},
\end{equation}
where $\sigma_{\mathrm{SM}}$ represents the purely SM contributions, and the second term is due to interference between
SM amplitudes and those generated by dimension-6 operator matrix elements.
For the Wilson coefficients $C^1_{tu}$ and $C^1_{td}$ considered in this work, we note that the interference term begins to
contribute only at next-to-leading order (NLO) in QCD and beyond. 
Lastly, the third term in Eq.~(\ref{eq:xsec}) arises from squared amplitudes generated by the 
various dimension-6 operators and can rival the interference contributions,
despite the suppression from higher powers of $\Lambda$.
In fact, SMEFT contributions to cross sections depend on $C(\mu_c)/\Lambda^2$ in its entirety, where $C(\mu_c)$ is the Wilson coefficient evaluated at an arbitrary matching scale, $\mu_c$.
The scale $\mu_c$ does not necessarily equal
$\Lambda$ and usually is chosen to be close to
the hard scale of the process to account for RG running effects.
In our calculations, we set $\mu_c\!=\!1$ TeV following conventions used in previous literature.
We present constraints for the full quantity $C(\mu_c)/\Lambda^2$ which can be interpreted in terms of new physics at an arbitrary scale $\Lambda$ if it is much larger than the hard scale(s) of the process.   
Further details of the theoretical calculations for different observables
in top-quark pair and jet production
are summarized in Table.~\ref{tab:theo-calc} and will be
explained further below. 
\begin{table}[h]
  \centering
  \begin{tabular}{|l|l|l|l|l|l|}
    \hline
    observable & $\mu_{0}$ & SM QCD & SM EW & SMEFT QCD & th. unc.\\
    \hline
    ${t\bar t}$ total & $m_t$ & NNLO+NNLL & no  & NLO & $\mu_{F,R}$ var.\\
    \hline
    ${t\bar t}$ $p_T$ dist. & ${m_T/2}$ & NNLO & NLO & NLO & $\mu_{F,R}$ var.\\
        \hline
    ${t\bar t}$ $m_{t\bar t}$ dist. & ${H_T/4}$ & NNLO(+NLP) & NLO & NLO & $\mu_{F,R}$ var.\\
    \hline
    ${t\bar t}$ 2D dist. & ${H_T/4}$ & NNLO & no & NLO & no \\
    \hline
    inc. jet & $p_{T,j}$ & NNLO & NLO & NLO & 0.5\% uncor.\\
    \hline
    dijet & $m_{jj}$ & NNLO & NLO & NLO & 0.5\% uncor\\
    \hline
  \end{tabular}
  \caption{Ingredients of theoretical calculations for different observables
  in top-quark pair and jet production, including the nominal scale choice, orders of
  perturbative calculations, and treatment of theoretical uncertainties.}
  \label{tab:theo-calc}
\end{table}

We summarize calculations for the SM contributions and additional contributions
from BSM separately, namely the first term and last two terms in Eq.~(\ref{eq:xsec}). 
The total cross sections for top-quark pair production in the SM are
calculated with {\sc Top++ v2.0}~\cite{1112.5675,1303.6254} program. 
These predictions thus include corrections at next-to-next-to-leading order (NNLO) and soft-gluon
resummation at next-to-next-to-leading logarithmic (NNLL) accuracy in QCD. 
Dependence of the total cross sections on the top-quark mass are also included exactly.
On the other hand, we have not included EW corrections for the {\it total} cross sections, as the effects
of these corrections are much smaller than the experimental uncertainties on the relevant data.
For SM distributions involving top-quark pair
production at the LHC, we use results calculated at NNLO in QCD~\cite{1606.03350,1704.08551}
and implemented in the {\sc fastNLO} interface~\cite{hep-ph/0609285,1109.1310}.
The dependence of NNLO predictions on the top-quark mass is approximated by
multiplicative factors derived from NLO predictions calculated via
{\sc MadGraph5\_aMC@NLO}~\cite{1405.0301}, since the {\sc fastNLO} tables at NNLO
are only available for a fixed top-quark mass.
While $\mathrm{EW}$ corrections are not available for double-differential cross sections,
they have been evaluated for $p_T$ and $m_{t\bar{t}}$
distributions in Ref.~\cite{1705.04105}, where all LO
EW [$\mathcal{O}(\alpha_{s} \alpha)$, $\mathcal{O} (\alpha^2)$] and
NLO EW [$\mathcal{O}(\alpha_{s}^2 \alpha)$,
$\mathcal{O} (\alpha_{s}\alpha^2)$, $\mathcal{O} (\alpha^3)$]
corrections have been considered.
We include these $\mathrm{EW}$ corrections multiplicatively on top of the NNLO QCD predictions
using bin-specific K-factors.
Moreover, for distributions in terms of the invariant mass of the top-quark pair close
to threshold, there exist higher-order Coulomb corrections from QCD which
are potentially large~\cite{0804.1014,0812.0919}.
These have been resummed to all orders in QCD at the next-to-leading power (NLP)
accuracy~\cite{Ju:2019mqc},
and can change the cross sections significantly, for instance in the first kinematic
bin of $m_{t\bar t}$ distributions measured at the LHC 13 TeV.
In our variant fits, we therefore include these soft-gluon
resummed corrections as calculated for theoretical predictions in Ref.~\cite{Ju:2019mqc}; 
we conservatively assign 50\% of these corrections as an additional uncertainty
to account for this effect.
In our nominal calculations, the dynamical renormalization and factorization scales
always take the same value, $\mu_R = \mu_F \equiv \mu_0$.
For top-quark pair production, the nominal scale $\mu_0$ is set to~\cite{1606.03350}
\begin{eqnarray}
  \mu_0 &=& m_t
          \quad \text{for the total cross section}, \nonumber \\
  \mu_0 &=& \frac{m_{T, t}}{2}
          \quad \text{for the $p_{T, t}$ distribution, } \nonumber\\
  \mu_0 &=& \frac{H_T}{4} \equiv \frac{1}{4}  (m_{T, t} + m_{T, \bar{t}}) 
          \quad \text{for the $m_{t \bar{t}}$  and $(1 / \sigma) d^2
  \sigma / d p_{T, t} d y_t$ distributions,} \nonumber
\end{eqnarray}
where $m_{T, t} \equiv \sqrt{m_t^2 + p_{T, t}^2}$ and $m_{T, \bar{t}} \equiv
\sqrt{m_t^2 + p_{T, \bar{t}}^2}$ are the transverse masses of the top quark and
anti-quark.
The theoretical/perturbative uncertainties for SM predictions of top-quark pair production
are estimated by varying the renormalization or factorization scale
$\mu_R = \xi_R \mu_0$, $\mu_F = \xi_F \mu_0$ by a factor of 2 up and
down.
To be specific, the renormalization and factorization scale
uncertainties $\delta_{\mu_R}$ and $\delta_{\mu_F}$ for arbitrary observable
$\hat{O}$ are defined as
\begin{eqnarray}
  \delta_{\mu_R}(\hat{O}) & \equiv & \frac{\hat{O} (\xi_R = 2, \xi_F = 1)
  -\hat{O} (\xi_R = 1 / 2, \xi_F = 1)}{2\hat{O} (\xi_R = 1, \xi_F =
  1)} \,, \\
  \delta_{\mu_F}(\hat{O}) & \equiv & \frac{\hat{O} (\xi_R = 1, \xi_F = 2)
  -\hat{O} (\xi_R = 1, \xi_F = 1 / 2)}{2\hat{O} (\xi_R = 1, \xi_F =
  1)} \,.
\end{eqnarray}
The scale uncertainties are assumed to be fully correlated among different
bins of the same observable, and are included in the fit by introducing two nuisance
parameters for each observable, similar to the experimental systematic uncertainties.
These nuisance parameters are not included in the published CT18 analysis where the theoretical uncertainties in top production are probed by exploring different choices of the central scale.
Contributions to the cross sections from EFT operators 
are calculated at NLO in QCD using {\sc MadGraph5\_aMC@NLO}~\cite{1405.0301}
together with NLO implementations of EFT models~\cite{2008.11743}.
We further link the calculations to the
{\sc PineAPPL} interface~\cite{christopher_schwan_2022_6394794,2008.12789} to generate the necessary
interpolation tables.
For each observable we need to generate several tables in order to reconstruct
full dependence of the cross sections on all the Wilson coefficients.
By doing so, we can calculate the BSM contributions exactly and efficiently
for arbitrary choices of the Wilson coefficients and PDFs.
We have not considered theoretical uncertainties or scale variations of these contributions
from new physics for simplicity.
Such effects are sub-leading but may change our final results of the extracted Wilson
coefficients slightly; this merits further study in future analyses.
SM cross sections for jet production have been calculated to NNLO accuracy in
QCD for limited selections of PDFs.
We first use the {\sc fastNLO} tables~\cite{hep-ph/0609285} at NLO in QCD to compute cross sections
for any prescribed PDFs, and then apply NNLO/NLO point-by-point $K$-factors
calculated by the {\sc NNLOJET}~\cite{1611.01460, 1807.03692, 1801.06415} program.
The {\sc fastNLO} tables at NLO are generated 
using the {\sc NLOJet++}~\cite{hep-ph/0110315, hep-ph/0307268} package.
$\mathrm{EW}$ corrections to jet production at hadron colliders
from Ref.~\cite{1210.0438} are also included on top of the NNLO QCD
predictions, again using multiplicative schemes.
For inclusive jet production, the nominal renormalization and
factorization scales are set to the transverse momentum of the individual jet, $p_{T,j}$. 
For dijet production, the nominal scale is set to the invariant mass of the dijet
system $m_{jj}$.
For both inclusive jet and dijet production, a 0.5\% uncorrelated theory uncertainty is
assumed for each bin to account for statistical fluctuations in
Monte Carlo calculations of the NNLO cross sections as well as residual perturbative
uncertainties as done in the CT18 analyses.
Contributions to jet production from quark contact interactions are calculated at NLO 
in QCD by {\sc CIJet} framework~\cite{1101.4611,1301.7263}.
We note that this program provides an interpolations interface with pre-calculated
tables, ensuring fast computations with arbitrary PDFs.
An interface to xFitter~\cite{2206.12465} is also available.
As with top-quark pair production, we have not assigned theoretical uncertainties for possible BSM contributions on similar grounds; again, we reserve this aspect for future study.

\section{Describing log-likelihood functions with neural networks}\label{sec:nn}
In this section, we describe our use of neural networks (NNs) and machine learning techniques
to model the profile of the log-likelihood function ($\chi^2$) in the
multi-dimensional parameter space of the combined SMEFT-PDF analysis.
This new and improved approach has a range of validity beyond the quadratic approximation commonly used in single $\chi^2$-minimization
studies based on the Hessian method and ensures efficient scans of the full parameter space.
The $\chi^2$ values are calculated for the full set of experimental data included
in our global analyses, where the perturbative QCD accuracy of the associated theoretical
predictions is consistently at NNLO (NLO) for the separate SM (SMEFT)
contributions.
Following a brief introduction to the configuration and settings of the NNs used in this study, we validate
their performance through several comparisons between the original, ``true'' $\chi^2$ and
the predictions obtained by NNs post-training.
We also include a short introduction to the method of LM scans for completeness,
which will be used in later sections.

\subsection{The log-likelihood function}
The quality of the agreement between experimental measurements and the
corresponding theoretical predictions for a given set of SM, SMEFT and PDF
parameters, $\{ a_\ell \}$, is quantified by the $\chi^2$ function, which is given
by~\cite{hep-ph/0201195}
\begin{equation}
  \chi^2 (\{a_\ell\}, \{\lambda\}) = \sum_{k = 1}^{N_{\tmop{pt}}} \frac{1}{s_k^2} 
	\left( D_k - T_k(\{a_\ell\}) - \sum_{\alpha = 1}^{N_{\lambda}} \beta_{k, \alpha}
  \lambda_{\alpha} \right)^2 + \sum_{\alpha = 1}^{N_{\lambda}}
  \lambda_{\alpha}^2,
\label{eq:chi2}
\end{equation}
$N_{\tmop{pt}}$ is the number of data points, $s^2_k$ are the total
{\it uncorrelated} uncertainties obtained by
adding statistical and uncorrelated systematic uncertainties in quadrature,
$D_k$ are the central values of the experimental measurements, and $T_k$ are the
corresponding theoretical prediction which depend on $\{ a_\ell \}$.
$\beta_{k, \alpha}$ are the {\it correlated} systematic uncertainties on the
$k^\mathit{th}$ datum from each of $\alpha$ sources.
We assume the nuisance parameters, $\lambda_{\alpha}$, respect a standard
normal distribution.
By minimizing $\chi^2 (\{ a_\ell \}, \{ \lambda \})$ with respect to the
nuisance parameters, we get the profiled $\chi^2$ function,
\begin{equation}
	\chi^2 (\{a_\ell\},\{\hat{\lambda}\}) = \sum_{i, j = 1}^{N_{\tmop{pt}}} \big(T_i(\{a_\ell\}) - D_i\big) [\tmop{cov}^{-
  1}]_{ij}  \big(T_j(\{a_\ell\}) - D_j\big)\ ,
\end{equation}
where $\tmop{cov}^{- 1}$ is the inverse of the covariance matrix
\begin{equation}
  (\mathrm{cov})_{ij} \equiv s_i^2 \delta_{ij} + \sum_{\alpha =
  1}^{N_{\lambda}} \beta_{i, \alpha} \beta_{j, \alpha}\ .
\label{eq:covmat}
\end{equation}
The experimental systematic errors are usually expressed as relative errors,
$\sigma_{i, \alpha}$, with respect to the data.
Correlated systematic errors are then calculated as
$\sigma_{i, \alpha} T_i$ (known as the `t' definition~\cite{1211.5142}) instead of $\sigma_{i, \alpha} D_i$
in order to avoid D'Agostini bias~\cite{0912.2276}.
As mentioned earlier, we include theoretical uncertainties into the covariance matrix of Eq.~(\ref{eq:covmat})
as well, assuming these to be fully correlated (uncorrelated) for top-quark pair (jet)
production, by using the `t' definition.

\subsection{Neural network architecture and training}
\begin{figure}[htbp]
  \centering
  \includegraphics[width=1\textwidth,clip]{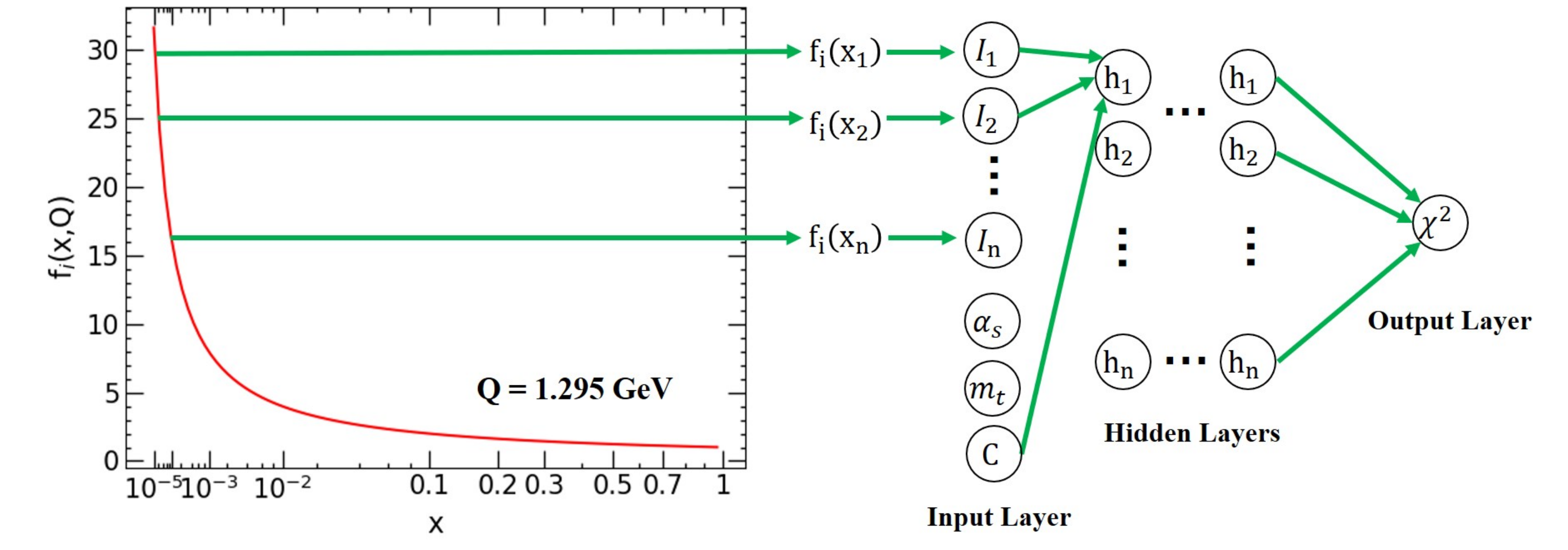}
  \hfill
  \caption{
  An example of the architecture of NNs adapted in this work, taking $\chi^2$ as the target function.
  Inputs to the NN include the flavor-dependent PDFs at discrete values of $x$ ({\it e.g.}, $x_l\! \in\! [0,1])$; pQCD parameters, namely, the top-quark mass, $m_t$, and strong coupling constant, $\alpha_s$; and the SMEFT Wilson
	coefficients.}
  \label{Fig:NN_archi}
\end{figure}

The NNs in this paper are constructed following the guidance provided in Ref.~\cite{2201.06586},
now extended to include the SM pQCD parameters and SMEFT coefficients, in addition to the initial-scale
PDFs at discrete values, $f_i(x_l)$; for the latter, we assume the CT18 parametric forms in the present study.
Note that, in this approach, we use PDF values as direct
inputs to the NNs rather than the PDF parameter themselves as explained in the Appendix of Ref.~\cite{2201.06586}.
The inputs at the outermost layer of the NN are then $m_t$, $\alpha_s(M_Z)$, values of the PDFs at finite $x_l$, and
the SMEFT Wilson coefficients, which are associated with the $\chi^2$ of individual data set as
target functions.
Of these, there are 45 experimental data sets considered in this analysis with individual $\chi^2$
modeled using slightly different setups of NNs as summarized in Table.~\ref{tab:architecture}.
For the 32 data sets involving DIS and DY processes, the only change with respect to the
proposal in Ref.~\cite{2201.06586} is to include the strong coupling on top of the PDF values as an additional input at
the primary layer of the NN.
Beyond DIS and DY, there are also 7 jet production data sets;
for these, we add another hidden layer (for 3 in total) to the architecture as well two more inputs,
the strong coupling constant, $\alpha_s$, and the jet-related Wilson coefficient, $C_1$, associated with Eq.~(\ref{eq:CI}).
The introduction of this additional layer improves the performance of the NNs significantly
due to the quartic dependence of $\chi^2$ on the Wilson coefficients. 
For the 6 remaining top-quark pair sets, we add one final input node for the top-quark pole mass, $m_t$.
As for the jet data, we again consider only a single Wilson coefficient at a time (among $C_{tu}^{1}$, $C_{tq}^{8}$ and $C_{tG}$)
for simplicity;
extending this calculation to include all three top-associated coefficients simultaneously is straightforward --- we
reserve this for future work, pending the availability of additional data.
An example of the architecture of our NNs is shown in Fig.~\ref{Fig:NN_archi}, where the inputs in the form of
initial-scale PDF values, $f_i(x_l)$, together with $\alpha_s$, $m_t$ and the aforementioned Wilson coefficients, are
explicitly shown.
The PDFs, $f_{i}(x_l, Q_0)$, are evaluated at an initial scale of $Q_0\! =\! 1.295$ GeV and with the momentum
fraction, $x_l$, selected among 14 different values from $3.3\times10^{-5}$ to 0.831, and $i \in \{g, u, d, \bar{u}, \bar{d}, s\}$ ---
{\it i.e.}, running over the gluon and all light-quark flavors.
We note that our assumption of the CT18 parametrization results in a symmetric strange sea, $s\!=\!\bar s$, at $Q\!=\!Q_0$. 
We also stress that the $x_l$-grid chosen for sampling the PDFs, which in total produces 84 values of $f_i(x_l)$, is more than sufficient to fully
describe the PDFs' shape and normalization as parametrized in CT18, given that this fit involved 28 free PDF parameters. 
In the end, these finite PDF values, when taken together with $\alpha_{s}$ for DIS and DY, another single SMEFT Wilson coefficient for
the jet and top data, as well as $m_t$ for top-pair production, collectively lead to NNs with input layers consisting of 85, 86, or 87 nodes for
DIS/DY, jet production, or $t\bar{t}$ experiments, respectively.
The three hidden layers consist of 60, 40, and 40 nodes, respectively, with different
activation functions as shown in Tab.~\ref{tab:architecture}.
Moreover, the $\chi^2$ likelihood function given as the final output is constrained to be positive-definite by requiring weights to be strictly positive in
the last layer.

\begin{table}[htpb]
  \centering
  \resizebox{\textwidth}{22.5mm}{
  \begin{tabular}{c|cccc}
  \hline
  \makecell*[c]{Process\\(No. of data sets)} & Inputs & Architecture & \makecell*[c]{Activation functions\\for each layer} & \makecell*[c]{No. of\\total params.}\\
  \hline
  \makecell*[c]{$t\bar{t}$ production\\(6)} & \makecell*[c]{\{PDFs, $\alpha_s$, $m_t$, $C_{tu}^{1}$\\ ($C_{tq}^{8}$, $C_{tG}$)\}} & 87-60-40-40-1 & $\tanh$, $(x^{2}+2)$, $(x^{2}+2)$, linear & 9401\\
  \hline
  \makecell*[c]{jet production\\(7)} & \makecell*[c]{\{PDFs, $\alpha_s$, $C_{1}$\}} & 86-60-40-40-1 & $\tanh$, $(x^{2}+2)$, $(x^{2}+2)$, linear & 9341\\
  \hline
  \makecell*[c]{Others\\(32)} & \{PDFs, $\alpha_s$\} & 85-60-40-1 & $\tanh$, $(x^{2}+2)$, linear & 7641\\
  \hline
  \end{tabular}
  }
  \caption{
  A summary of the different NN architectures used in this paper for the various processes considered. 
  A total of 45 NNs have been constructed in our nominal fit.
  }
  \label{tab:architecture}
\end{table}

A training sample consisting of 12000 PDF replicas with different $\alpha_s$ values, and associated
inputs of $m_t$, $C_1$, and one of $\{C_{tu}^{1}, C_{tq}^{8}$, $C_{tG}\}$ is generated.
The $\alpha_s$, $m_t$, $C_1$, and $\{C_{tu}^{1}, C_{tq}^{8}$, $C_{tG}\}$ values are generated randomly
from uniform distributions defined over reasonably-chosen domains of interest.
Details on the generation of PDF replicas are described in Ref.~\cite{2201.06586}.
We compute the $\chi^2$ of all the data sets for each of the replicas according to the theoretical choices
described above.
We train each NN for 12 hours on a single CPU-core (2.4 GHz) which is sufficient
to obtain the necessary accuracy, as we discuss below.

\subsection{Validating the neural network}
\begin{figure}[htbp]
  \centering
  \includegraphics[width=0.5\textwidth,clip]{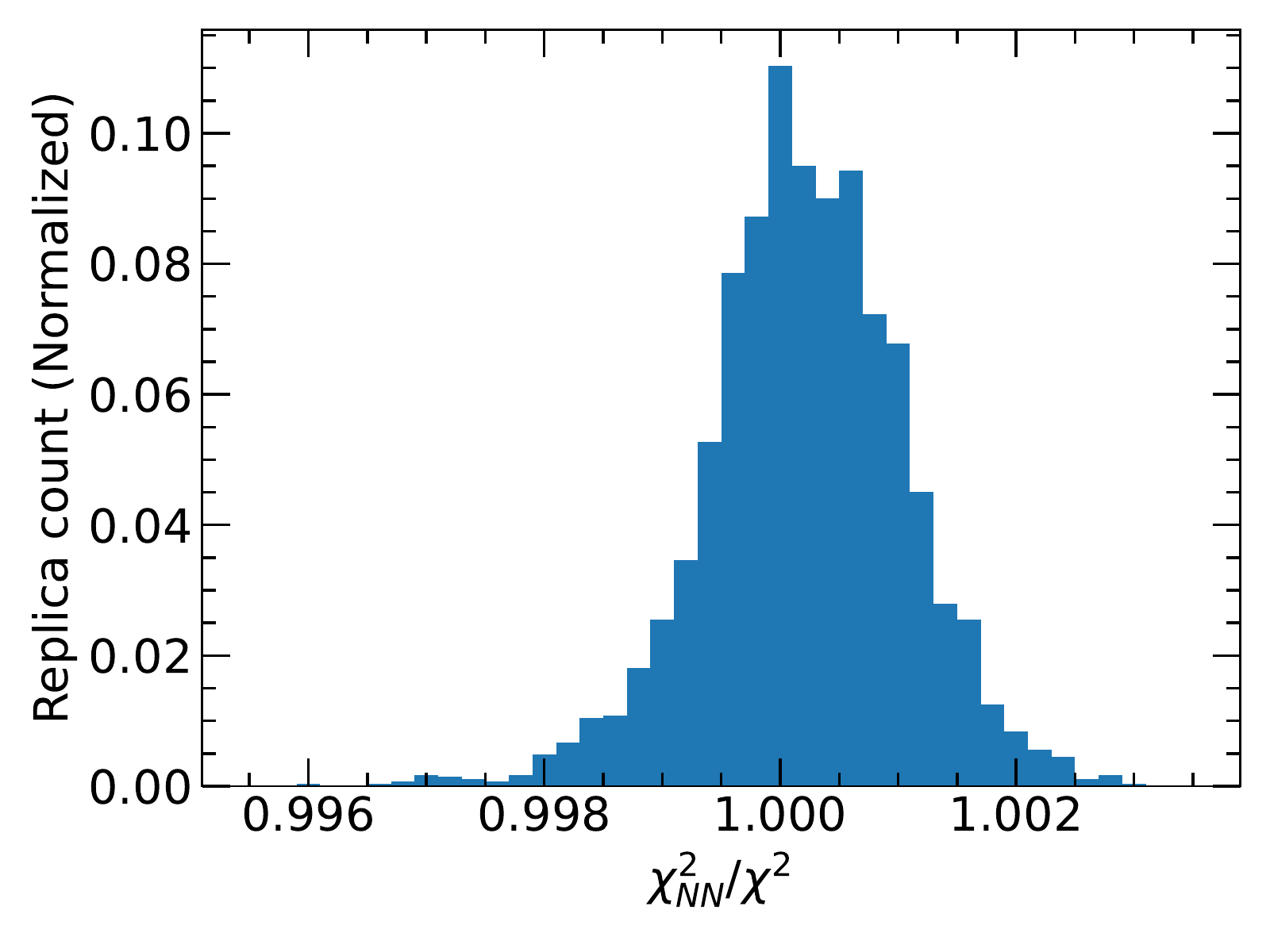}
  \caption{
  A histogram giving the ratio of $\chi^2_{t\bar t}$ as predicted by NNs to the true value
	from direct calculation for all 4000 PDF replicas in the test sample.
  The distribution is normalized to the total number of the replicas included in the test sample.
  }
  \label{Fig:NN_validation_hist}
\end{figure}

\begin{figure}[htbp]
  \centering
    \subcaptionbox{}[7cm] 
    {\includegraphics[width=7cm]{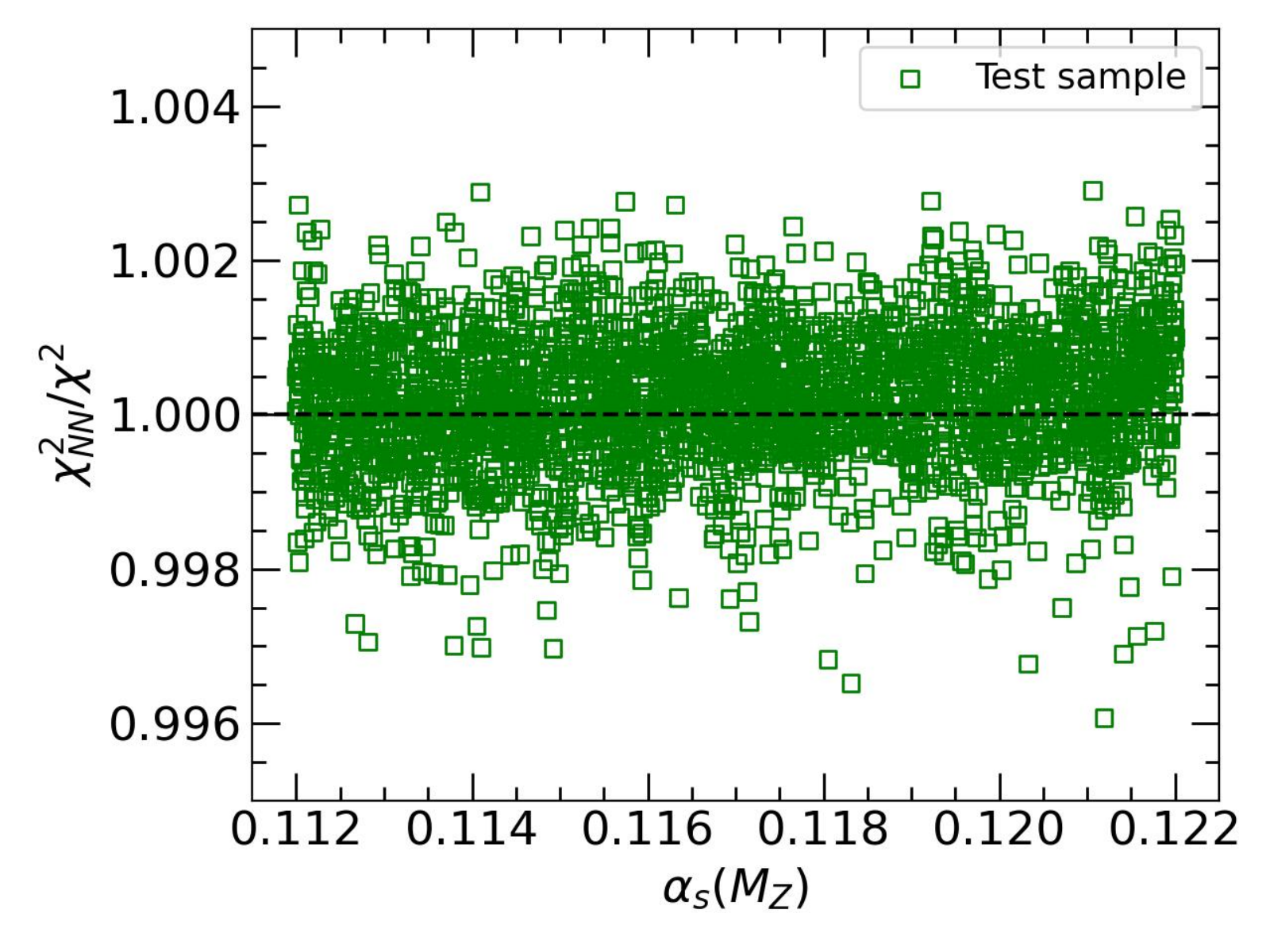}}
  \subcaptionbox{}[7cm] 
    {\includegraphics[width=7cm]{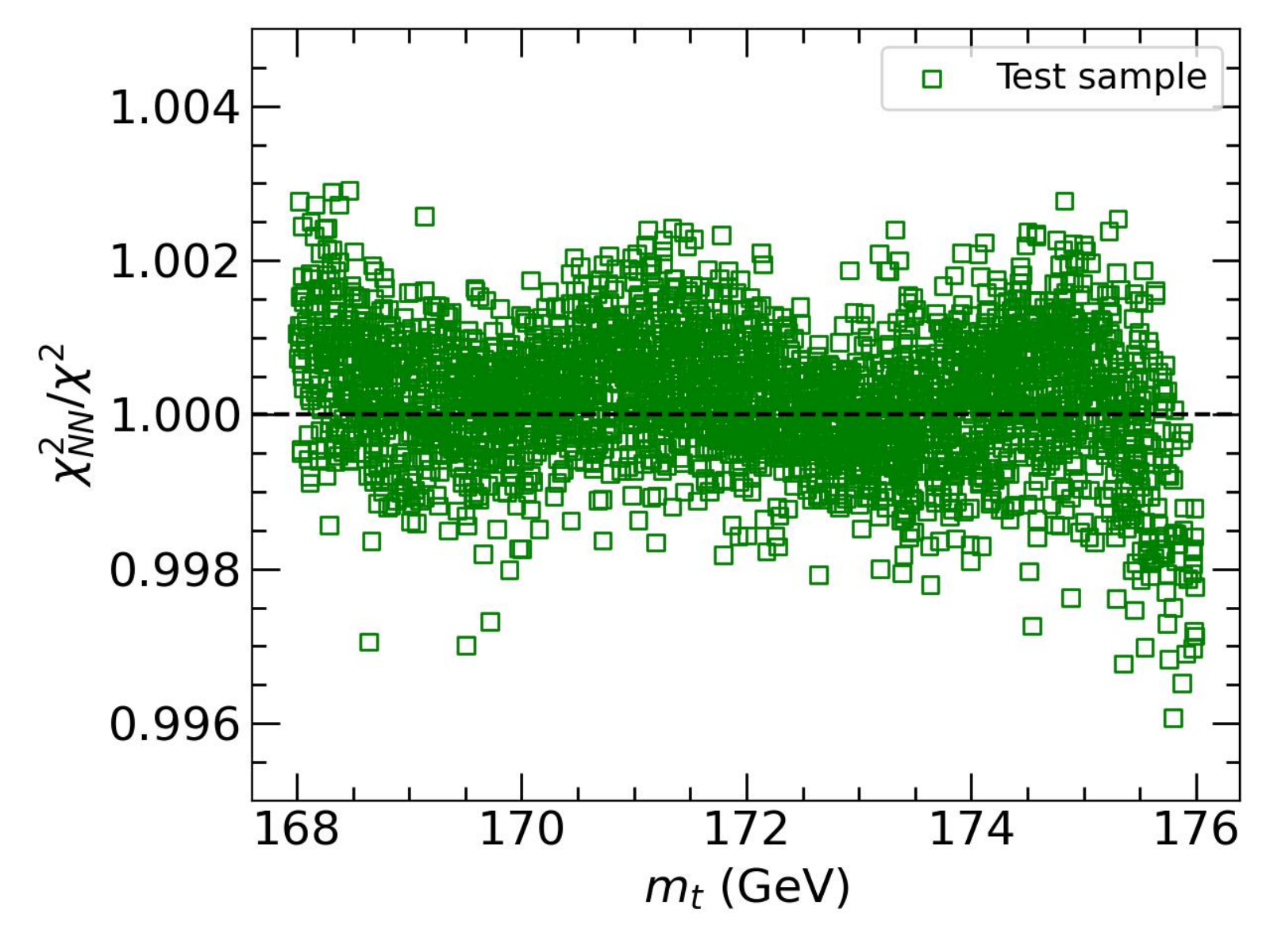}}
  \subcaptionbox{}[7cm]
    {\includegraphics[width=7cm]{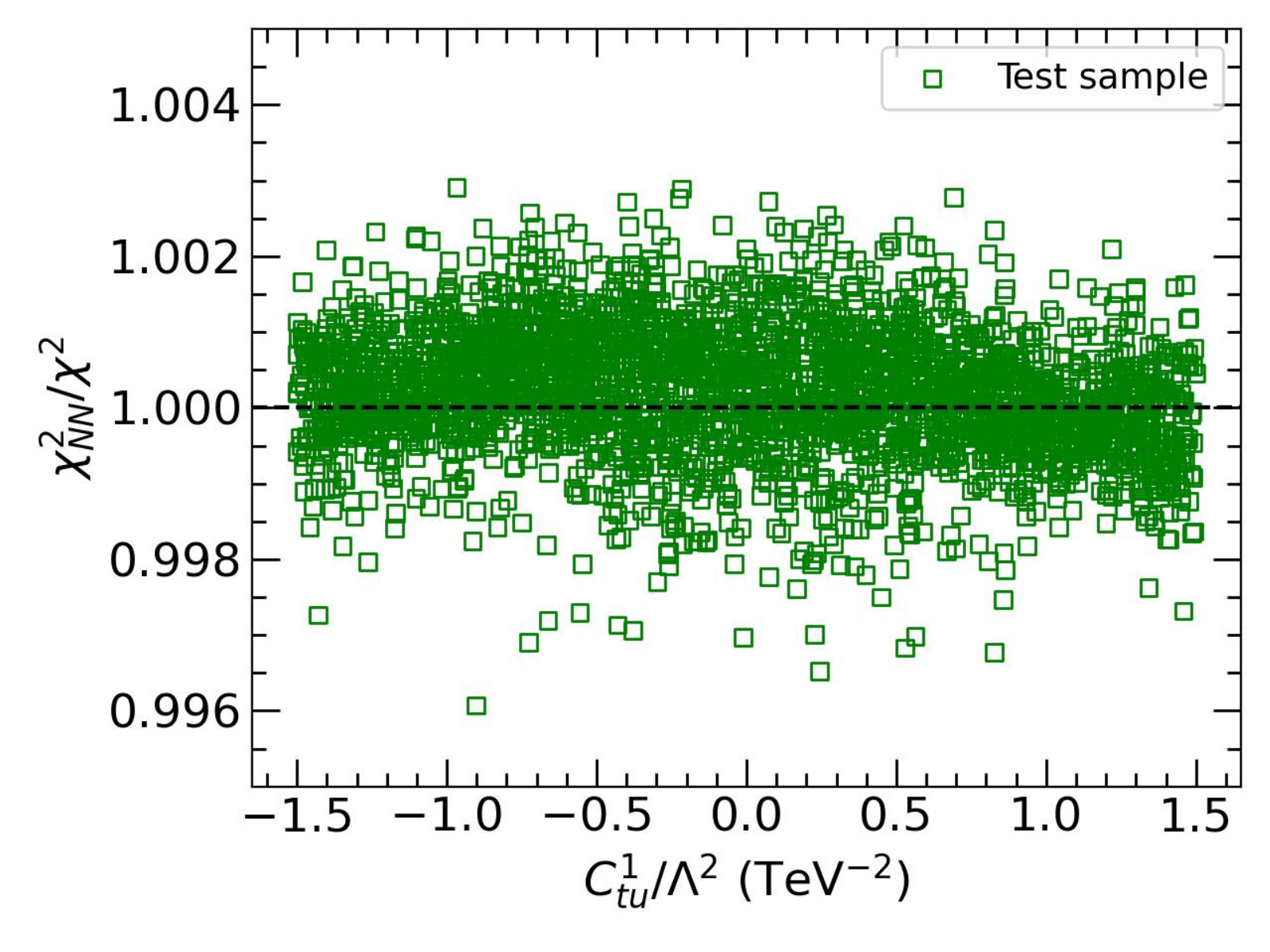}}
   \subcaptionbox{}[7cm]
     {\includegraphics[width=7cm]{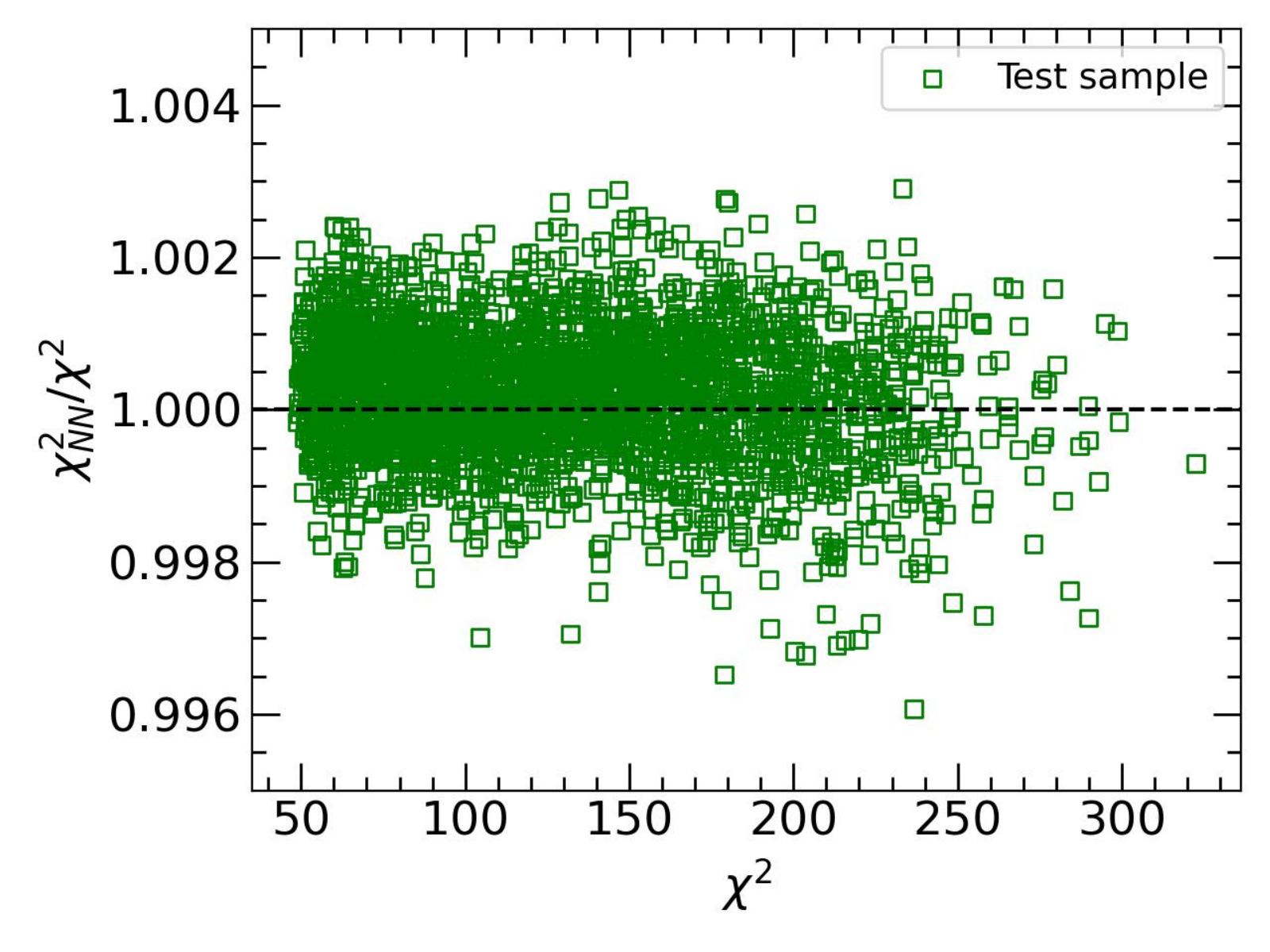}}
  \caption{
  The ratio of the $\chi^2_{t\bar t}$ prediction by NNs to its truth for each of the PDF
  replica in the test sample, distributed in the variable $\alpha_s$, $m_t$, $C_{tu}^{1}$ and
	$\chi^2_{t\bar{t}}$, respectively.
  }
  \label{Fig:NN_validation}
\end{figure}

The accuracy of the prescribed NNs used in this study has already been validated thoroughly in Ref.~\cite{2201.06586}
when using only PDF values as inputs at the first layer.
We generate an independent 4000-replica test sample to validate the performance of the NN so
as to prevent over-training and crosscheck the statistical agreement of predictions based on
the NN output with the true parametric shape of the underlying $\chi^2$ function.
In the end, we find equally good performance for all data sets considered in this work using the updated
architecture outlined above with the additional SM parameters and SMEFT inputs.
For example, we consider the data sets involving top-quark pair production and define
$\chi^2_{t\bar{t}}$ to be the sum of the individual $\chi^2$ values for each of the 5 data sets used
in our nominal fit as summarized in Tab.~\ref{tab:exp}.
In Fig.~\ref{Fig:NN_validation_hist}, we show histograms based on the 4000 PDF replicas in the test sample
giving the ratio of the $\chi^2_{t\bar t}$ prediction from the trained NNs to the true value from direct computation.
We note that the $\chi^2$ predicted for the full data set is obtained by summing the respective outputs of the NNs.
The resulting distribution is then normalized to the total number of the replicas included in the test sample.
Post-training, the NN predictions agree with the true $\chi^2$ calculation to much better than sub-percent accuracy --- within 4 per-mille;
the deviations from $\chi^2_\mathrm{NN}/\chi^2\! =\! 1$ exhibit an approximately Gaussian distribution.
In Fig.~\ref{Fig:NN_validation}, we show the ratio of the $\chi^2_{t\bar t}$
prediction based on the NN output to its true value for each of the PDF replicas in the test sample,
in this case plotted against select SM parameters and the true $\chi^2$;
namely, every PDF replica is associated with a corresponding value of $\alpha_s$, $m_t$ and $C_{tu}^{1}$,
and the associated $\chi^2_{t\bar t}$.
While there is an a sub-permille shift in the direction of $\chi^2_\mathrm{NN}\! >\!\chi^2$ as well as extremely
soft oscillations in the $m_t$ plot, the panels in Fig.~\ref{Fig:NN_validation} otherwise reveal no significant
dependence of the $\chi^2_\mathrm{NN}/\chi^2$ ratio upon the input parameters. This behavior confirms that the NNs
indeed reproduce the local dependence of the likelihood function on the SM parameters introduced in this study,
with no evidence of systematic, parameter-dependent deviations from the true $\chi^2$.
The distribution of $\chi^2_{t\bar{t}}$ for the 4000-replica training set is bounded within [40, 330] $\chi^2$-units,
and the absolute deviations of the NN predictions are generally within 0.1 unit, especially when close to the global
minimum. This level of agreement is sufficiently accurate for a global analysis.

\subsection{Method of LM scans}
\label{Secsub:LM}
The Lagrange multiplier (LM) method~\cite{hep-ph/0008191,hep-ph/0101051} is a robust approach for estimating
the uncertainty of any dependent variable $X(\{a_\ell\})$, where $\{a_\ell\}$ represent the free parameters
in the global analysis as before.
In this method, the $\chi^2$ of the global fit is modified by introducing the
derived variable, $X(\{a_\ell\})$, of the underlying fit parameters as a LM constraint.
The new function to be minimized in the global fit is then given by the sum of two parts,
\begin{equation}
\Psi\left(\lambda,\left\{a_\ell\right\}\right) \equiv \chi^{2}\left(\left\{a_\ell\right\}\right)+\lambda X\left(\left\{a_\ell\right\}\right),
\label{Eq:LM}
\end{equation}
where $\lambda$ is the Lagrange multiplier, which can be continuously varied.
For each value of $\lambda$, one can determine a set of $\{a_\ell\}$, $X(\{a_\ell\})$ and $\chi^2(\{a_\ell\})$
by minimizing $\Psi$,
such that the corresponding $\chi^2(\{a_\ell\})$ then represents the lowest possible $\chi^2$ for
the corresponding value of the $X(\{a_\ell\})$.
The best-fit value of $X(\{a_\ell\})$ and the global minimum, $\chi^2_{min}$, correspond
to the choice $\lambda$ = 0 in Eq.~(\ref{Eq:LM}).
By repeating the minimization for different values of $\lambda$, one can determine
sets of $\{a_\ell\}$, $X(\{a_\ell\})$ and $\chi^2(\{a_\ell\})$.
With this information, it is possible to determine the profiled $\chi^2$ as a function of
the variable $X$,
and the PDF uncertainty of $X$ at the 90\% CL may be evaluated against a tolerance
criteria, $\Delta\chi^2\! +\! P\! \le\! 100$, following the CT18 default analysis.
The penalty term $P$, called the Tier-2 penalty~\cite{1709.04922}, is introduced to ensure the
tolerance is saturated as soon as any data set shows disagreement at the 90\% CL.
We point out that in the special case in which $X(\{a_\ell\})$ is simply taken to be one of the input parameters, $a_\ell$,
of the global fit, the LM scan is equivalent to repeating the fit with one of $a_\ell$ systematically
fixed to different values.
The correlation between two dependent variables, $X_1(\{a_\ell\})$ and $X_2(\{a_\ell\})$, can be assessed with
two-dimensional (2D) LM scans~\cite{2201.06586}, which can be achieved by
simultaneously introducing both $X_1(\{a\})$ and $X_2(\{a\})$ via Lagrange multipliers.
The new function to be minimized in this case is a straightforward generalization of Eq.~(\ref{Eq:LM}), becoming
\begin{equation}
\Psi\left(\lambda_1,\lambda_2,\left\{a_\ell\right\}\right) \equiv \chi^{2}\left(\left\{a_\ell\right\}\right)+\lambda_{1} {X_{1}}\left(\left\{a_\ell\right\}\right)+\lambda_{2} {X_{2}}\left(\left\{a_\ell\right\}\right) ,
\label{eq:LM_2d}
\end{equation}
where $\lambda_1$ and $\lambda_2$ are the specified LM constants as before.
One can determine the profiled $\chi^2$ as a function
of the variables $X_1$ and $X_2$.
The resulting 2D manifold for $\Delta\chi^2$ in the plane of $X_1$ vs.~$X_2$ can be read as
a traditional contour plot, quantifying the correlation between $X_1$ and $X_2$.

\section{Experimental data}
\label{sec:exp}
In this section, we briefly summarize the relevant experimental data sets in
our simultaneous global analysis of QCD and SMEFT.
We start with the CT18 NNLO fit as a baseline by including all 39 default data sets from this study,
consisting of DIS and DY as well as top-quark pair and jet production.
We then include several additional LHC experiments at 8 and 13 TeV --- specifically, distributions from
top-quark pair and jet production; we also incorporate total cross section data for
top-quark pair production at both the Tevatron and LHC.
BSM scenarios parametrized through SMEFT are directly constrained in particular by the 13 data sets on top-quark pair
and jet production as summarized in Tab.~\ref{tab:exp}.
Additional detail regarding the other 32 data sets on DIS and DY production can be found in Ref.~\cite{1912.10053}. 
In Tab.~\ref{tab:exp}, sets marked with a star (dagger) are included in our nominal PDF+SMEFT fits of
top-quark pair (jet) production data.
We further summarize key aspects of these experiments, including their respective kinematical coverages, in the
subsections below.
\begin{table}[th]
  \centering
  \begin{tabular}{|l|l|l|ll|l|}
    \hline
    Experiments & $\sqrt s$(TeV) & $\mathcal{L} (\text{fb}^{- 1})$ & observable & &
    $N_{\text{pt}}$\\\hline\hline
    $^{*\dag}$ LHC(Tevatron)  & 7/8/13(1.96) & --- & $t\bar t$ total cross section & \cite{1309.7570, 1406.5375, 1208.2671, 1603.02303, ATLAS-tot-13,1611.04040}  & 8\\\hline
    $^{*\dag}$ ATLAS $t\bar t$ & 8 & 20.3 & 
    1D dis. in $p_{T,t}$ or $m_{t \bar{t}}$ & \cite{1511.04716} & 15\\\hline
    $^{*\dag}$ CMS $t\bar t$ & 8 & 19.7 & 2D dis. in $p_{T,t}$ and $y_t$ &
    \cite{1703.01630} & 16\\\hline
        CMS $t\bar t$ & 8 & 19.7 & 
      1D dis. in $m_{t \bar{t}}$ & \cite{1505.04480} & 7\\\hline
    $^{*\dag}$ ATLAS $t\bar t$ & 13 & 36 & 1D dis. in $m_{t \bar{t}}$  & \cite{1908.07305} &
    7\\\hline
    $^{*\dag}$ CMS $t\bar t$ & 13 & 35.9 &  1D dis. in $m_{t \bar{t}}$ & \cite{1811.06625} &
    7\\\hline
    $^{*\dag}$ CDF II inc. jet & 1.96& 1.13 &  2D dis. in $p_T$ and $y$ & \cite{0807.2204} &72\\\hline
    $^{*\dag}$ D0 II inc. jet & 1.96& 0.7 &  2D dis. in $p_T$ and $y$ & \cite{0802.2400} &110\\\hline
    $^{*\dag}$ ATLAS inc. jet & 7 & 4.5 &  2D dis. in $p_T$ and $y$ & \cite{1410.8857} &140\\\hline
     $^{*\dag}$ CMS inc. jet & 7 & 5 &  2D dis. in $p_T$ and $y$ & \cite{1406.0324} &158\\\hline
    $^*$ CMS inc. jet & 8  & 19.7 & 2D dis. in $p_T$ and $y$ & \cite{1609.05331} &185\\\hline
    $^{\dag}$ CMS dijet & 8  & 19.7 & 3D dis. in $p_T^{ave.}$, $y_b$ and $y^*$ & \cite{1705.02628} &122\\\hline
    $^{\dag}$ CMS inc. jet & 13  & 36.3 &  2D dis. in $p_T$ and $y$ & \cite{2111.10431} &78\\\hline
  \end{tabular}
  \caption{\label{tab:exp}
Experimental data sets on top-quark pair and jets production included in
the global analyses. 
$N_{\text{pt}}$ indicates the total number of data points in each data set.
The data sets marked with star (dagger) are included in our nominal fits for
study of SMEFT in top-quark pair (jet) production.
Other data sets on DIS and DY productions are the same as in CT18 analyses and are not shown
here for simplicity.}
\end{table}

\subsection{Top-quark pair production}
The ATLAS and CMS collaborations at the LHC have measured differential cross
sections for top-quark pair production in several kinematic variables at ${\sqrt{s}}=8$ TeV.
With the exception of ATLAS, we avoid including multiple distributions from the same experiment due to the
complicated and hard-to-control statistical correlations which would exist among these data sets.
For ATLAS, however, we include one-dimensional distributions in both the invariant mass of the
top-quark pair, $d\sigma/dm_{t\bar t}$, and the transverse momentum of the top quark,
$d\sigma/dp_{T,t}$, corresponding to an integrated luminosity of
$\mathcal{L}\! =\! 20.3\, \mathrm{fb}^{-1}$~\cite{1511.04716}; this results in 8 and 7 data points,
respectively.
For CMS, with $\mathcal{L}\! =\! 19.7\, \mathrm{fb}^{-1}$ of integrated luminosity, we
nominally fit the normalized double-differential
cross section, $(1 / \sigma) \times d^2 \sigma \big/ d p_{T, t}\, d y_t$~\cite{1703.01630}.
These data sets were included in the CT18 fit as Exp.~ID\# 580 and 573, respectively.
As mentioned earlier, in our variant fit the CMS measurement of the normalized
differential cross section, $(1 / \sigma) \times d \sigma / d m_{t\bar t}$~\cite{1505.04480},
is used instead.
We also note that the kinematic reach of these data was restricted to $p_{T,t}\! <\! 600$ GeV and $m_{t\bar t}\! <\! 1600$ GeV.
For $\sqrt{s}\! =\! 13$ TeV, we select the distribution on
$m_{t \bar{t}}$ on the logic that the large
$m_{t \bar{t}}$ region is more sensitive to BSM physics.
In this case, the ATLAS data we fit~\cite{1908.07305} were collected in 2015 and 2016
with $\mathcal{L}\!=\!36$~$\text{fb}^{-1}$ in the
lepton+jet decay channel of the top-quark pair.
For CMS~\cite{1811.06625}, the data were collected in 2016,
corresponding to an integrated luminosity of 35.9~$\text{fb}^{-1}$
in the dilepton decay channel.
The two collaborations provide measurements based on the same binning
scheme over $m_{t \bar{t}}$ with bin edges located at
$[300, 380, 470, 620, 820, 1100, 1500, 2500]~\text{GeV}$; this therefore
results in $N_\mathrm{pt}\!=\!7$ for both experiments.
Lastly, we have also taken into account measurements of the $t\bar{t}$ total cross section from
the Tevatron and LHC (for the latter, with $\sqrt{s}\! =\! 7, 8, 13$~TeV), leading to a combined total
of 8 more data points.
To be specific, we include the data from
CDF and D0 at $\sqrt{s}=1.96$ TeV~\cite{1309.7570};
ATLAS~\cite{1406.5375} ($e\mu$ channel) and CMS~\cite{1208.2671} (dilepton channel) at $\sqrt{s}=7$~TeV;
ATLAS~\cite{1406.5375} ($e\mu$) and CMS~\cite{1603.02303} ($e\mu$) at $\sqrt{s}=8$~TeV;
and ATLAS~\cite{ATLAS-tot-13} ($e\mu$) and CMS~\cite{1611.04040} ($e\mu$) at $\sqrt{s}=13$ TeV.
The precision of these measurements ranges from 2\% for ATLAS at 13 TeV to 8\% for D0. 

\subsection{Inclusive jet and dijet production}
For inclusive jet production we include data on the double-differential cross section in
the transverse momentum and rapidity of the jet, $d^2\sigma/(dp_T\,dy)$, 
as measured by the CDF and D0 experiments during Run-II of the Tevatron.
The CDF experiment measured the inclusive jet cross section in
$p\bar{p}$ collisions at $\sqrt{s}=1.96$ TeV with data corresponding
to an integrated luminosity of $1.13~\text{fb}^{-1}$~\cite{0807.2204}.
This measurement used the cone-based midpoint jet-clustering algorithm
in the jet rapidity region of $|y|<2.1$,
with a cone radius $R\equiv \sqrt{(\Delta y)^2+(\Delta\phi)^2} = 0.7$
in rapidity $y$ and azimuthal angle $\phi$, resulting in 72 data points in total.
Meanwhile, the D0 experiment collected a data sample at a center-of-mass-energy of
$\sqrt{s}=1.96$ TeV, corresponding to an integrated luminosity of 0.70
$\text{fb}^{-1}$~\cite{0802.2400}.
Cross sections on inclusive jet production with jet transverse momenta from 50 to 600 GeV
and jet rapidities of up to 2.4 were divided into 110 bins.
Again, the midpoint jet algorithm with radius $R = 0.7$ was adopted.
For ATLAS, measurements of inclusive jet production at $\sqrt{s}=7$ TeV based on the
anti-$k_T$ jet algorithm~\cite{0802.1189} with radius $R=0.6$ are included in our fit,
where these data have an integrated luminosity of 4.5 $\text{fb}^{-1}$~\cite{1410.8857}. 
This set covers a jet rapidity range of $0\le |y| \le 3.0$ and associated transverse
momentum range of $74\le p_T \le 1992$ GeV, for a total of 140 data points.
For CMS, we fit inclusive jet data measured at $\sqrt{s}=7$ TeV~\cite{1406.0324},
8 TeV~\cite{1609.05331} and 13 TeV~\cite{2111.10431}, with these corresponding to
$\mathcal{L}\! =\! 5$, 19.7 and 36.3
$\text{fb}^{-1}$, respectively.
The 7 TeV set contains 158 data points,
covering a phase-space region with
jet transverse momentum 56 $\le p_T \le$ 1327 GeV, 
and rapidity $0\le |y| \le 3.0$.
For the 8 TeV data, there are a total of 185 points,
in this case covering transverse momenta 74 $\le p_T \le$ 2500
GeV and jet rapidities over $0\le |y| \le 3.0$.
Finally, the 13 TeV data set involves 78 points, covering a phase space region
with jet transverse momentum from 97 GeV up to 3.1 TeV and rapidity $|y| \le 2.0$.
We stress that the CMS 8 TeV jet data are only used in the variant SMEFT fit for jet
production.
In our nominal fit, however, we instead use the CMS dijet measurements recorded at
8 TeV~\cite{1705.02628}.
These measurements provide 122 data points on the triple-differential
cross section, which is dependent on the average transverse momentum of the two leading jets,
$p_{T,\text{avg}}\equiv (p_{T,1}+p_{T,2})/2$;
half of their rapidity separation, $y^* \equiv|y_1-y_2|/2$; and the
rapidity of the dijet system, $y_b \equiv |y_1+y_2|/2$.
In this case, the average transverse momentum can reach 1600~GeV in the central rapidity region,
and we point out that, for all CMS measurements, we select those data which were measured using the anti-$k_T$
algorithm with a jet radius $R=0.7$.

\section{LM scans with top-quark pair production}
\label{sec:tt_np}

In this section, we investigate the determination of the top-associated SMEFT Wilson coefficients [corresponding
to the operators of Eq.~(\ref{eq:top_SMEFT})]
in our combined analysis with PDF degrees-of-freedom; we quantify constraints on the SMEFT coefficients
via LM scans as discussed in Sec.~\ref{Secsub:LM}.
Before doing this, however, we first examine the impact of our fitted data on the purely SM input parameters,
namely, the top-quark mass, $m_t$, and the strong coupling constant, $\alpha_s(M_Z)$, given that these quantities
are strongly correlated with top-quark pair production.
We note that the other default data sets for jet production and DIS/DY are
always included in our global analyses as well, although these information do not impose as direct
constraints, particularly with respect to $m_{t}$.

\subsection{Impact of strong coupling and top-quark mass}
\label{Sec:top_mass}
\begin{figure}[htbp]
  \centering
  \subcaptionbox{}[7.7cm]
    {\includegraphics[width=7.7cm]{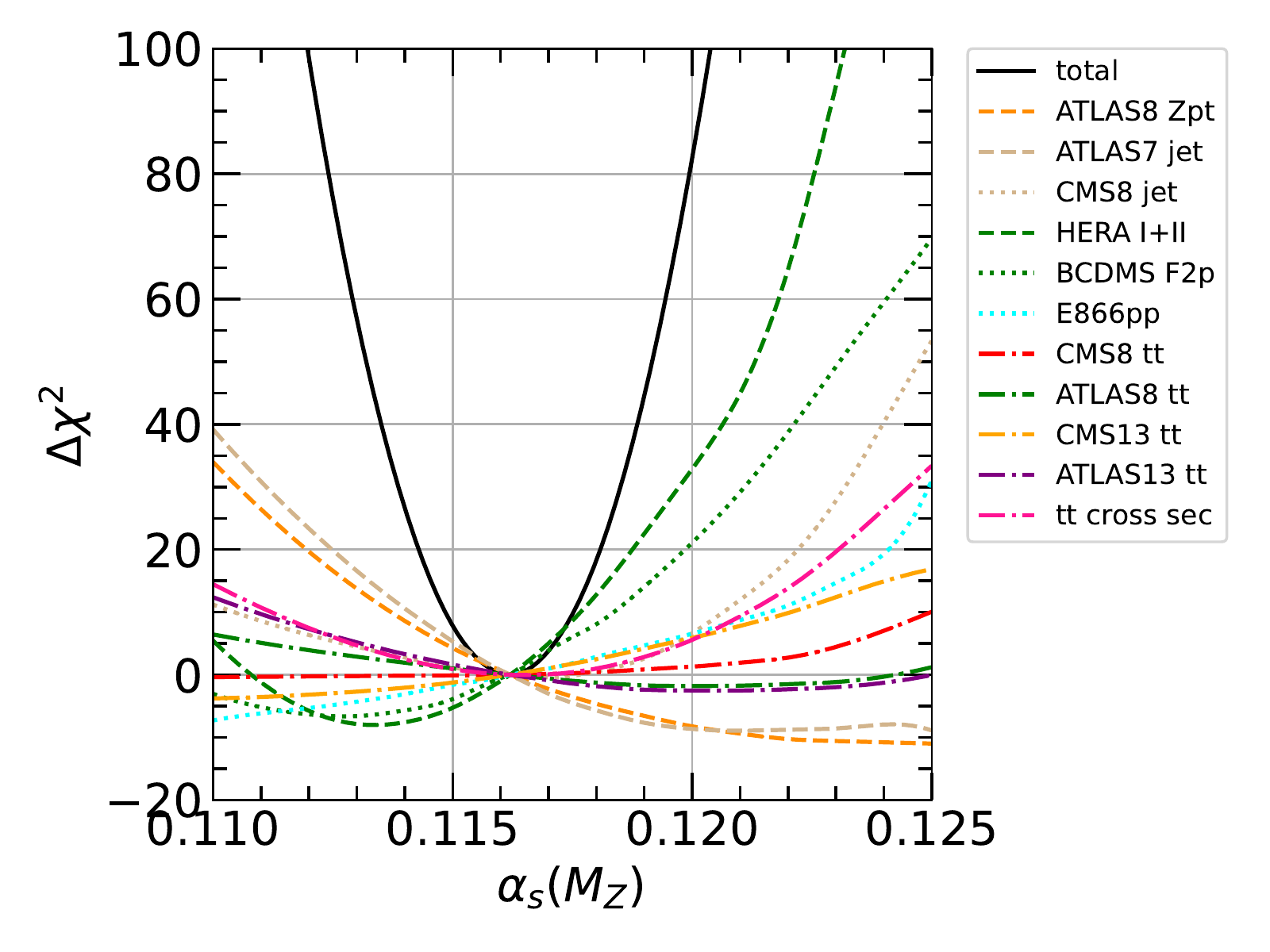}}
  \subcaptionbox{}[7.7cm]
    {\includegraphics[width=7.7cm]{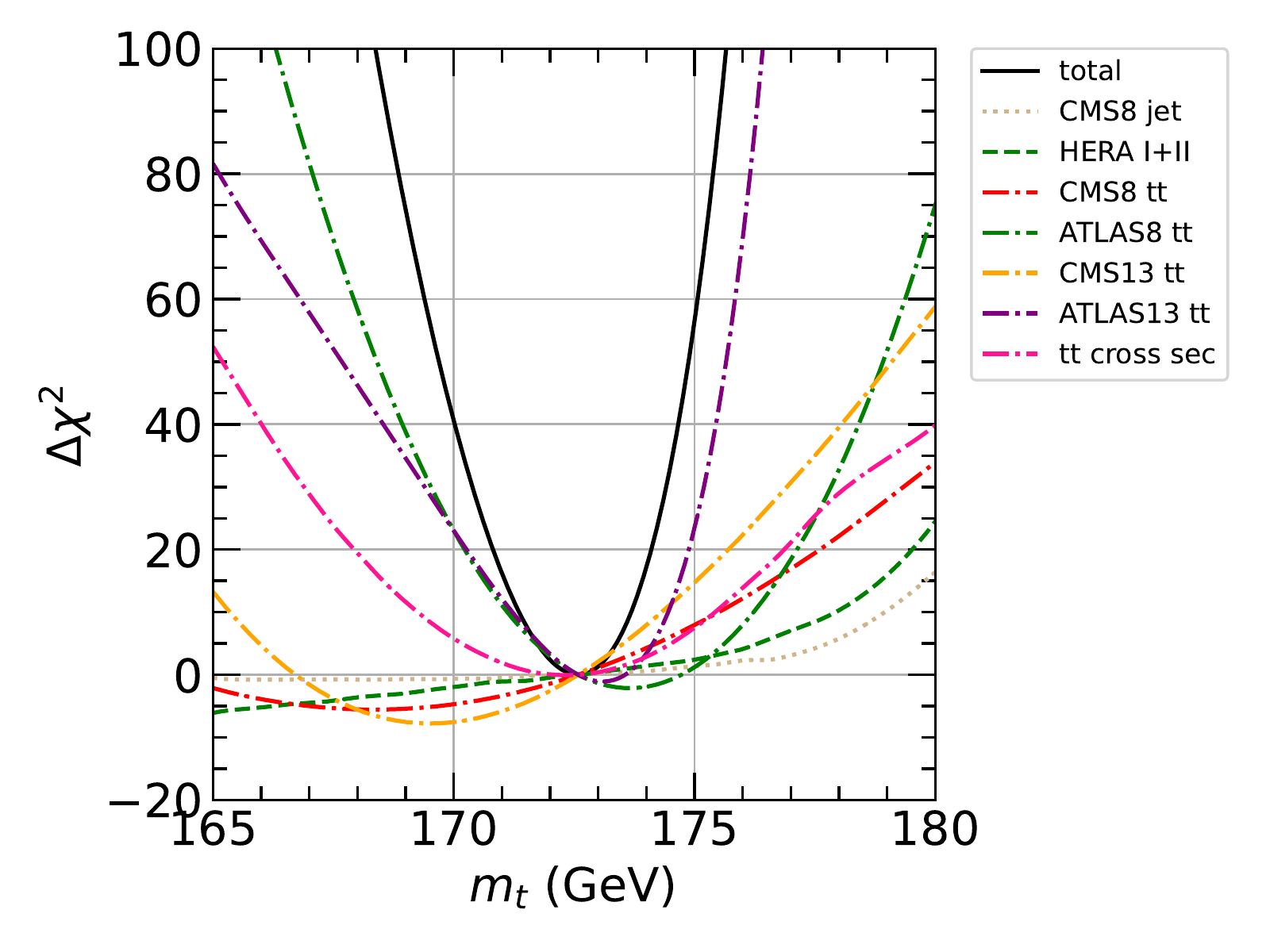}}
  \caption{LM scans on $\alpha_s(M_Z)$ (left panel) and $m_t$ (right panel).
The black-solid lines represent the total $\Delta \chi^2$ of the global fit.
The dotted, dashed, and dot-dashed curves represent the contributions to
$\Delta \chi^2$ from individual experimental data sets.}
  \label{Fig:LM_sm_free}
\end{figure}

We first carry out a series of LM scans on a joint fit of PDFs, $m_t$ and $\alpha_s(M_Z)$ ---
without considering SMEFT contributions.
In Fig.~\ref{Fig:LM_sm_free} (a) and (b), we show the profiled $\chi^2$ as a function of $\alpha_s(M_Z)$ and $m_t$,
respectively.
The black-solid line(s) represent the change, $\Delta \chi^2$, in the global likelihood function relative to the best fit,
$\Delta \chi^2\! =\! 0$, whereas the various colorful dot, dash and dot-dash curves represent the contributions to
$\Delta \chi^2$ from individual experimental data sets.
We find that both the global and individual experimental $\Delta\chi^2$ curves show
an almost quadratic dependence on the variables in the neighborhood of the global minimum.

In Fig.~\ref{Fig:LM_sm_free} (a) for $\alpha_s(M_Z)$,
we see that, as expected, the combined HERA DIS data stand out as providing an especially important constraint due to both the
high experimental precision and large volume of data for this set.
The LM scans predict a value of $\alpha_s(M_Z) = 0.1162 $, which is slightly smaller than, but consistent with,
the world average of $\alpha_s(M_Z) = 0.1179\pm0.001$~\cite{Workman:2022}.
These results on $\alpha_s(M_Z)$ are also consistent with those reported in the CT18 analysis
and serve as a crosscheck on the accuracy of our new approach based on NNs.
Meanwhile, in Fig.~\ref{Fig:LM_sm_free} (b) the LM scans over $m_t$
show that the $t\bar{t}$ data offer the dominant constraint(s), again as expected.
The LM scans predict a central value and uncertainty of $m_t = 172.58 $~GeV, which
is slightly larger than the world average of $m_t = 172.4\pm 0.7$ GeV~\cite{Workman:2022} from
measurements of cross sections; still, up to uncertainties, these values are nicely consistent.
We find that both the CMS 8 and 13 TeV $t\bar{t}$ data prefer a smaller value of $m_t$ compared to
the ATLAS $t\bar{t}$ data, which prefer larger $m_t$, much as was reported
in Ref.~\cite{1904.05237}, which attributed these preferences as mainly coming from constraints provided by
the first kinematic bin of the $m_{t\bar{t}}$ distribution close to the threshold region. 
A detailed study on determination of the top-quark mass will be presented elsewhere.

\begin{figure}[htbp]
  \centering
  \includegraphics[width=0.47\textwidth,clip]{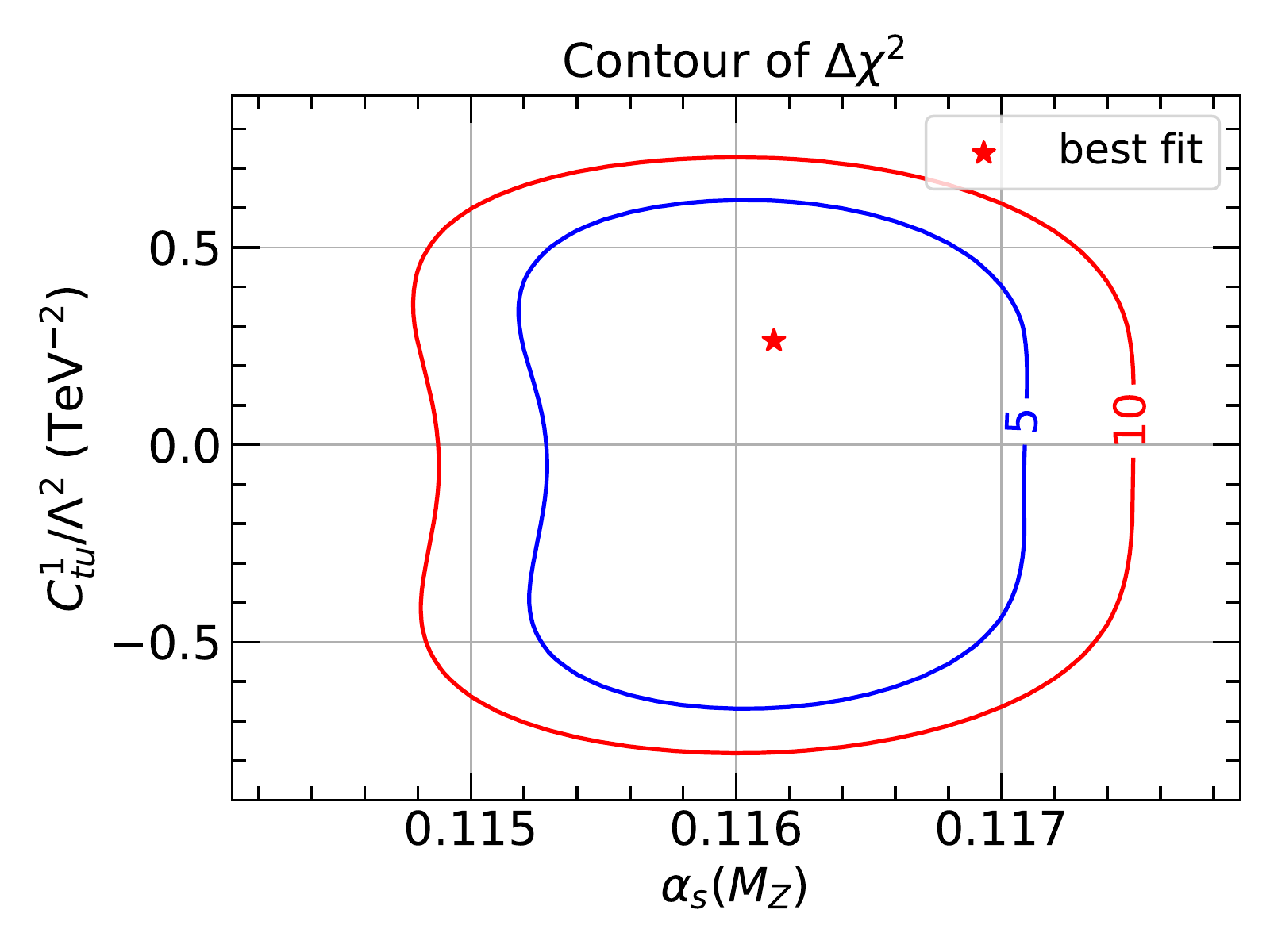}
  \hspace{0.2in}
  \includegraphics[width=0.47\textwidth,clip]{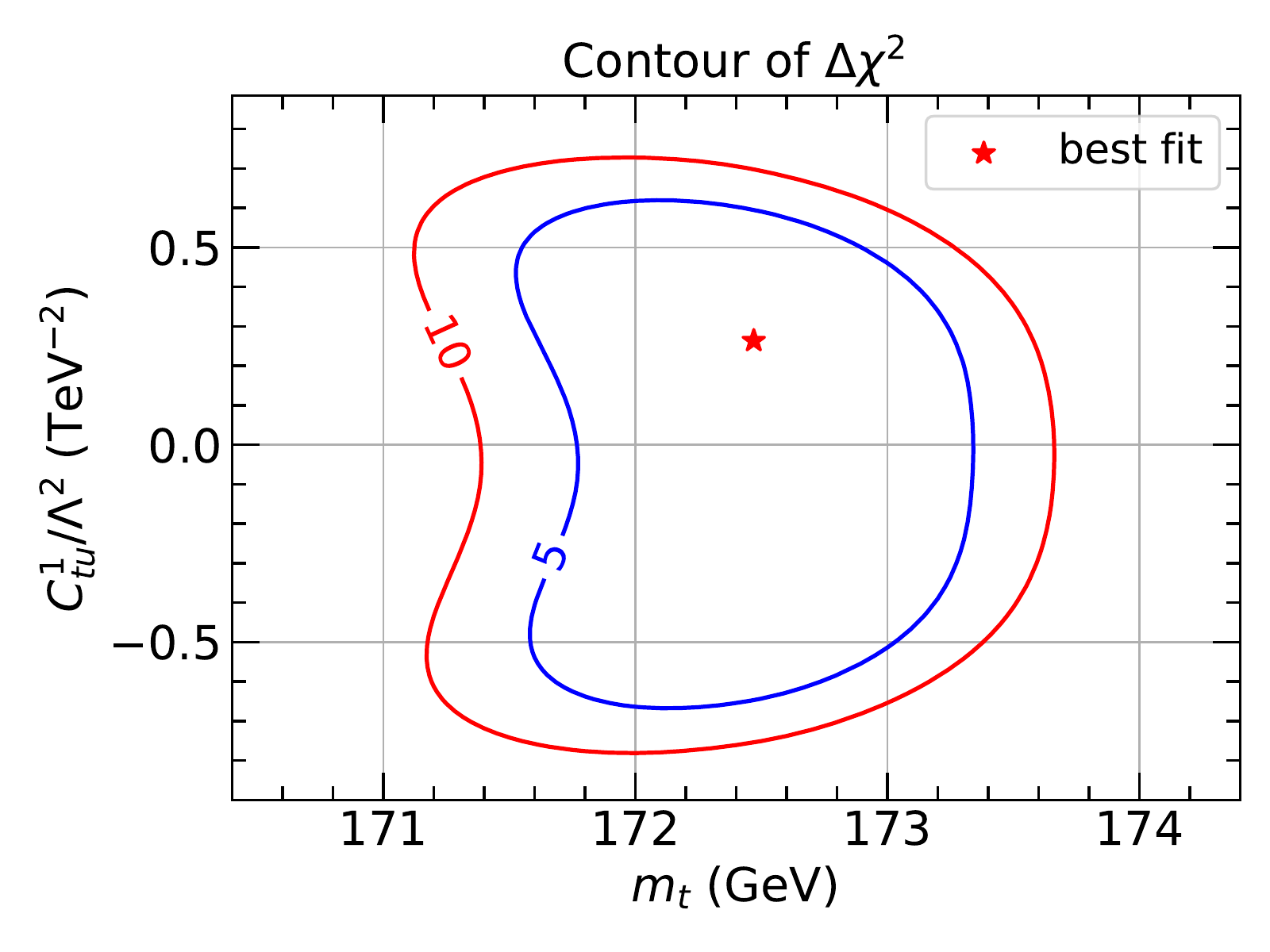}
	\caption{Contour plot of $\Delta \chi^2$ in the plane of $C^1_{tu}$ vs.~$\alpha_s(M_Z)$ [left] and $C^1_{tu}$ vs.~$m_t$ [right],
	as determined according to the 2D LM scan technique of Eq.~(\ref{eq:LM_2d}) and surrounding text.}
  \label{Fig:2d_alp_mt}
\end{figure}

\begin{table}[htpb]
  \centering
  \begin{tabular}{c|ccccc|c}
  \hline
  $\chi^2$ (nominal) & tot. cross sect. & CMS 8 & ATLAS 8 & CMS 13 & ATLAS 13 & global\\
  \hline
  all free & 5.08 & 16.70 & 11.41 & 14.24 & 4.73  & 4278.62\\
  \hline
  $m_t$ fixed & 5.09 & 16.76 & 11.30 & 14.30 & 4.71  & 4278.63\\
  \hline
  $\alpha_s$ fixed & 6.13 & 17.27 & 9.85 & 15.96 & 4.13  & 4297.38\\
  \hline
  $\alpha_s$ and $m_t$ fixed  & 6.91 & 16.52 & 10.95 & 15.06 & 4.47  & 4297.97\\
  \hline
  all fixed & 6.90 & 16.49 & 11.03 & 14.62 & 4.96  & 4298.03\\
  \hline
  \end{tabular}
  \caption{Central values of $\chi^2$ for the individual $t\bar{t}$ data sets as well as the full global fit assuming
  various fixing conditions on the parameters $\alpha_s(M_Z)$, $m_t$, and $C_{tu}^1$.} 
  \label{tab:chi2_alp118}
\end{table}

Following these preliminaries on the purely SM fits, we now move to joint SMEFT/SM fits, taking $C^1_{tu}$ as a first
example.
{\it A priori}, it is possible that significant parametric correlations might exist among the PDF and SM input parameters
illustrated above and the SMEFT operator coefficients.
To investigate this potential interplay, we simultaneously consider the Wilson coefficient and $\alpha_s(M_Z)$ or $m_t$
in the two panels of Fig.~\ref{Fig:2d_alp_mt}, showing the contours of $\Delta \chi^2$ in the plane of $C^1_{tu}$
vs.~$\alpha_s(M_Z)$ (left panel) and $C^1_{tu}$ vs.~$m_t$ (right) based on the 2D LM scans of Eq.~(\ref{eq:LM_2d}).
Here, the blue and red contours represent $\Delta \chi^2 =$ 5 and 10, respectively.
For the fitted data set, we find only very minimal correlations between $C^1_{tu}$ and $\alpha_s(M_Z)$ or $m_t$,
consistent with the SM, corresponding to $C^1_{tu}=0$.
In fact, the best-fit values of $\alpha_s(M_Z)$ and $m_t$ are almost identical (and well within a small $\Delta \chi^2$
interval) to those shown in Fig.~\ref{Fig:LM_sm_free}, which corresponded to fitting without SMEFT contributions. 
We note that the shapes of the contours are mildly asymmetric because of the non-quadratic dependence of $\chi^2$ on
the underlying parameters.
These conclusions also hold for the other top-associated Wilson coefficients of Eq.~(\ref{eq:top_SMEFT}).

To further disentangle possible correlations among the SM/EFT input parameters impacting the description of the $t\bar{t}$ data, we
perform a series of global fits with either $\alpha_s(M_Z)$ fixed to 0.118 or $m_t$ fixed to 172.5~GeV,
with both of these fixed, or with $C^1_{tu}$ fixed to 0.
The resulting $\chi^2$ values under these scenarios for the various individual $t\bar{t}$ data sets as well as the total at the global minimum
are summarized in Tab.~\ref{tab:chi2_alp118}.
The total $\chi^2$ is elevated by about 18 units if $\alpha_s(M_Z)$ is fixed
to 0.118, consistent with results shown in Fig.~\ref{Fig:LM_sm_free}, while it changes by less than
one unit when fixing $m_t$ or $C^1_{tu}$ as call be deduced from the ``all fixed'' scenario of Tab.~\ref{tab:chi2_alp118}.
The $\chi^2$ of individual $t\bar{t}$ data sets only change slightly with shifts in opposing directions depending on the specific data set.
The sum of $\chi^2$ from all $t\bar t$ data sets varies within 2 units as a consequence.
We conclude that varying $\alpha_s(M_Z)$ and $m_t$ away from their respective world averages
has little impact on extractions of SMEFT Wilson coefficients, provided these variations
remain within present uncertainties. 
Below, we present LM scans to further explore constraints on the SMEFT coefficients; for these, we perform joint fits of PDFs and Wilson
coefficients only, fixing $\alpha_s(M_Z) = 0.118$ and $m_t = 172.5$~GeV.
Lastly, we have also checked the impact of the resummed Coulomb corrections mentioned in Sec.~\ref{sec:the_cal}, and we
find these corrections have a negligible effect upon determinations of the Wilson coefficients.  

\subsection{Four-quark and gluonic operators}
\label{sec:tt_contact}
\begin{figure}[htbp]
  \centering
  \subcaptionbox{}[7.7cm]
    {\includegraphics[width=7.7cm]{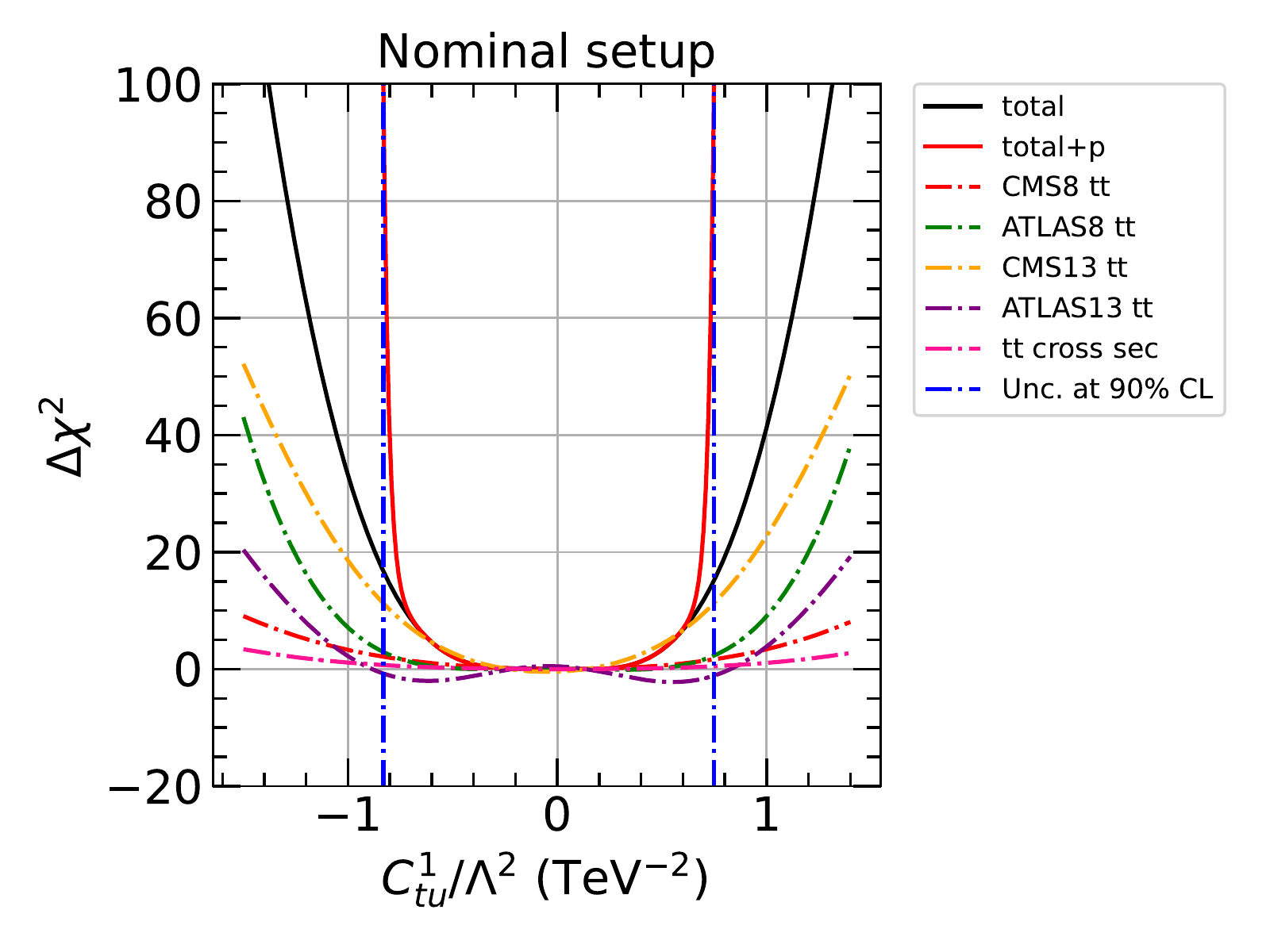}}
  \subcaptionbox{}[7.7cm]
    {\includegraphics[width=7.7cm]{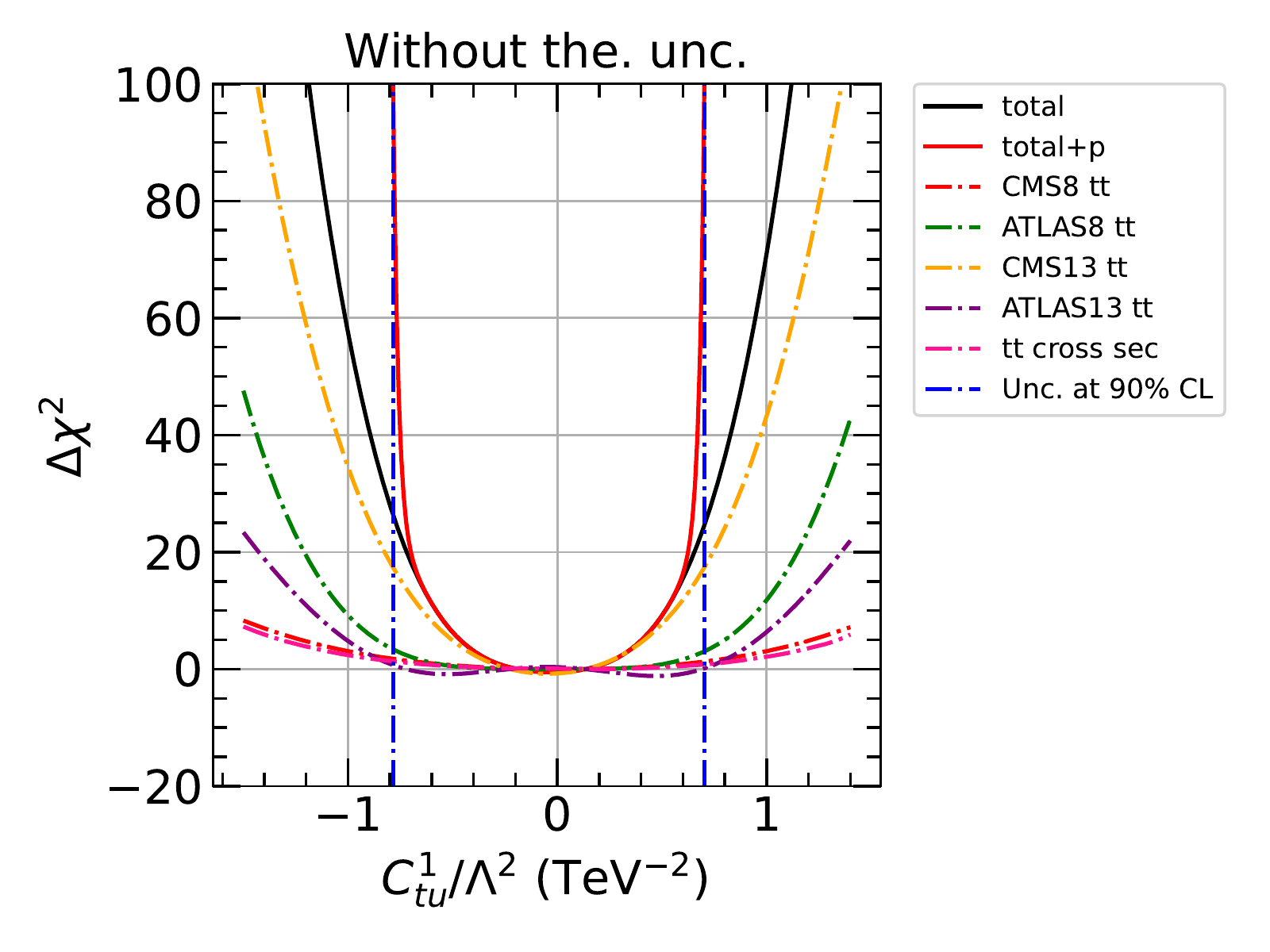}}
  \caption{
LM scans on $C_{tu}^{1}/\Lambda^2$ ($=\!C_{td}^{1}/\Lambda^2$) under the nominal fit (left panel) as well as excluding theoretical scale-choice uncertainties (right panel).
   The solid black and red lines represent $\Delta \chi^2$ and $\Delta \chi^2 + P$, respectively.
The dot-dashed curves represent the contributions to $\Delta \chi^2$ from individual experimental data sets.
The blue vertical dot-dashed lines indicate the 90\% CL uncertainties as determined by requiring $\Delta\chi^2 + P = 100$.}
  \label{Fig:np_c}
\end{figure}

We first show results for the four-quark operators, $O_{tu}^{1}$ and $O_{td}^{1}$.
We perform LM scans on a single effective Wilson coefficient, assuming $C_{tu}^{1}=C_{td}^{1}$, with all other SMEFT
coefficients set to zero as discussed earlier.
The profiled $\chi^2$ as a function of $C_{tu}^{1}$ is shown in Fig.~\ref{Fig:np_c}.
We find that both the global $\Delta\chi^2$ and the $\Delta\chi^2$ curves for individual experiments
show a predominantly quartic dependence on $C_{tu}^{1}$, which is expected since the interference
between the SM and the SMEFT operators, $O_{tu}^{1}$ and $O_{td}^{1}$, starts at NLO in QCD.

In the left panel of Fig.~\ref{Fig:np_c} we present the nominal calculation --- {\it i.e.}, with default
scale choices and uncertainties as discussed in Sec.~\ref{sec:theory} --- finding that the 13 TeV CMS $t\bar{t}$ data impose the strongest
constraint under this scenario, showing the most rapid growth in $\Delta \chi^2$, especially for larger values of $|C_{tu}^{1}|$. Intriguingly,
the 13 TeV ATLAS $t\bar{t}$ data suggest a tiny preference for nonzero $|C_{tu}^{1}/\Lambda^2|$, with a $\sim\!2$-unit $\Delta \chi^2$ dip
in the neighborhood of $|C_{tu}^{1}/\Lambda^2|\! \sim\! (0.5-0.6)$ TeV$^{-2}$ relative to the zero-SMEFT baseline; this, coupled with the comparatively slow growth in
$\Delta \chi^2$ seen for the other $t\bar{t}$ sets, has the effect of broadening the total uncertainty allowed for this SMEFT Wilson coefficient.
Still, the uncertainty range is mostly determined by the penalty term of this data set.
The LM scans ultimately predict a result of $C_{tu}^{1}/\Lambda^2 = 0.14^{+0.61}_{-0.97}$ TeV$^{-2}$ at 90\% CL, which is consistent with the SM.
Analogously, in the right panel we show the corresponding results determined without theoretical scale-choice uncertainties.
We find that the behaviors of both global $\Delta\chi^2$ and individual experimental $\Delta\chi^2$ are very similar to those shown in
the left panel, but with a modest increase in the takeoff of $\Delta \chi^2$, particularly in the tails of the profiled experiments.
This implies a slight reduction in the uncertainties for $C_{tu}^{1}/\Lambda^2$, as is to be expected.
In addition to Fig.~\ref{Fig:np_c}, we also perform LM scans on $C_{tu}^{1}$ under the scenario that
the PDF parameters are fixed to values at the global minimum.
This leads to $C_{tu}^{1}/\Lambda^2 = 0.14^{+0.60}_{-0.95}$ TeV$^{-2}$ at 90\% CL,
which is very close to the results with the nominal setup,
indicating that correlations between the PDFs and $C_{tu}^{1}$ are indeed weak when fitted to present data.

\begin{figure}[htbp]
  \centering
  \subcaptionbox{}[7.7cm]
    {\includegraphics[width=7.7cm]{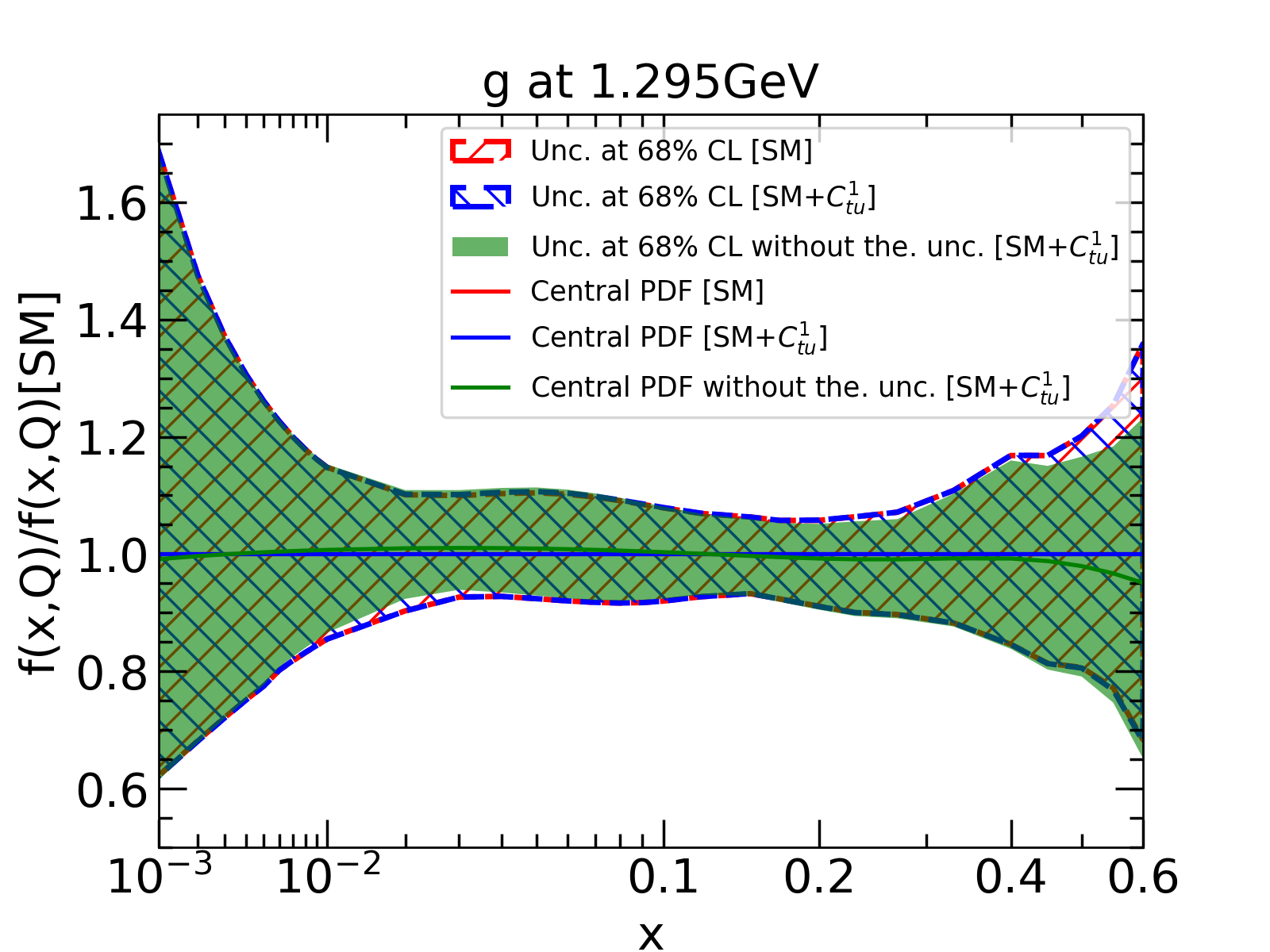}}
  \subcaptionbox{}[7.7cm]
    {\includegraphics[width=7.7cm]{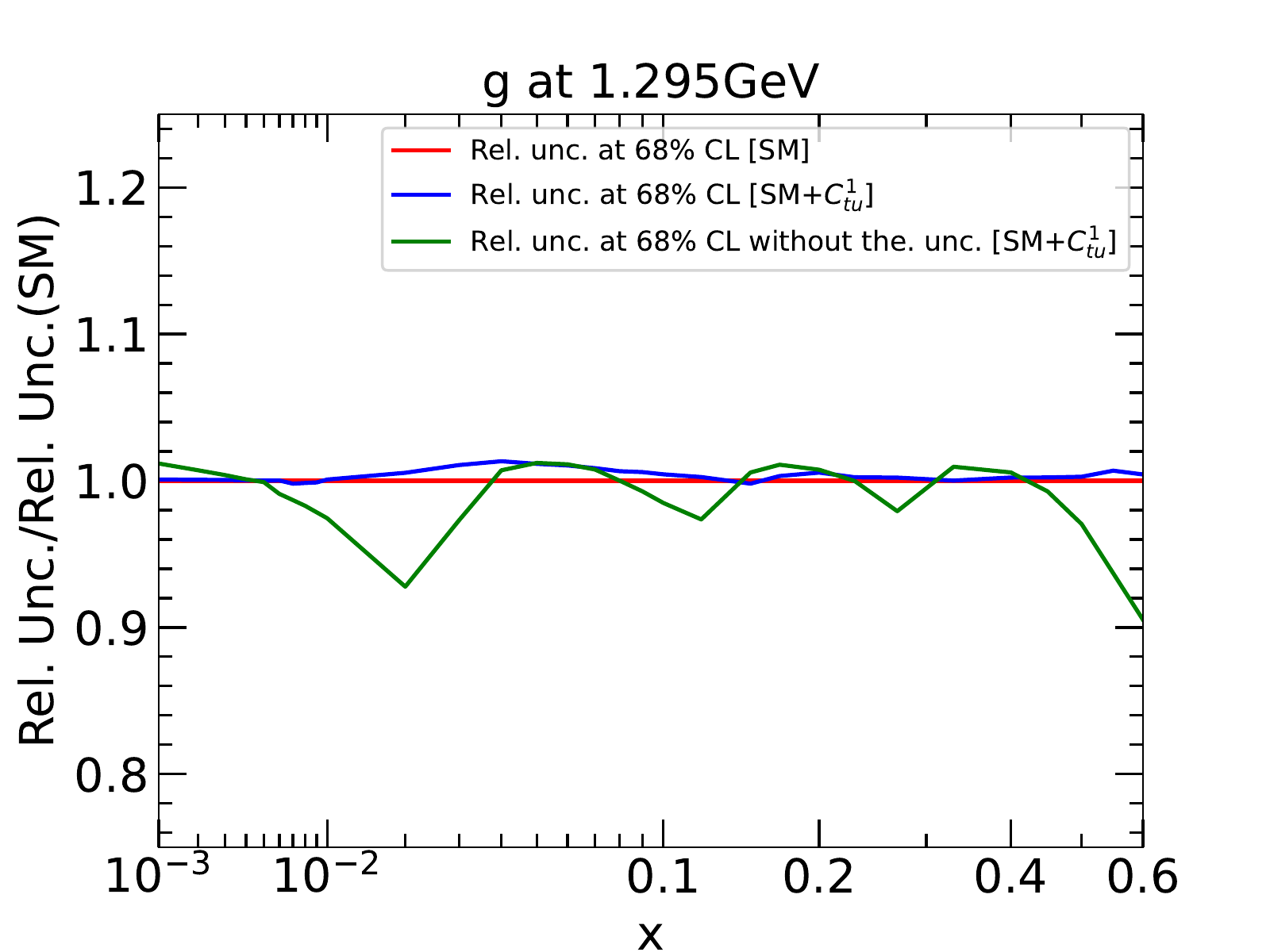}}    
  \caption{The gluon PDFs, $g(x,Q)$, determined by fitting with and without BSM contributions from $O_{tu}^{1}$ and $O_{td}^{1}$ at $Q = 1.295$ GeV are shown in the left panel. The blue and red solid lines represent the central values determined by fitting with and without SMEFT contributions respectively, and the green solid line represents the central value determined by fitting with SMEFT contributions and without theoretical uncertainties. The PDF uncertainties at 68\% CL are shown through hatched areas with relevant colors. The relative uncertainties are shown in the right panel with the same colors.}
  \label{Fig:c_g}
\end{figure}

In Fig.~\ref{Fig:c_g}, we compare the gluon PDF, $g(x,Q_0)$, at $Q_0$ = 1.295 GeV determined by fitting
with and without BSM SMEFT contributions from the inclusion of $O_{tu}^{1}\! =\! O_{td}^{1}$.
We also show the $g$-PDF determined without theoretical uncertainties as also explored in Fig.~\ref{Fig:np_c}.
In the left panel, the blue and red solid lines represent the central values of the gluon PDF determined with [SM+$C^1_{tu}$] and without [SM] nonzero
SMEFT contributions, respectively.
The green-solid line represents the central value of the $g$-PDF when determined in the presence of nonzero SMEFT but without theoretical uncertainties.
The PDF uncertainties at 68\% CL are shown as hatched areas in the various relevant colors.
We find that the fitted gluon PDFs obtained with and without BSM as parametrized by SMEFT are almost indistinguishable in terms of both the central value and uncertainty.
We note a very slight upward shift in the central value of the PDF, and corresponding $\sim\! 7\%$ reduction in the uncertainty, for $x\! \sim\! 0.02$ once
theoretical uncertainties are removed.
In addition, a slight downward shift in the central gluon PDF, of relative magnitude $\lesssim\!5\%$ and with a $\sim\! 10\%$ narrowing of the uncertainty,
occurs near $x\!\sim\! 0.5$.
As a companion plot, in the right panel of Fig.~\ref{Fig:c_g} we show the relative PDF uncertainties at 68\% CL for each of the curves discussed above, now
normalized to the nominal SM fit so as to more clearly illustrate the effect on the size of the PDF errors of incorporating SMEFT coefficients and (not) including theoretical
uncertainties.

\begin{figure}[htbp]
  \centering
  \subcaptionbox{PDF free}[7.7cm]
    {\includegraphics[width=7.7cm]{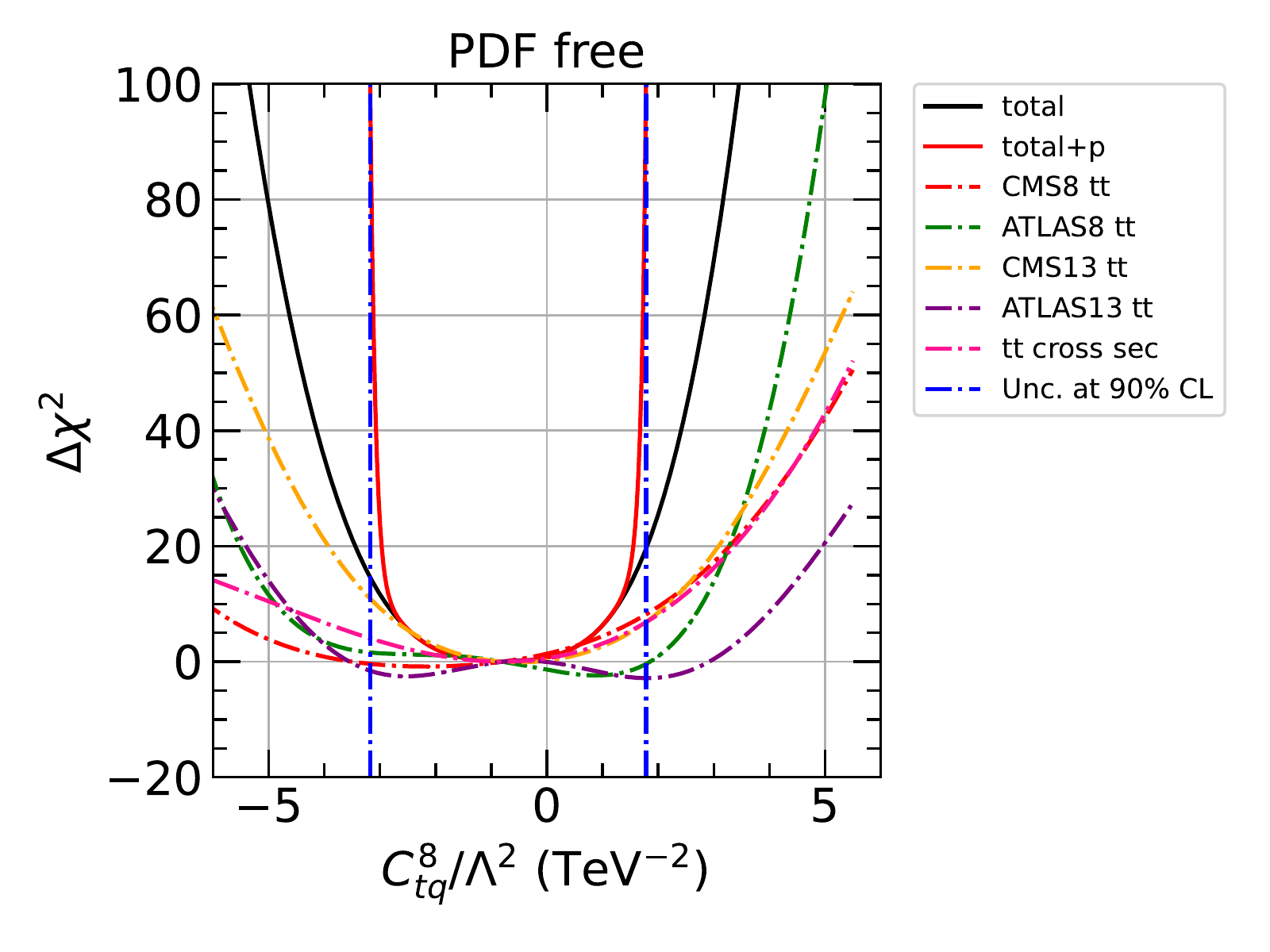}}
  \subcaptionbox{PDF fixed}[7.7cm]
    {\includegraphics[width=7.7cm]{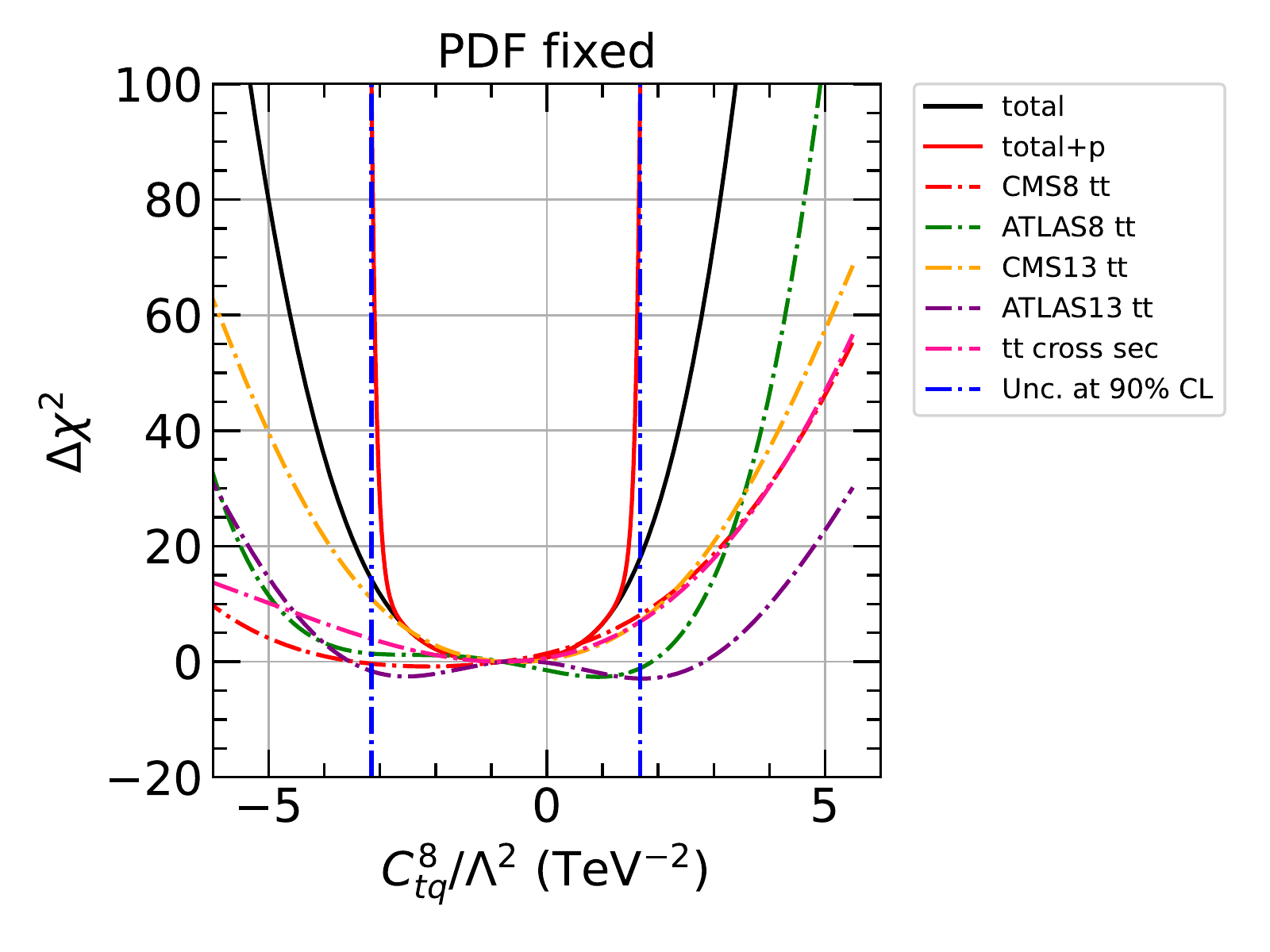}}
  \caption{LM scans on $C_{tq}^{8}/\Lambda^2$ with the PDF parameters allowed to free vary (left panel) or fixed at the global minimum (right panel).}
  \label{Fig:np_ctq8}
\end{figure}

We next turn our focus to another of the four-quark operators of Eq.~\ref{eq:top_SMEFT}, $O_{tq}^{8}$.
In Fig.~\ref{Fig:np_ctq8} we show the results of LM scans on $C_{tq}^{8}$.
As before, in the left panel we show the result of the calculation based on our nominal fit configuration, in this case finding strong constraints from the 8 and 13 TeV
CMS data as well as the total $t\bar{t}$ cross section measurements. 
In the nominal fit, we treat PDF parameters and EFT coefficients on the same footing and thus consistently obtain bounds on EFT coefficients with PDF uncertainties inside the framework of CT18.
As had been the case for $C_{tu}^{1}$, there is a very small hint of a preference for a nonzero
SMEFT Wilson coefficient from both the 8 and 13 TeV ATLAS data, but these largely lie at values suppressed by the penalty term of the likelihood function, which again
plays a decisive role in the full uncertainty on $C_{tq}^{8}/\Lambda^2$.
The LM scans predict a result of $C_{tq}^{8}/\Lambda^2 = -0.80^{+2.58}_{-2.38}$ TeV$^{-2}$ at 90\% CL.
In the right panel of Fig.~\ref{Fig:np_ctq8}, the PDF parameters are fixed to their values at the global minimum.
Hence, the PDF uncertainties are not included in the bounds on EFT coefficients.
The LM scans predict a result of $C_{tq}^{8}/\Lambda^2 = -0.80^{+2.48}_{-2.35}$ TeV$^{-2}$ at 90\% CL with
uncertainties slightly smaller than those shown in the left panel. As before, this suggests only very mild correlations
between SMEFT and PDF parameters at the current time, but with a mild possibility of slightly underestimating the
full uncertainty on SMEFT Wilson coefficients in analyses with fixed PDF degrees-of-freedom.

The SMEFiT collaboration presents a global interpretation of Higgs, diboson, and top-quark production and decay measurements from the LHC in the framework of the SMEFT~\cite{2105.00006}.
The 95\% CL bound associated with the one-parameter EFT fits for $C_{tq}^{8}/\Lambda^2$ is [-0.483, 0.393] TeV$^{-2}$.
In our PDF fixed case, the 95\% CL bound determined with the same parameter-fitting criterion as Ref.~\cite{2105.00006} is [-2.285, 0.701] TeV$^{-2}$, which is slightly weaker than the SMEFiT result.
The main reason for this is that the SMEFiT study included more experimental data sets with larger luminosity.
In this work, we only consider 5 data sets involving top-quark pair production, amounting to a total integrated luminosity of $\mathcal{L}\!\sim$111.9 fb$^{-1}$.
The total number of top quark-pair sets included in the SMEFiT study is 9, 
and the total integrated luminosity is $\sim\! 193$ fb$^{-1}$.
Also relevant is the fact that the scale uncertainties on the $t\bar{t}$ production cross sections were not considered in Ref.~\cite{2105.00006}.

\begin{figure}[htbp]
  \centering
  \subcaptionbox{}[7.7cm]
    {\includegraphics[width=7.7cm]{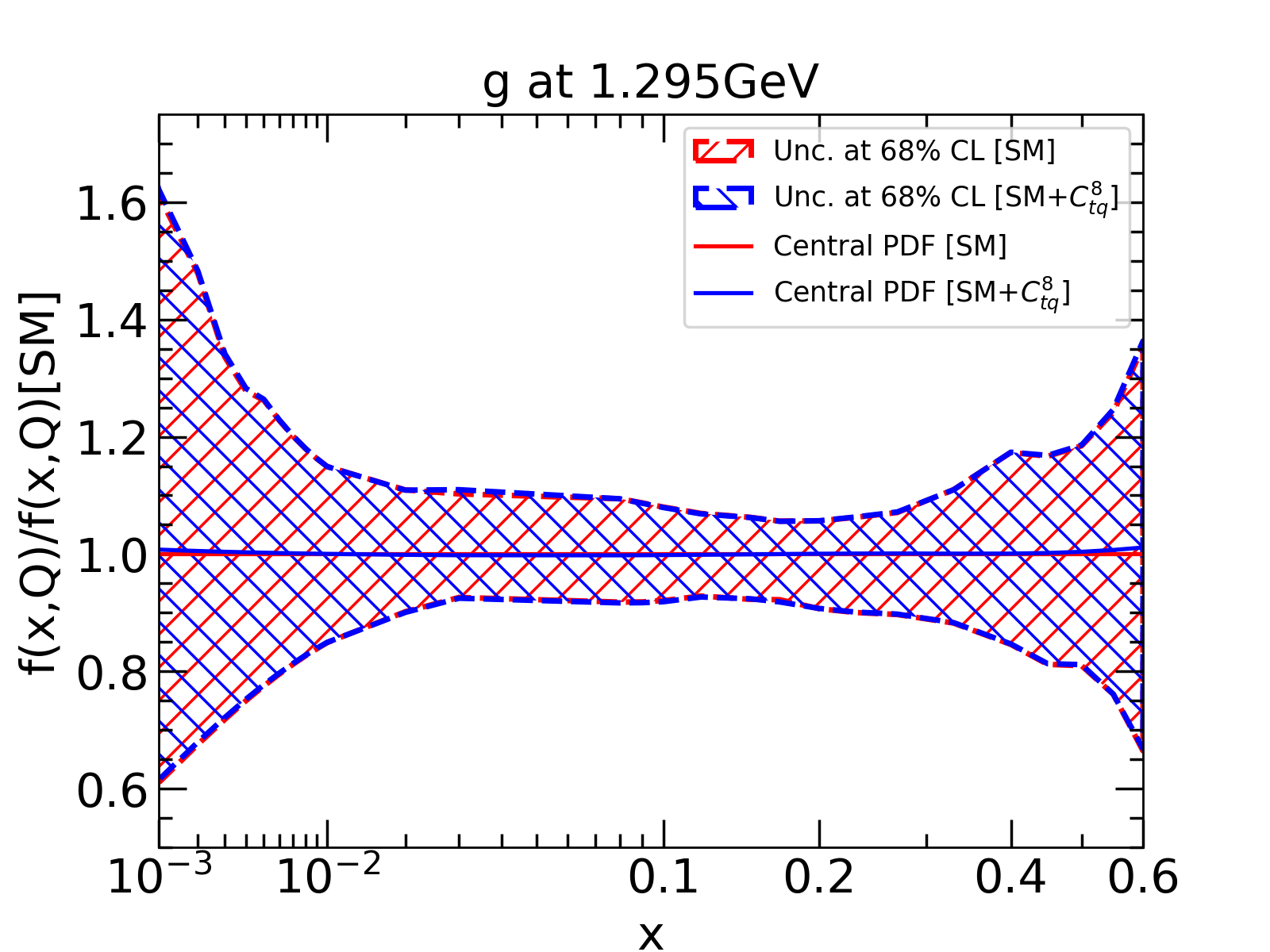}}
  \subcaptionbox{}[7.7cm]
    {\includegraphics[width=7.7cm]{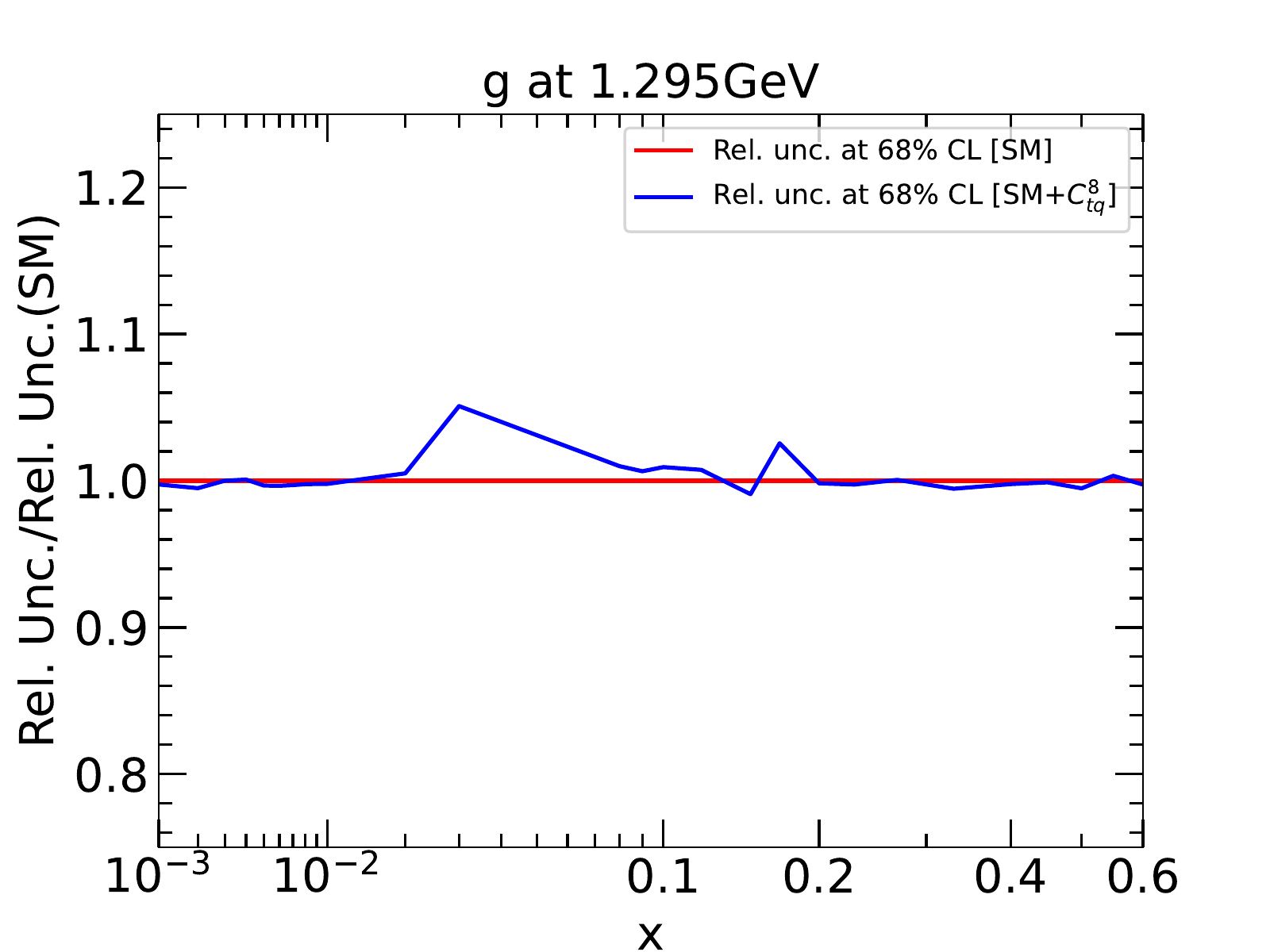}}
  \caption{The $g$ PDFs determined by fitting with and without SMEFT contributions from $O_{tq}^{8}$ at $Q_0 = 1.295$ GeV are shown in the left panel. The blue and red solid lines represent the central values determined by fitting with and without BSM contributions, respectively. The PDF uncertainties at the 68\% CL are shown through hatched areas in the relevant colors. The relative 68\%-level uncertainties, normalized to the purely SM scenario, are shown in the right panel in matching colors.}
  \label{Fig:ctq8_g}
\end{figure}

In Fig.~\ref{Fig:ctq8_g}, we compare gluon PDFs at $Q_0$ = 1.295 GeV as determined by fits with and without the freely-varying SMEFT contributions from
$O_{tq}^{8}$.
The PDF uncertainties at 68\% CL are shown through hatched areas with relevant colors.
In the left panel, we find that the PDFs from the two fits are almost indistinguishable for both the central value and the uncertainty region.
A negligible upward shift smaller than 1\% on the central value can be seen in the endpoint regions of $x \lesssim 3 \times 10^{-3}$ and $x\gtrsim 0.5$ after
including possible BSM contributions via the SMEFT coefficient.
In the right panel, the size of the relative PDF uncertainty is modestly enlarged at the $\sim\!5\%$ level around $x\!\sim\! 0.03$ following the inclusion the fitted
SMEFT coefficient.

\begin{figure}[htbp]
  \centering
  \subcaptionbox{}[7.7cm]
    {\includegraphics[width=7.7cm]{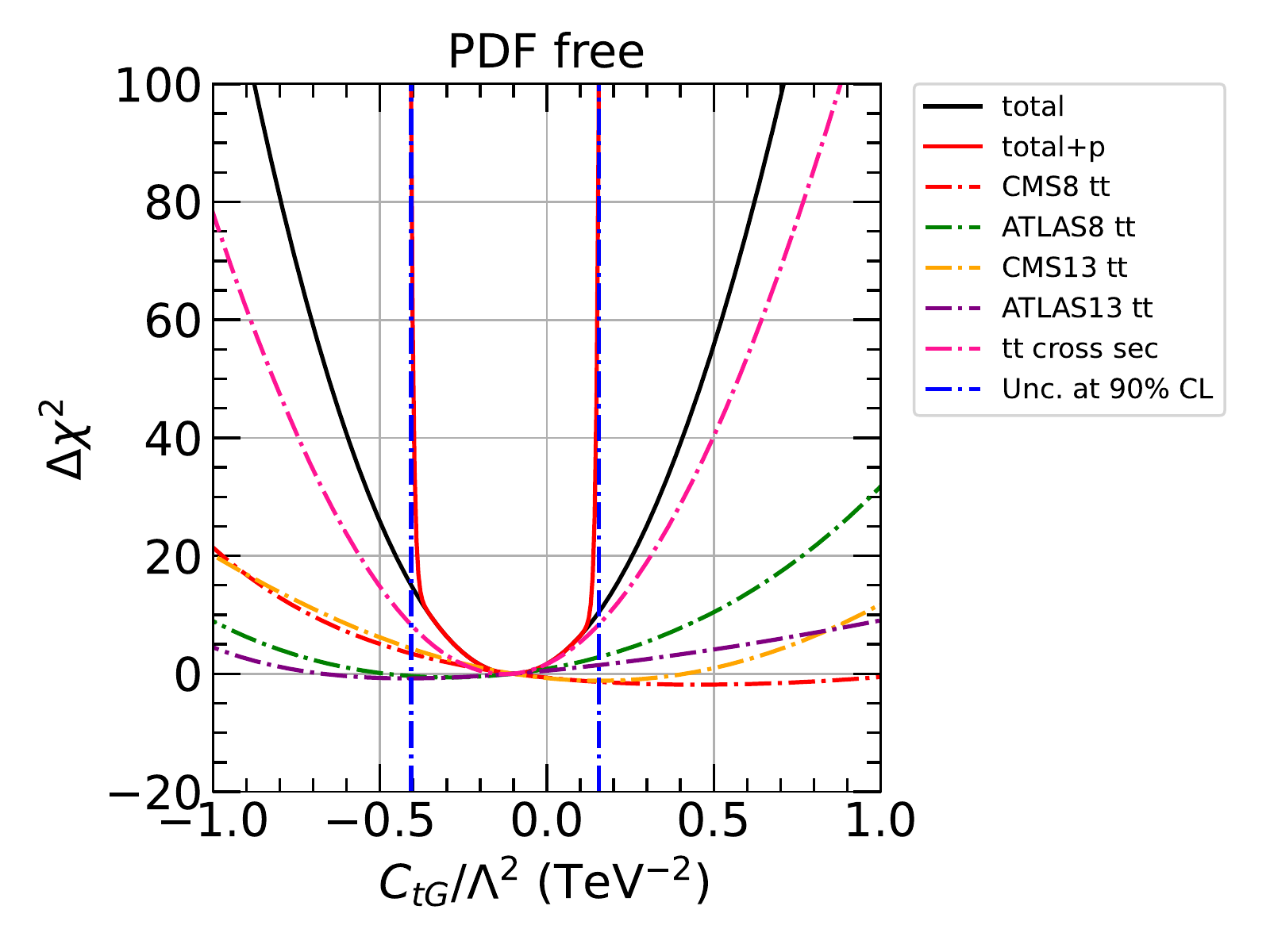}}
  \subcaptionbox{}[7.7cm]
    {\includegraphics[width=7.7cm]{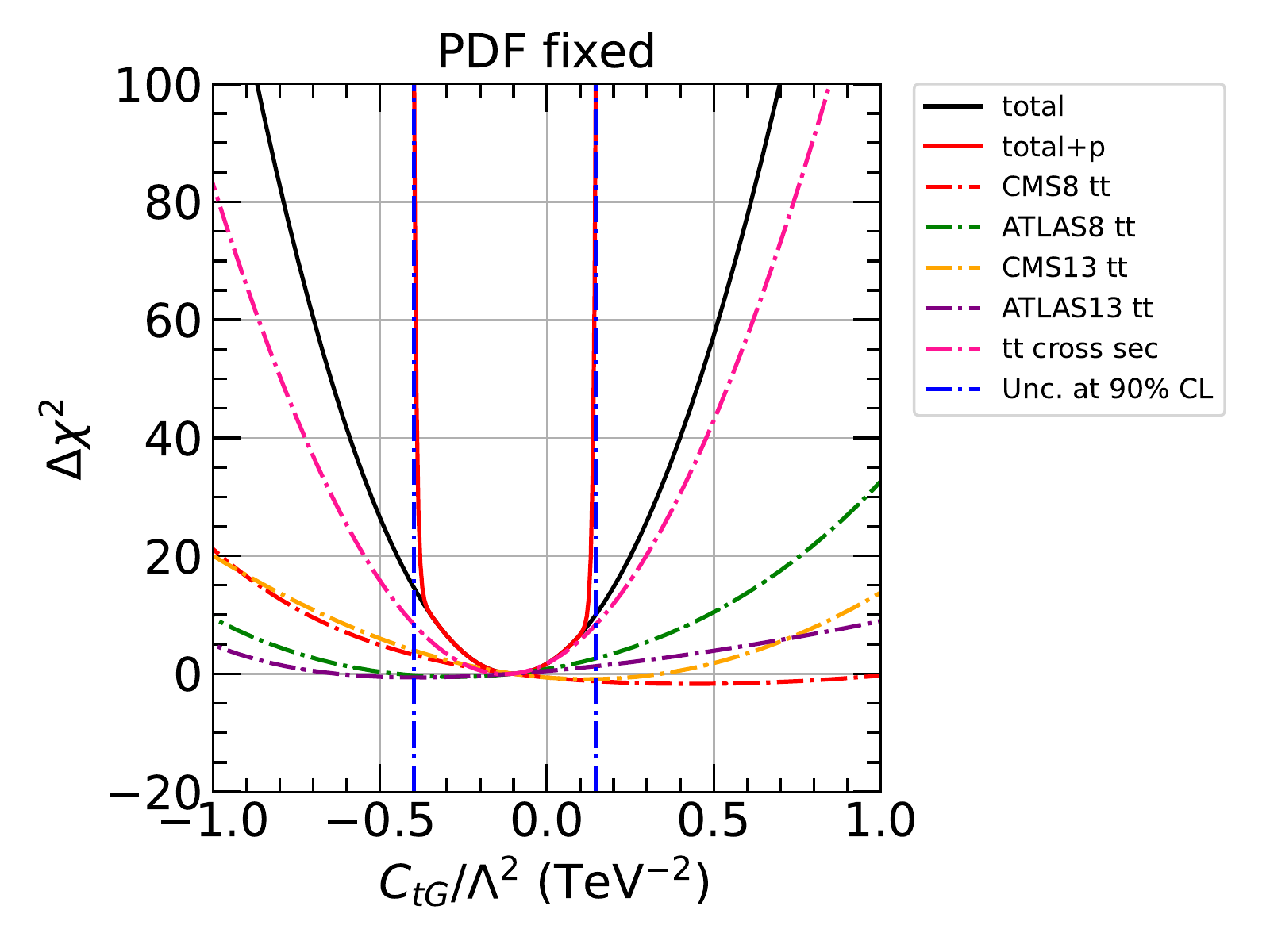}}
  \caption{LM scans over $C_{tG}/\Lambda^2$ with the PDF parameters freely varying (left panel) or fixed to their best-fit
	values (right panel).}
  \label{Fig:np_ctg}
\end{figure}

Lastly, we also consider the other top-relevant (gluonic) operator, $O_{tG}$, showing in Fig.~\ref{Fig:np_ctg} the analogous LM scan results
on $C_{tG}$.
For the nominal setup, given in the left panel, we find that the constraint from
the total cross section measurements predominate among the various fitted experiments;
as before, the uncertainty on $C_{tG}/\Lambda^2$ mostly comes from the penalty term of this data set, such that
the LM scans predict a result of $C_{tG}/\Lambda^2=-0.10^{+0.26}_{-0.30}$ TeV$^{-2}$ at 90\% CL.
In the right panel, the PDF parameters are fixed to their values at the global minimum.
The result in this case, $C_{tG}/\Lambda^2=-0.10^{+0.25}_{-0.30}$ TeV$^{-2}$ at 90\% CL,
is again closely to that obtained under the nominal setup.
This once again indicates a weak correlation between PDFs and $C_{tG}$ in the global fit.
We also note that, unlike the corresponding LM scans for $C^1_{tu}/\Lambda^2$ and $C^8_{tq}/\Lambda^2$
shown earlier, $\Delta \chi^2$ mostly grows monotonically away from the $C\!=\!0$ SM scenario, with
minima of only extremely shallow depth for, {\it e.g.}, the 8 and 13 TeV ATLAS $t\bar{t}$
experiments.
The SMEFiT collaboration reports a 95\% CL bound on $C_{tG}/\Lambda^2$ of [0.006, 0.107] TeV$^{-2}$ from the one-parameter EFT fits~\cite{2105.00006}.
In our case when using the same criterion for the PDF fixed case, we obtain a bound of [-0.255, 0.052] TeV$^{-2}$ at 95\% CL, which is again weaker for the reasons summarized before.

\begin{figure}[htbp]
  \centering
  \subcaptionbox{}[7.7cm]
    {\includegraphics[width=7.7cm]{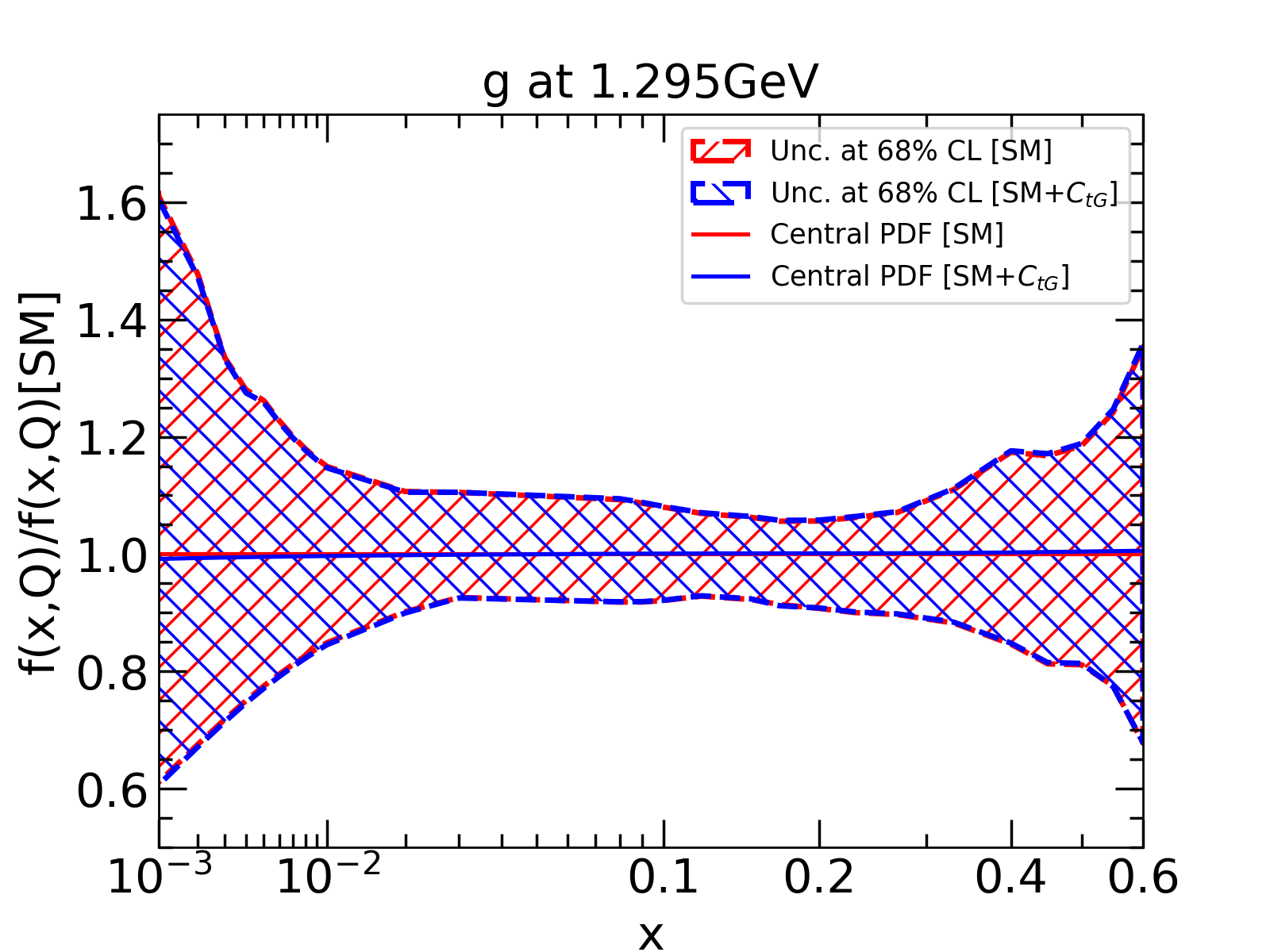}}
  \subcaptionbox{}[7.7cm]
    {\includegraphics[width=7.7cm]{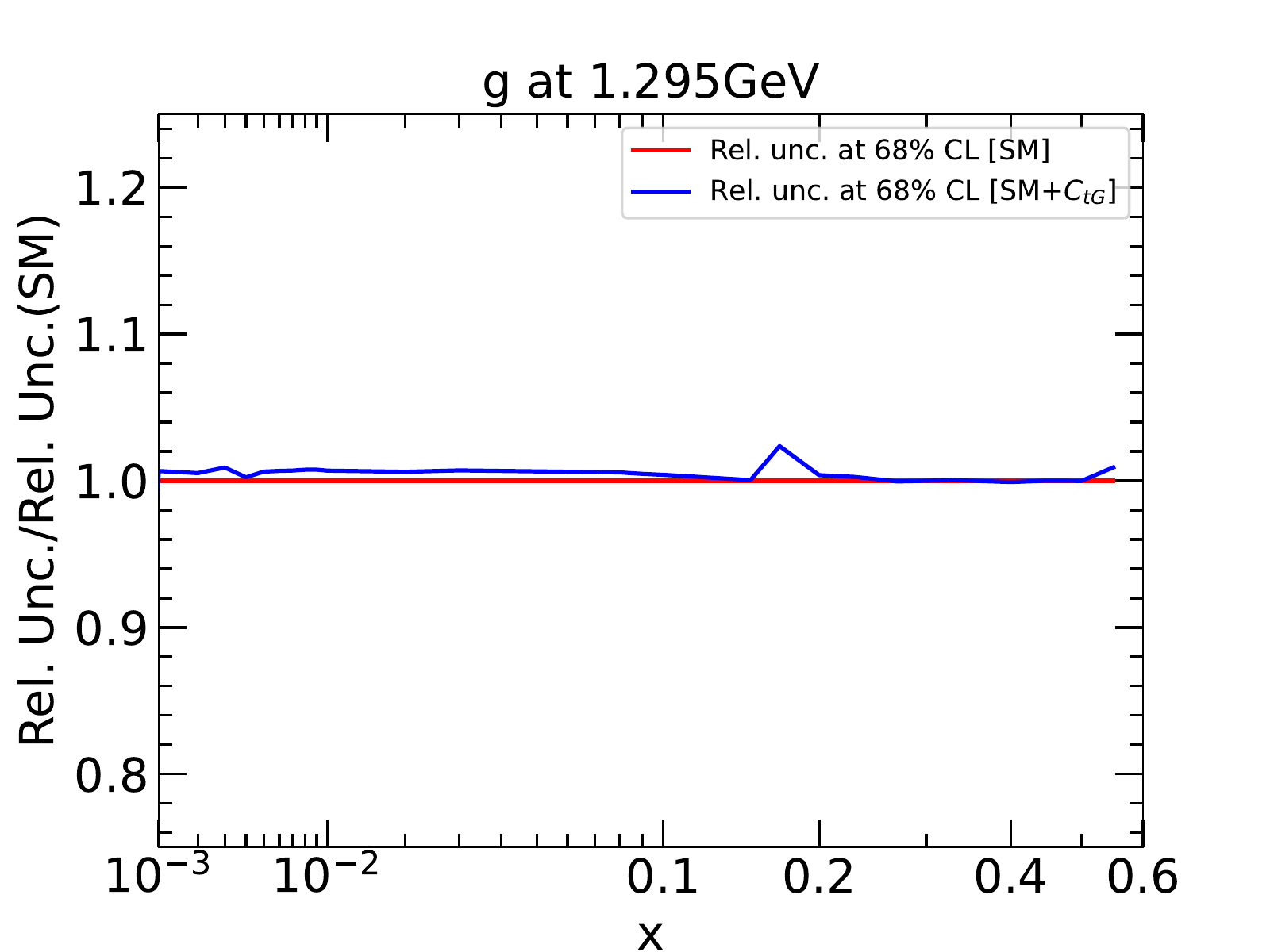}}
  \caption{The $g$-PDFs determined by fitting with and without SMEFT contributions via $O_{tG}$ at $Q_0 = 1.295$ GeV are shown in the left panel. The blue and red solid lines represent the central values determined by fitting with and without BSM contributions respectively. The PDF uncertainties at 68\% CL are shown through hatched areas with the corresponding colors. The relative uncertainties at 68\% CL are shown in the right panel, normalized to the SM curve.}
  \label{Fig:ctg_g}
\end{figure}

We summarize the results we obtain for all three Wilson coefficients under different fitted assumptions
in Tab.~\ref{tab:np_constraint}.
In Fig.~\ref{Fig:ctg_g}, we compare $g$ PDFs at $Q_0$ = 1.295 GeV determined by fitting with and
without SMEFT contributions from $O_{tG}$, finding that these fits are essentially indistinguishable.
We therefore conclude that, at present, SMEFT-PDF correlations in the $t\bar{t}$ sector are effectively absent
for $O^1_{tu}$ (=$O^1_{td}$) and $O_{tG}$, while nonzero but very weak for the octet operator, $O^8_{tq}$.

\begin{table}[htpb]
  \centering
  \begin{tabular}{c|ccc}
  \hline
   TeV$^{-2}$& nominal & PDF fixed & no the. unc.\\
  \hline
  $C_{tu}^{1}/\Lambda^2$ & $0.14^{+0.61}_{-0.97}$ & $0.14^{+0.60}_{-0.95}$ & $0.14^{+0.57}_{-0.92}$ \\
    \hline
  $C_{tq}^{8}/\Lambda^2$ & $-0.80^{+2.58}_{-2.38}$ & $-0.80^{+2.48}_{-2.35}$ & -\\
  \hline
  $C_{tG}/\Lambda^2$ & $-0.10^{+0.26}_{-0.30}$ & $-0.10^{+0.25}_{-0.30}$ & -\\
  \hline
  \end{tabular}
  \caption{Constraints on new physics at 90\% CL.} 
  \label{tab:np_constraint}
\end{table}

We evaluate the constraints on the Wilson coefficients from each $t\bar t$ experiment
by repeating the global fit, retaining only a single data set at a time.
The results of this procedure are listed in Tab.~\ref{tab:np_individual}.
The second column shows the results obtained through the full data set fitted under nominal settings.
For $C_{tu}^{1}/\Lambda^2$, given in the first row, the CMS 13 TeV data the give strongest constraint,
consistent with the $\Delta \chi^2$ profiles shown in Fig.~\ref{Fig:np_c}.
In addition, the ATLAS 8 and 13 TeV data both prefer a positive $C_{tu}^{1}/\Lambda^2$.
For $C_{tq}^{8}/\Lambda^2$ (second row), the total cross section measurements and the CMS 13 TeV data give the strongest constraints.
The ATLAS 8 and 13 TeV data both prefer a positive $C_{tq}^{8}/\Lambda^2$, in contrast to the other three data sets, which all prefer negative values.
Finally, for $C_{tG}/\Lambda^2$, appearing in the last row, the total cross section measurements predominate over the other data sets.
Both the 8 and 13 TeV ATLAS data prefer negative values of $C_{tG}/\Lambda^2$, while both the CMS 8 and 13 TeV data prefer positive ones.

\begin{table}[htpb]
  \centering
  \begin{tabular}{c|c|ccccc}
  \hline
  TeV$^{-2}$ & nominal & tot. cross sect. & CMS 8 & ATLAS 8 & CMS 13 & ATLAS 13  \\
  \hline
  $C_{tu}^{1}/\Lambda^2$  & $0.14^{+0.61}_{-0.97}$ & $0.0^{+1.84}_{-1.84}$ & $0.01^{+1.33}_{-1.39}$ & $0.35^{+0.71}_{-1.46}$ & $-0.05^{+0.76}_{-0.75}$ & $0.54^{+0.58}_{-1.72}$  \\
  \hline
  $C_{tq}^{8}/\Lambda^2$ & $-0.80^{+2.58}_{-2.38}$ & $-0.81^{+2.59}_{-3.40}$ & $-2.16^{+3.74}_{-3.51}$ & $0.92^{+1.88}_{-5.85}$ & $-0.57^{+2.71}_{-2.58}$ & $1.72^{+2.07}_{-6.16}$  \\
  \hline
  $C_{tG}/\Lambda^2$&  $-0.10^{+0.26}_{-0.30}$ & $-0.13^{+0.28}_{-0.28}$ & $0.43^{+1.45}_{-1.02}$ & $-0.28^{+0.82}_{-0.79}$ & $0.12^{+0.75}_{-0.78}$ & $-0.38^{+1.30}_{-0.74}$  \\
  \hline
  \end{tabular}
  \caption{Constraints on the SMEFT Wilson coefficients at the 90\% CL from the individual $t\bar{t}$ data sets examined in this study.} 
  \label{tab:np_individual}
\end{table}

\begin{figure}[htbp]
  \centering
  \includegraphics[width=0.495\textwidth,clip]{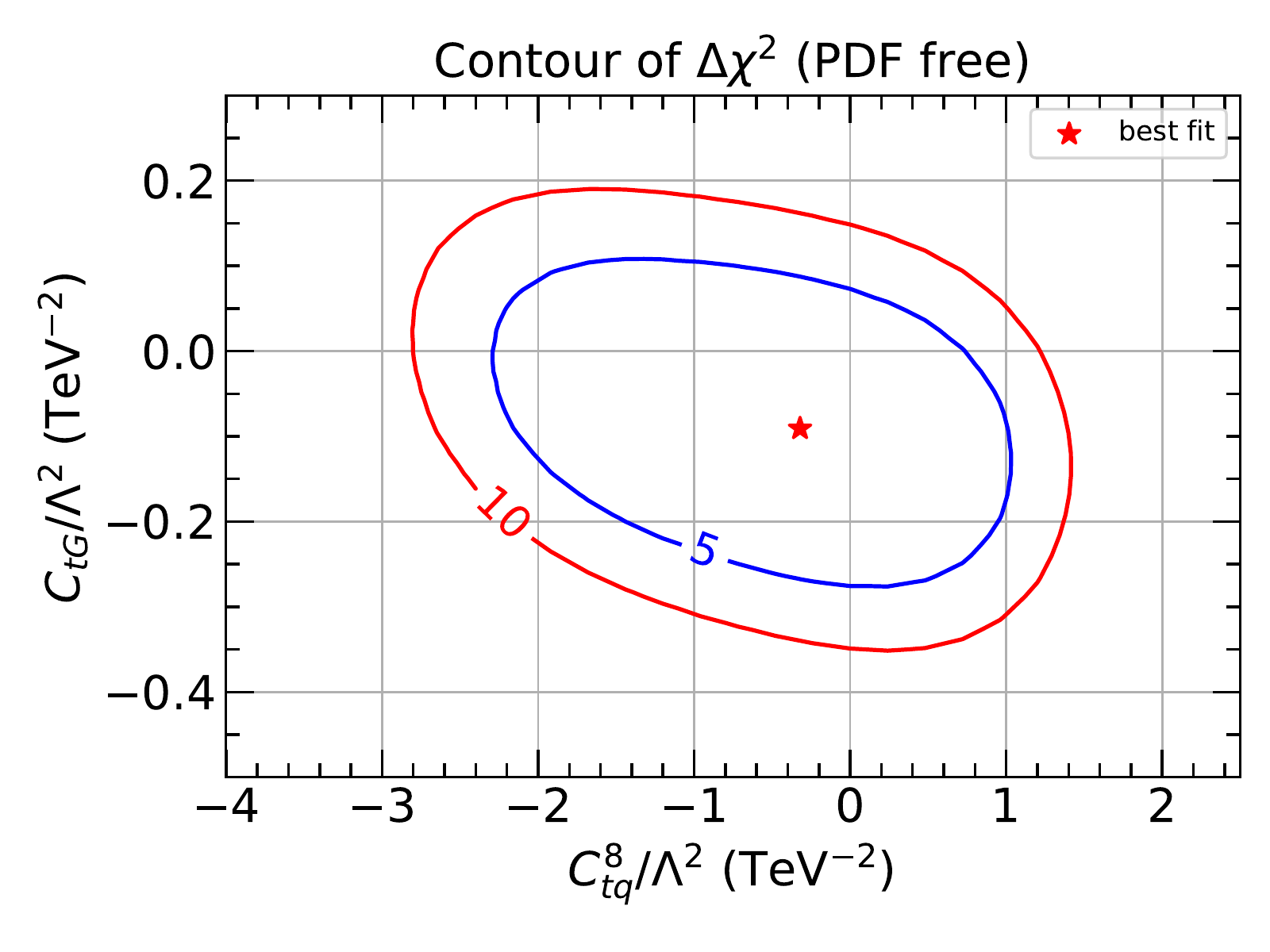}
  \includegraphics[width=0.495\textwidth,clip]{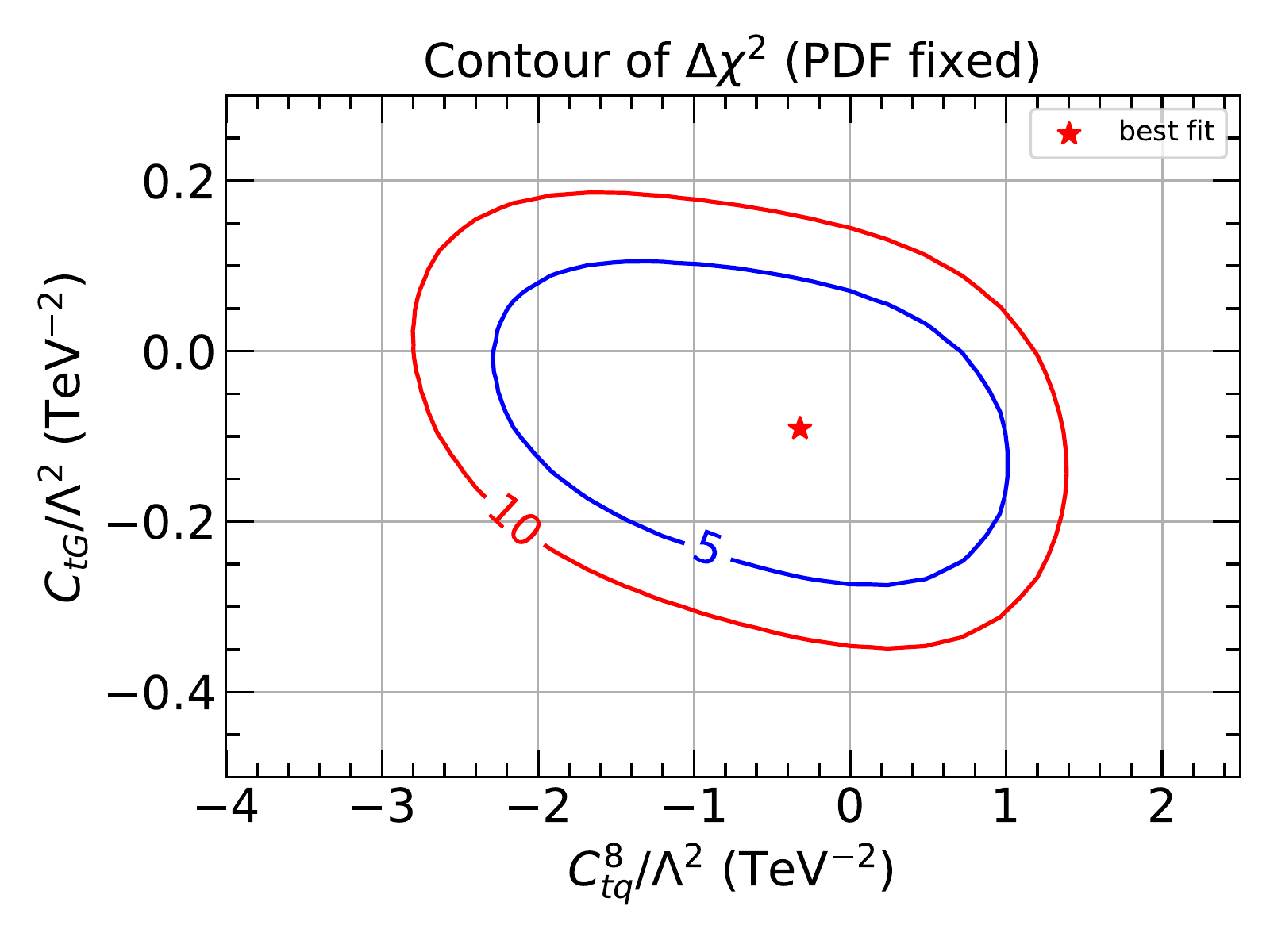}
	\caption{Contour plot of $\Delta \chi^2$ on the plane of $C_{tq}^8/\Lambda^2$ vs.~$C_{tG}/\Lambda^2$ with the PDF parameters freely varying (left panel) or fixed to their best-fit values (right).
	}
  \label{Fig:2d_ctq8_ctg}
\end{figure}

We further study the interplay between the Wilson coefficients relevant for $O_{tq}^8$ and $O_{tG}$.
In this case, the new NNs are built by adding both $C_{tq}^8/\Lambda^2$ and $C_{tG}/\Lambda^2$ into the input layer.
With the new NNs, an association between \{PDFs, $C_{tq}^8/\Lambda^2$, $C_{tG}/\Lambda^2$\} and $\chi^2$ is constructed.
The interference between $O_{tq}^8$ and $O_{tG}$ is not considered here for simplicity.
We test the possible correlations between $C_{tq}^8$ and $C_{tG}$ through simultaneous fits of PDFs, $C_{tq}^8$, and $C_{tG}$.
We perform 2D LM scans on $C_{tq}^8$ and $C_{tG}$, and the results are shown in Fig.~\ref{Fig:2d_ctq8_ctg}.
The blue and red contours represent surfaces of constant $\Delta\chi^2 = 5$ and 10, respectively.
In the left panel, the shape of the contours shows a moderate correlation between $C_{tq}^8$ and $C_{tG}$.
In the right panel, the PDF parameters are fixed to their values at the global minimum.
The contours are slightly narrower than those shown in the left panel, 
which indicates weak correlations between SMEFT and PDF parameters.
For the current framework, in principle we can carry out a global marginalised analysis with more Wilson coefficients (for instance, tens of parameters) fitted simultaneously, since the current NNs already have many more inputs than this.
Furthermore, the dependence of the $\chi^2$ on 
the EFT coefficients is much simpler in general.
It is also possible to proceed along the lines of Ref.~\cite{2211.02058}; namely, separating different combinations of EFT coefficients in $\chi^2$, and constructing and training a NN for each of the resulting terms.  

\section{LM scans with jet production}
\label{sec:jet_contact}
Following the exploration of $t\bar{t}$ data and top-associated SMEFT operators in the section above, we now turn our attention
to the determination of the Wilson coefficient $C_{1}$ from Eq.~(\ref{eq:CI}) via LM scans with a special focus on jet production measurements.
In addition, we study the interplay between the Wilson coefficients primarily associated with jet production ($C_1$) and
the top-associated Wilson coefficient most correlated with the gluon PDF ($C_{tG}$).

\subsection{Contact interactions}
\label{sec:jet_contact_inter}
\begin{figure}[htbp]
  \centering
  \subcaptionbox{}[7.7cm]
    {\includegraphics[width=7.7cm]{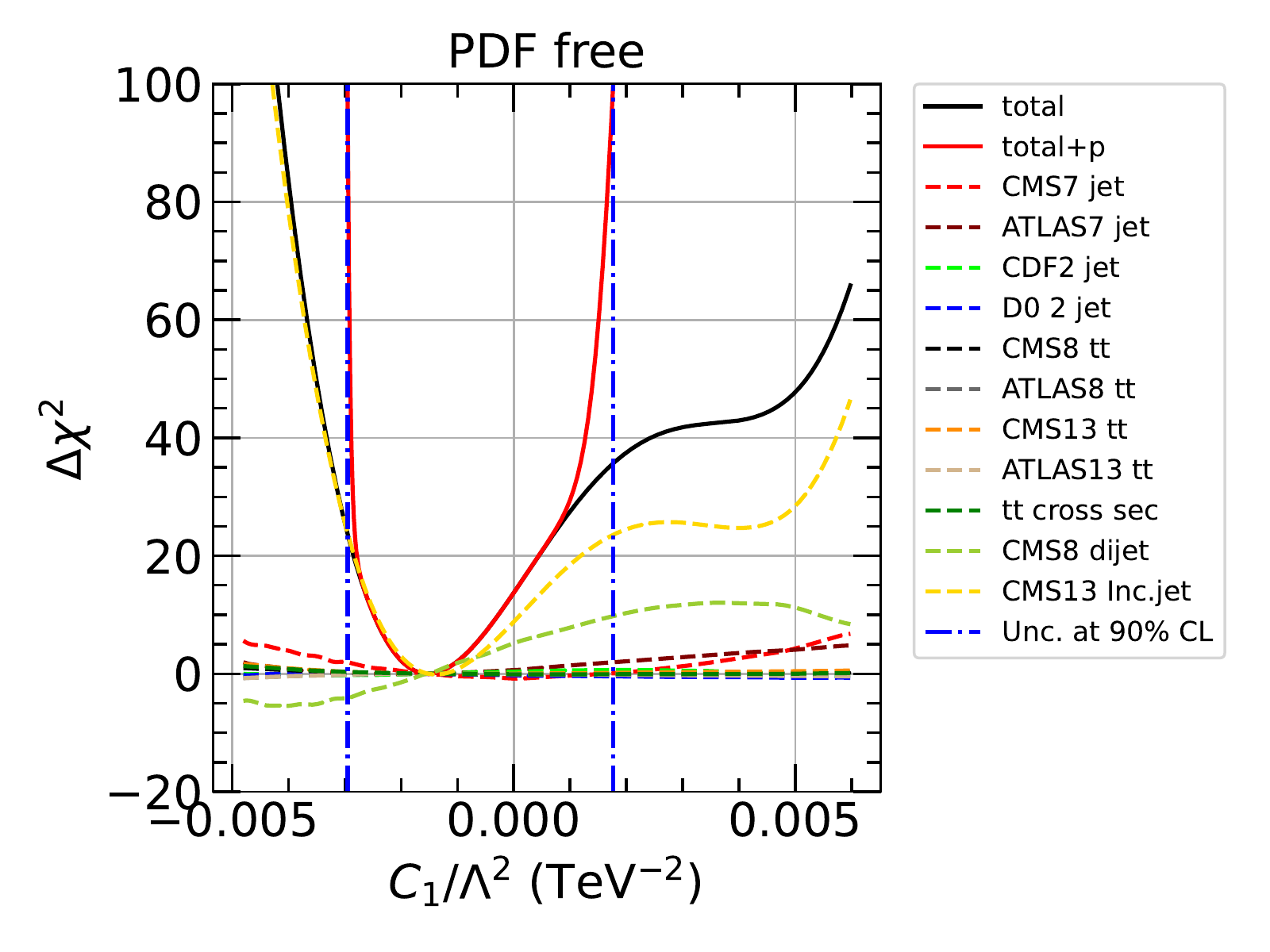}}
  \subcaptionbox{}[7.7cm]
    {\includegraphics[width=7.7cm]{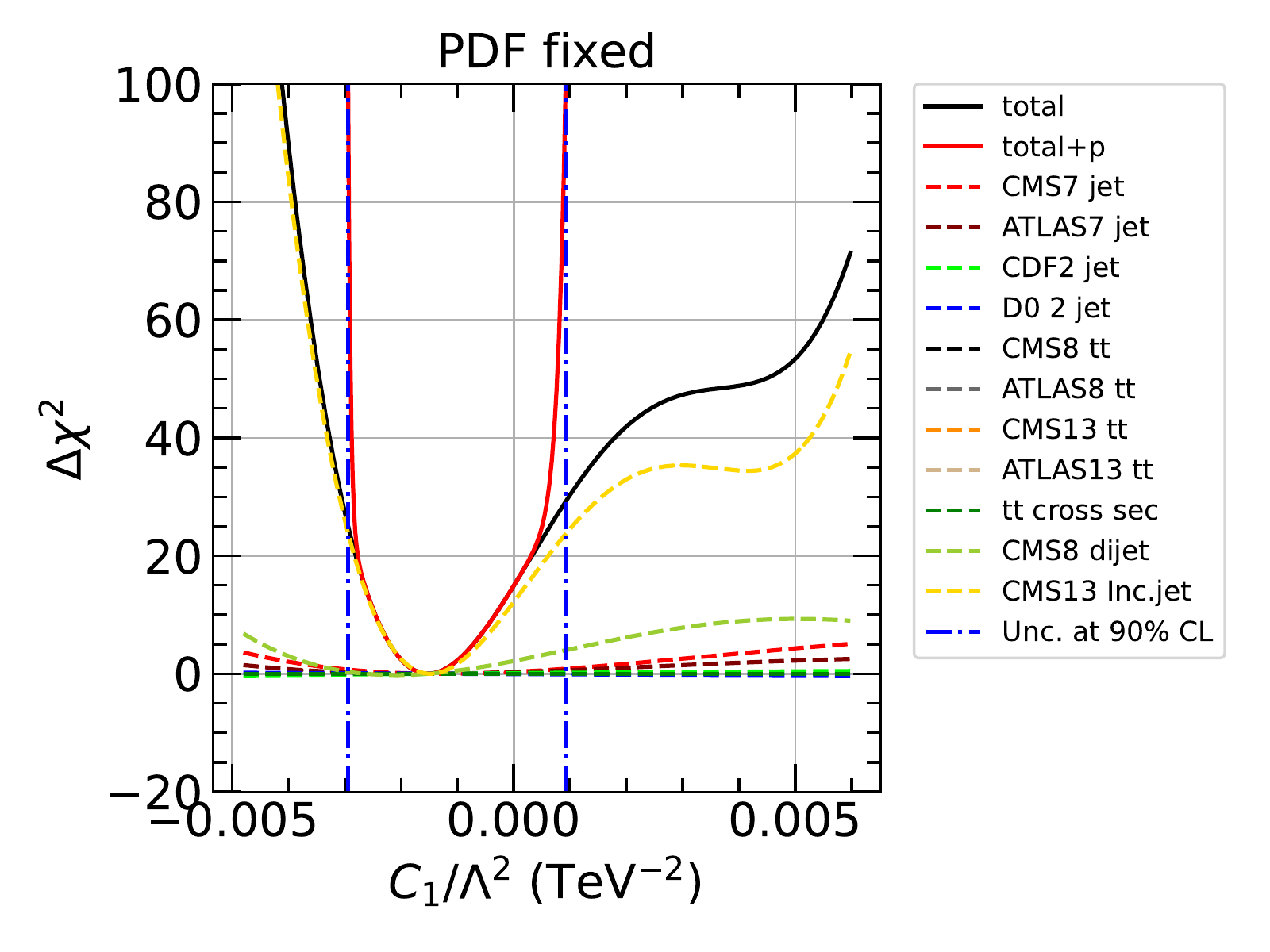}}
	\caption{LM scans over $C_{1}/\Lambda^2$ in our nominal setup allowing the PDF parameters to freely float (left panel)
	or be fixed at the global minimum (right panel).}
  \label{Fig:np_jet5}
\end{figure}

For the additional studies shown below, we include 6 jet production data sets in our nominal fits as indicated in Tab.~\ref{tab:exp},
including 3D distributions from the CMS 8 TeV dijet measurement and 2D distributions from the CMS 13 TeV inclusive jet measurement.
In a variant fit, the CMS 8 TeV dijet data are replaced by corresponding data on inclusive jet production. 
Furthermore, we include all top-quark pair production experiments used in the nominal fits of Sec.~\ref{sec:tt_np} above as well as the
other 32 baseline DIS/DY data sets, such the fits here represent the fullest accumulation of data considered in this work.
We note, however, that these other experiments do not directly constrain the contact-interaction Wilson coefficient, $C_1$.
In keeping with our nominal choices, the values of $\alpha_s(M_Z)$ and $m_t$ are set to their respective world averages, $\alpha_s(M_Z) = 0.118$ and $m_t = 172.5$ GeV.
Also, contributions from the other Wilson coefficients associated with top production are not included here unless otherwise specified.
The results of the LM scans over $C_{1}$ according to our nominal setup are shown in Fig.~\ref{Fig:np_jet5},
where the left panel shows the CMS 13 TeV inclusive jet data and CMS 8 TeV dijet data to have the tightest 
constraint, in addition to exhibiting a more subtle nonlinear dependence on $C_{1}/\Lambda^2$ as one moves away
from the best fit.
As in the previous section, the uncertainty range is mostly determined by the penalty term of these two leading data sets.
In comparison, the sensitivity of the other jet data to $C_1$ is much weaker. 
Much as expected, there are almost no constraints from the data sets on top-quark pair production.
The LM scans predict $C_{1}/\Lambda^2 = -0.0015^{+0.0033}_{-0.0014}$ TeV$^{-2}$ at 90\% CL,
consistent with the SM.
In the right panel, the PDF parameters are fixed to their values at the global minimum.
We find that the behaviors of both global $\Delta\chi^2$ and individual $\Delta\chi^2$ are
very similar to that shown in the left panel.
The LM scans predict a result of $C_{1}/\Lambda^2 = -0.0015^{+0.0024}_{-0.0014}$ TeV$^{-2}$ at 90\% CL
which has smaller uncertainties comparing with including PDF variations.

The total values of $\chi^2$, as well as those for individual jet data sets, are listed
in Tab.~\ref{tab:chi2_jet} for the global minimum determined with and without the BSM SMEFT contributions.
We find that the inclusion of SMEFT contributions causes the global $\chi^2$ to diminish by 13.7 units.
Of this, the individual $\chi^2$ values for the CMS 8 TeV dijet data and 13 TeV inclusive
jet data are reduced by 5.2 and 8.8 units, respectively, which is consistent with the left
panel of Fig.~\ref{Fig:np_jet5}.
Meanwhile, the individual $\chi^2$ values of the other jet data sets are largely unaltered.
Despite this apparent insensitivity, these experiments are important nonetheless for pinning down 
uncertainties in the gluon PDF and thus reducing correlations between the Wilson coefficient and PDFs
in the global analyses.
Notably, in the right panel of Fig.~\ref{Fig:np_jet5}, we see evidence of somewhat more significant PDF-SMEFT correlations,
a feature which can be deduced by comparing the dependence of the $\Delta \chi^2$ profiles on $C_1/\Lambda^2$; in
particular, the size and shape of the $\Delta \chi^2$ curves for the CMS 8 TeV dijet and 13 TeV inclusive jet experiments
are noticeably modified near $C_1/\Lambda^2\! \sim\!0.001\!$ TeV$^{-2}$ once PDF parameters are frozen in the right panel. These modifications
lead to a moderate increase in the growth of $\Delta \chi^2$ at higher $C_1/\Lambda^2$ and a corresponding underestimate in
the Wilson coefficient when not simultaneously fitted alongside the PDFs.

\begin{table}[htpb]
  \centering
  \begin{tabular}{c|cccccc|c}
  \hline
  $\chi^2$ (nominal) & D0 & CDF & ATLAS 7 & CMS 7 & CMS 8 & CMS 13 & global\\
  \hline
  $C_{1}=0$ & 112.91 & 113.35 & 198.90 & 203.95 & 184.15 & 119.55 & 4388.93 \\
  \hline
  $C_{1}$ free & 113.21 & 113.01 & 198.29 & 204.77 & 178.96 & 110.77 & 4375.23\\
  \hline
  \end{tabular}
  \caption{The total $\chi^2$ and the $\chi^2$ for individual jet experiments at the global minimum determined with and without SMEFT contributions parametrized by
	$C_1$.} 
  \label{tab:chi2_jet}
\end{table}

\begin{figure}[htbp]
  \centering
  \subcaptionbox{}[7.7cm]
    {\includegraphics[width=7.7cm]{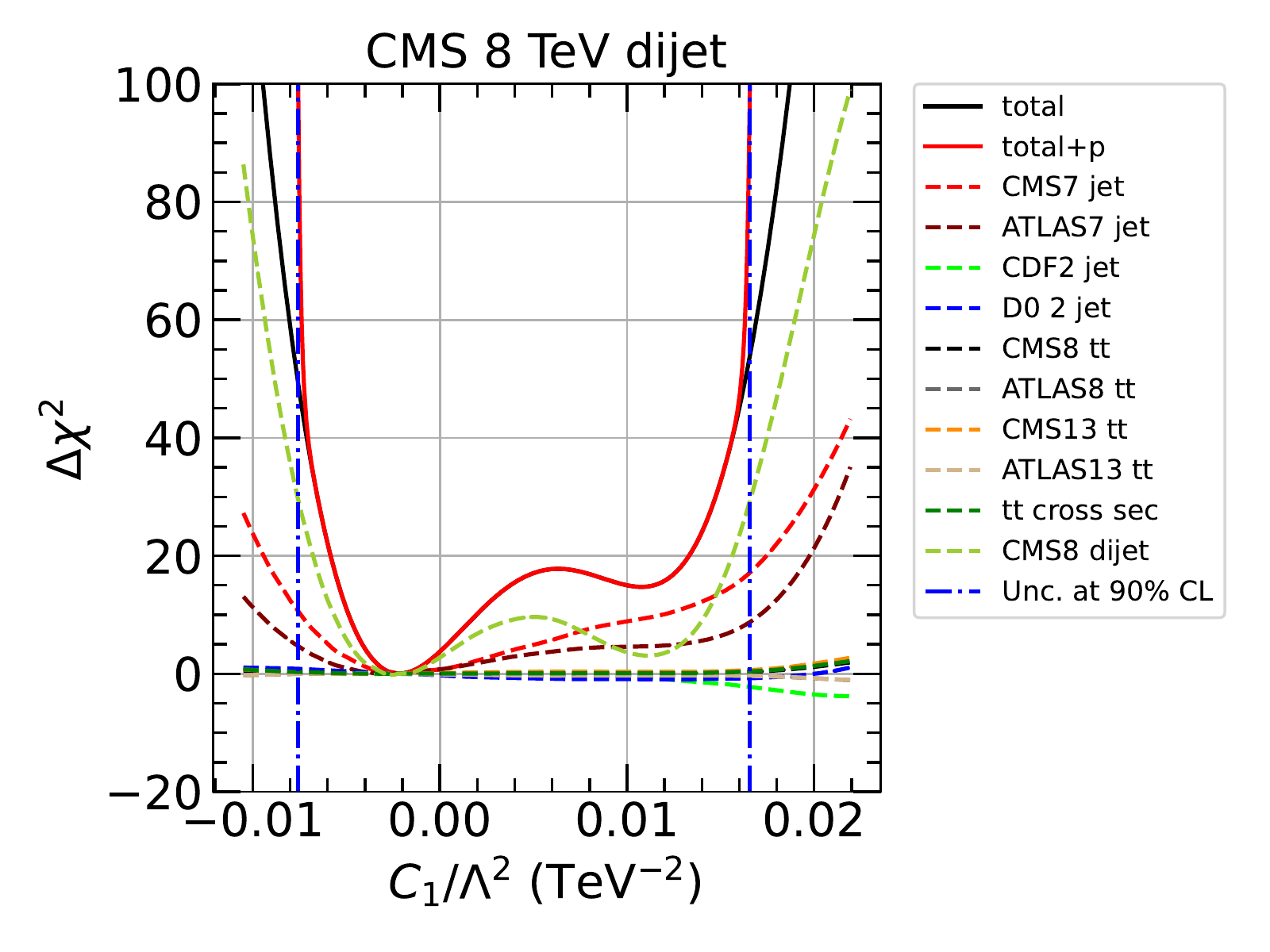}}
  \subcaptionbox{}[7.7cm]
    {\includegraphics[width=7.7cm]{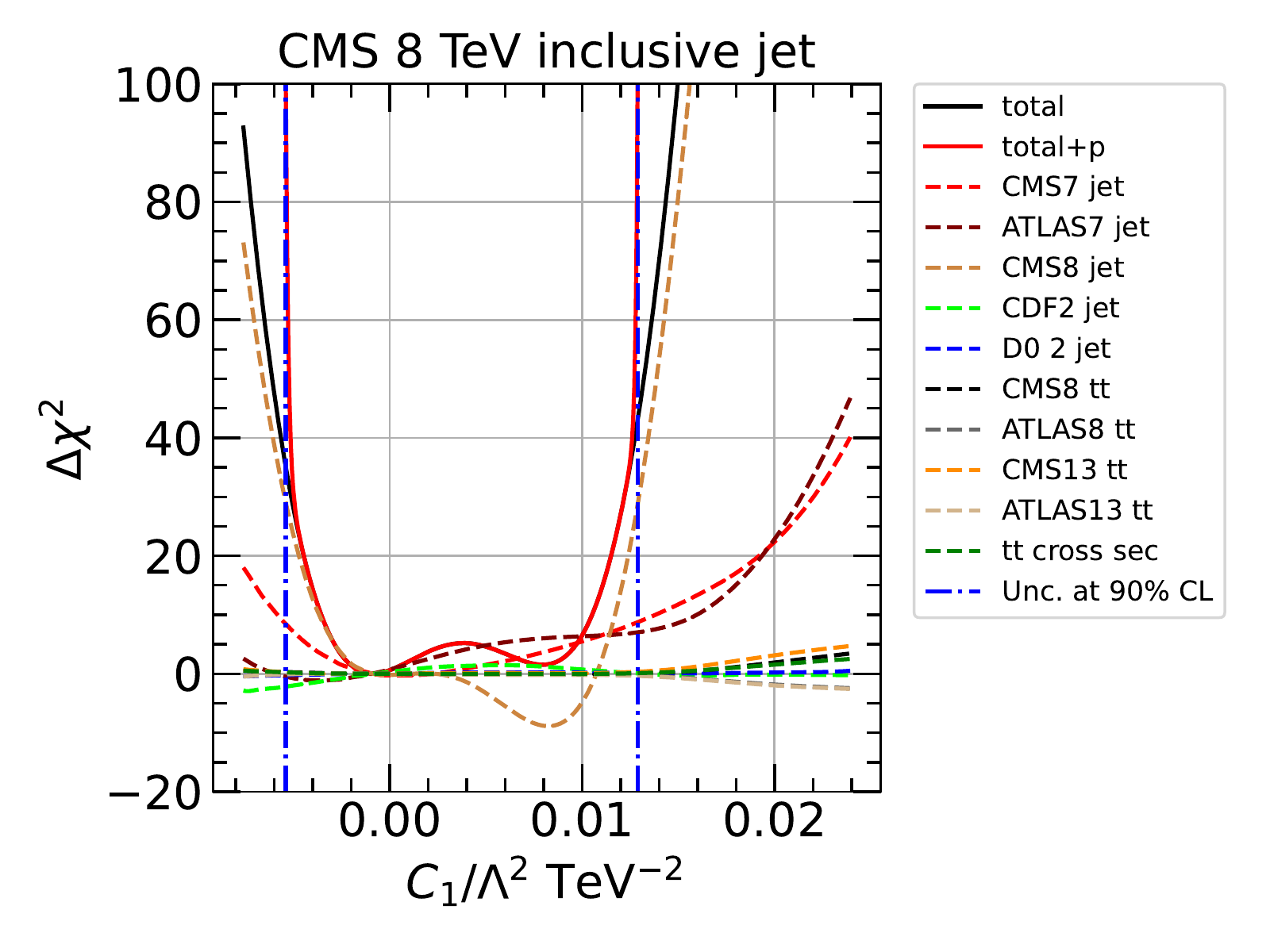}}
  \caption{LM scans on $C_{1}/\Lambda^2$ with the inclusion of the CMS 8 TeV dijet data (left panel) or CMS 8 TeV inclusive jet data (right panel).}
  \label{Fig:LM_jet23}
\end{figure}

It has been suggested~\cite{1101.4611} that contact interactions might be constrained
by dijet or inclusive jet production.
Apart from a modified energy dependence, contact interactions may also
induce a different angular distribution in dijet production relative to purely SM predictions. 
To explore this point, we compare constraints from the CMS 8 TeV dijet and inclusive jet data directly.
They are from the same data sample and differ only by the experimental observable.
We perform LM scans on $C_{1}$ with the inclusion of either the CMS 8 TeV dijet or inclusive jet data
respectively, meanwhile excluding the CMS 13 TeV inclusive jet data in the fit.
The results of doing this are shown in Fig.~\ref{Fig:LM_jet23}.
For the case of the CMS 8 TeV dijet data, in the left panel, we find that the CMS 8
TeV data together with the CMS and ATLAS 7 TeV jet data give the leading constraint.
It predicts a result of $C_{1}/\Lambda^2 = -0.0022^{+0.0187}_{-0.0054}$ TeV$^{-2}$ at 90\% CL that
has larger uncertainties than the result determined with our nominal setup, which is expected
since the CMS 13 TeV jet data are not included here.
In the right panel, we show the corresponding results for the CMS 8 TeV inclusive jet data.
The LM scans predict $C_{1}/\Lambda^2 = -0.0009^{+0.0138}_{-0.0045}$ TeV$^{-2}$ at 90\% CL,
that is, with reduced uncertainties relative to those determined from the CMS dijet data.
By themselves, however, the CMS 8 TeV inclusive data prefer a larger value of
$C_{1}/\Lambda^2 \approx 0.008$ TeV$^{-2}$, with $\chi^2$ lowered by about 10 units
relative to the global minimum.

The results on the two CMS 8 TeV jet data are listed in Tab.~\ref{tab:jet_indi}.
We also show the results from LM scans on $C_{1}$ with inclusion of only the CMS 13 TeV jet
data for comparison.
Note that, in all cases, the 7 TeV jet data, as well as the jet data from the Tevatron, are included in the fit.
We further compare results obtained by fixing the PDF parameters to their values at the respective global
minimum.
The CMS 13 TeV jet data give the strongest constraint, more so than the two 8 TeV data, which is consistent
with Fig.~\ref{Fig:np_jet5}.
Spuriously, the uncertainties on $C_{1}$ can be reduced significantly if the PDF parameters are fixed,
as is seen, {\it e.g.}, for the fit with CMS 13 TeV data alone.

\begin{table}[htpb]
  \centering
  \begin{tabular}{c|c|ccc}
  \hline
 TeV$^{-2}$   & nominal  & CMS 8 dijet & CMS 8 jet & CMS 13 jet\\
  \hline
    PDF free  & $-0.0015^{+0.0033}_{-0.0014}$ & $-0.0022^{+0.0187}_{-0.0054}$& $-0.0009^{+0.0138}_{-0.0045}$ & $-0.0013^{+0.0059}_{-0.0016}$\\
  \hline
   PDF fixed & $-0.0015^{+0.0024}_{-0.0014}$ & $-0.0022^{+0.0180}_{-0.0051}$& $-0.0009^{+0.0131}_{-0.0049}$ & $-0.0013^{+0.0026}_{-0.0015}$  \\
  \hline
  \end{tabular}
  \caption{Constraints on $C_{1}/\Lambda^2$ in ${\rm TeV}^{-2}$ at 90\% CL with individual data set on jet production.} 
  \label{tab:jet_indi}
\end{table}

\begin{figure}[htbp]
  \centering
  \subcaptionbox{}[7.7cm]
    {\includegraphics[width=7.7cm]{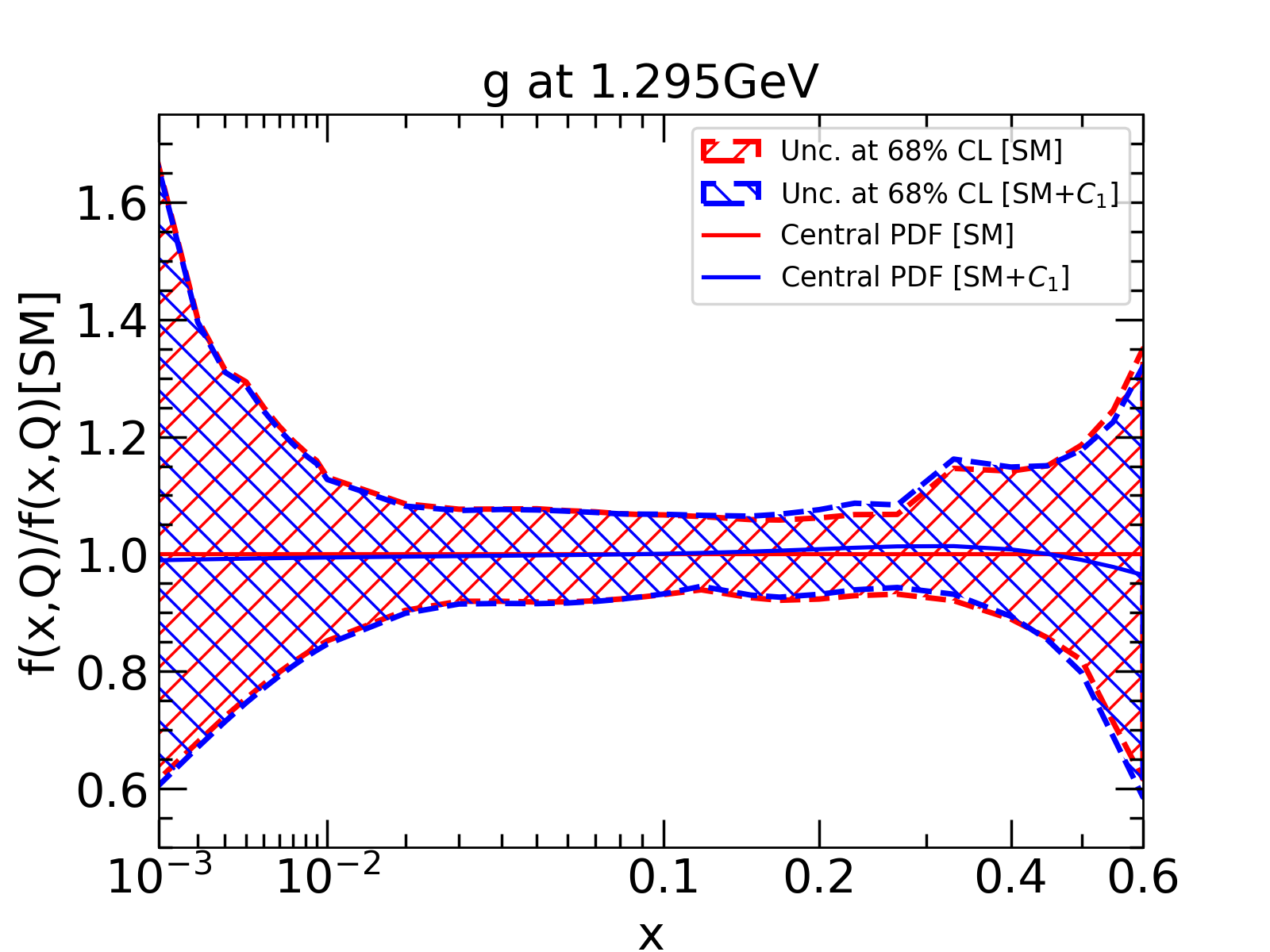}}
%  \subcaptionbox{}[7.7cm]
%    {\includegraphics[width=7.7cm]{figure/total_g_jet5_err.pdf}}
  \subcaptionbox{}[7.7cm]
    {\includegraphics[width=7.7cm]{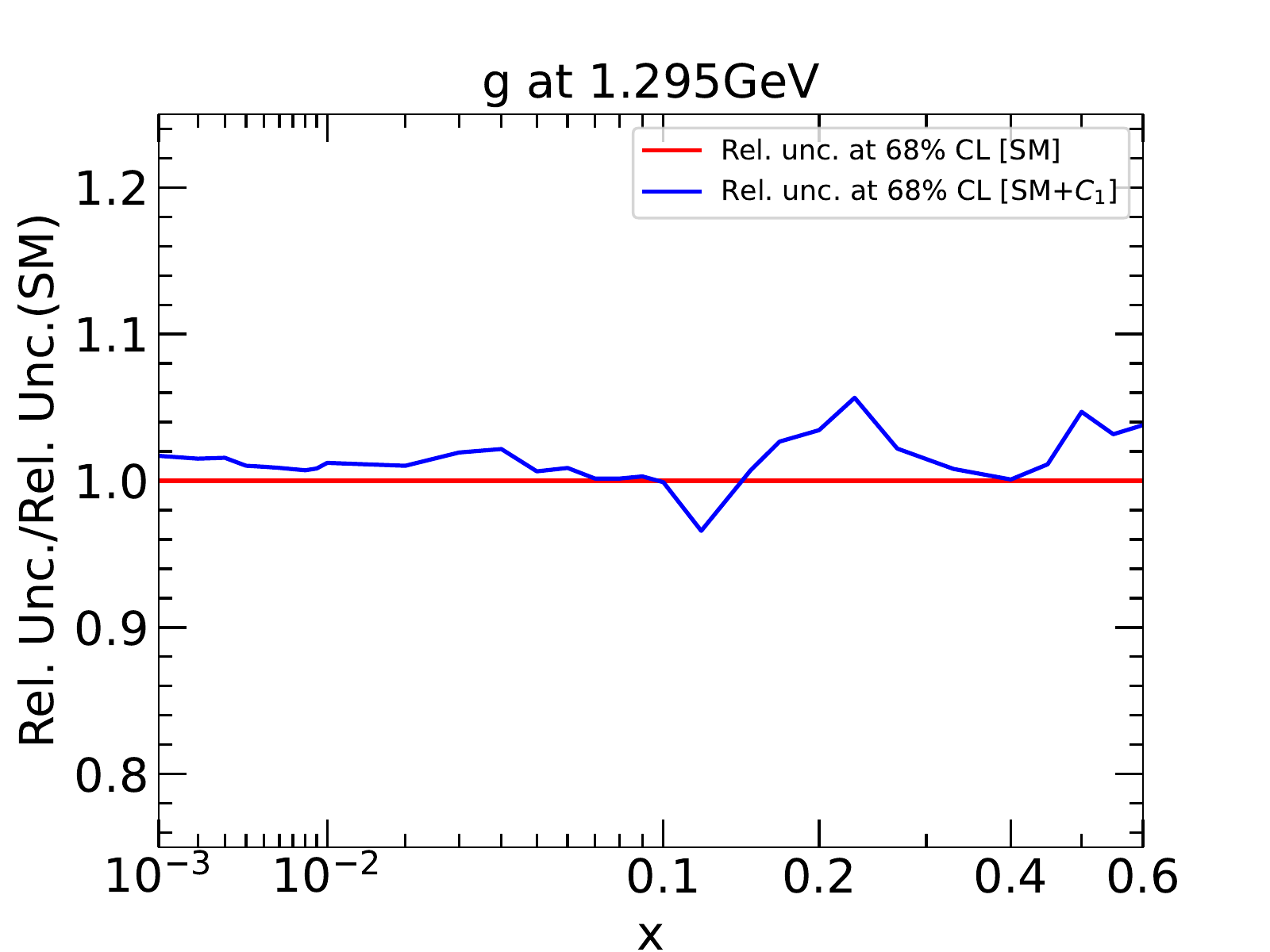}}
  \caption{The gluon PDFs $g(x,Q)$, as determined by fitting with and without SMEFT contributions at $Q_0 = 1.295$ GeV, are shown in the left panel. The blue and red solid lines represent the central values determined by fitting with and without SMEFT contributions, respectively. The PDF uncertainties at 68\% CL are indicated by hatched areas in the relevant colors. The relative uncertainties are shown in the right panel the same colors, normalized to the SM calculation.}
  \label{Fig:jet_g}
\end{figure}

In Fig.~\ref{Fig:jet_g}, we compare the gluon PDFs at $Q_0 = 1.295$ GeV determined by fitting with and without SMEFT.
In the left panel, we find almost no change for $x<0.1$, and an upward shift smaller than 2\% around $x \sim 0.3$,
due to the active inclusion of SMEFT.
In addition, a slight downward shift on both the central value and the uncertainty region
can be found in the region of $x>0.5$.
In the right panel, the relative PDF uncertainties are shown to slightly increase in the regions
of $x\sim 0.25$ and $x\gtrsim 0.4$ for the combined PDF+SMEFT analysis.
We conclude from Fig.~\ref{Fig:jet_g} that $C_{1}$ 
is moderately correlated with
the gluon PDF at large $x$.

\subsection{Interplay with top-quark production}
\begin{figure}[htbp]
  \centering
  \includegraphics[width=0.495\textwidth,clip]{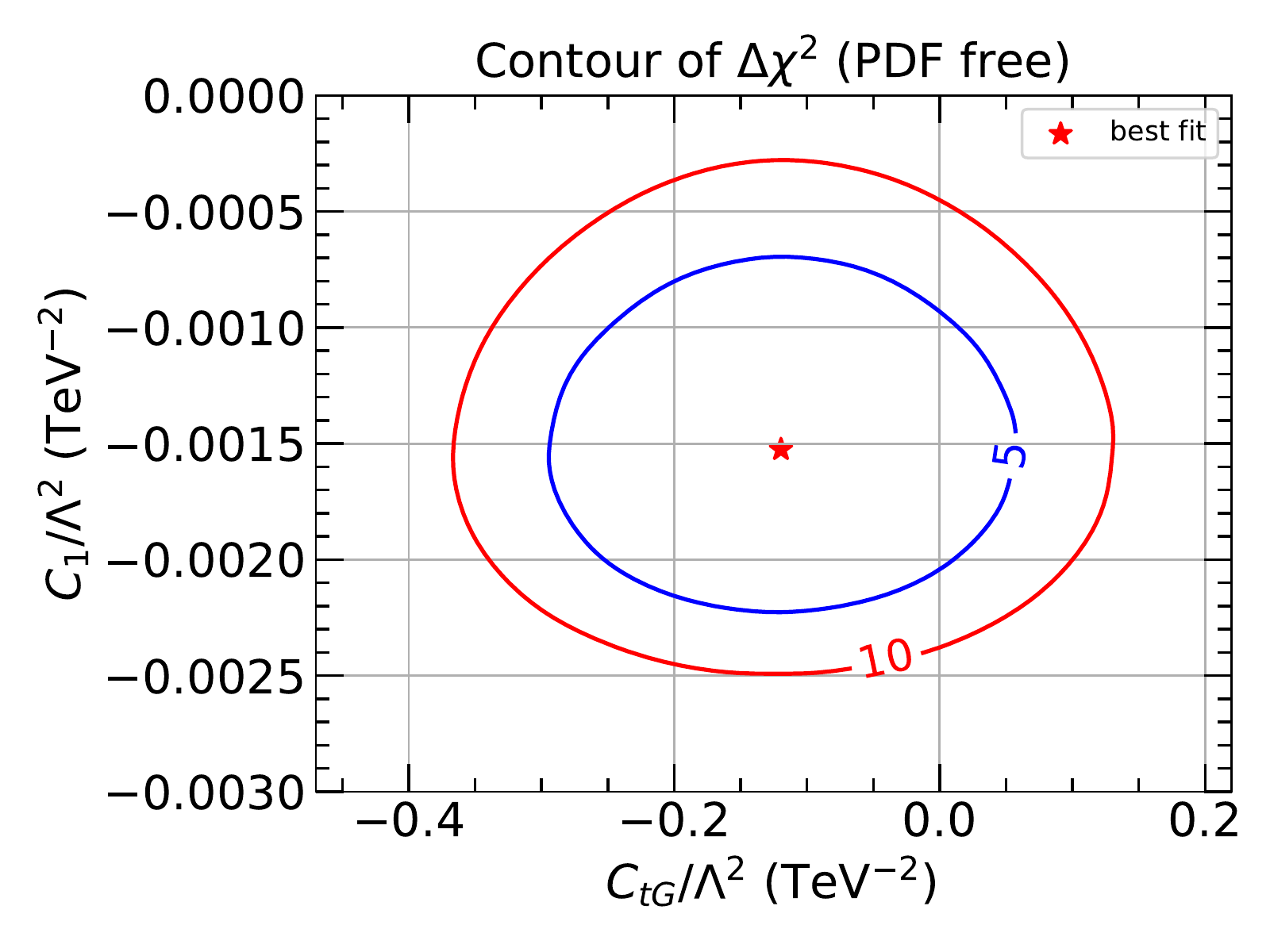}
  \includegraphics[width=0.495\textwidth,clip]{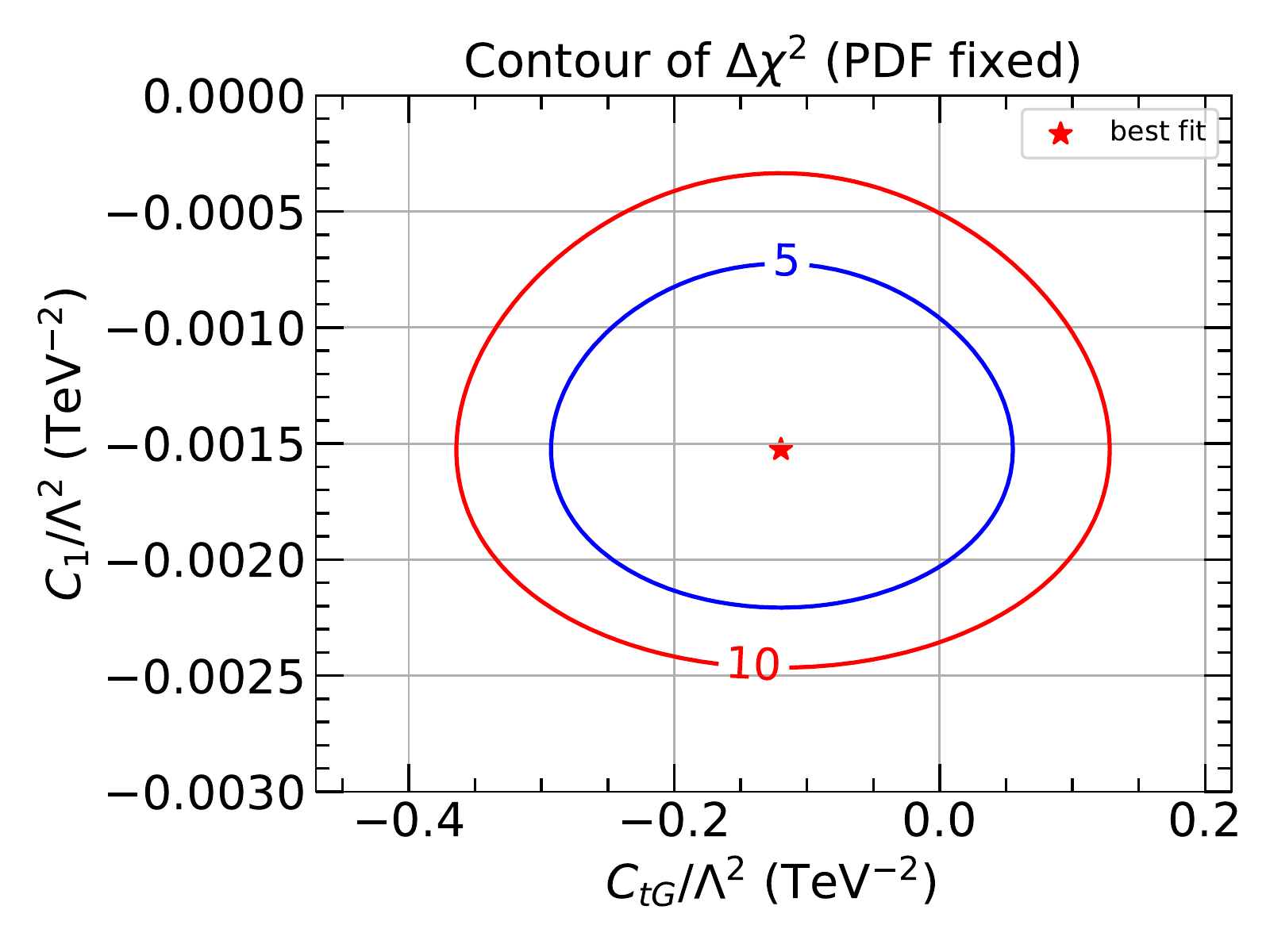}
	\caption{Contour plot of $\Delta \chi^2$ on the plane of $C_{tG}/\Lambda^2$ vs.~$C_{1}/\Lambda^2$ with the PDF parameters freely varying (left panel) or fixed to their best-fit values (right).
	}
  \label{Fig:2d_cj_ctg}
\end{figure}

It is interesting and important to study the interplay between the Wilson coefficients relevant
for jet production and those for top-quark pair production in the global analysis.
It is reasonable to expect some level of correlations between these since both jet and
top-quark pair production are ostensibly sensitive to the gluon PDF.
In this subsection we test these possible correlations through simultaneous fits of PDFs, $C_1$, and
$C_{tG}$. 
Specifically, we perform a series of LM scans on the individual coefficients by
fixing either $C_{1}$, $C_{tG}$, or neither, with the fitted results for these coefficients summarized in Tab.~\ref{tab:np_jet}.
In the first column, both $C_{1}$ and $C_{tG}$ are free, whereas
in the second and third columns, either $C_{1}$ or $C_{tG}$ is fixed to 0.
We find that the best-fit value and uncertainty on both coefficients are practically unchanged
when fixing either one coefficient or the other.
This indicates no direct correlation between $C_{1}$ and $C_{tG}$, and is reinforced
by the corresponding 2D LM scans in Fig.~\ref{Fig:2d_cj_ctg}.
In Fig.~\ref{Fig:2d_cj_ctg}, which explicitly plots the 2D LM scans correlating $C_{1}$ and $C_{tG}$, the blue and
red contours represent surfaces of constant $\Delta\chi^2 = 5$ and $10$, respectively.
In the left panel, the very weak correlation between $C_{1}$ and $C_{tG}$ is realized in the robust rotational symmetry of the
contour plot, especially in light of the fact that neither SMEFT coefficient exhibited particularly strong correlation with the
gluon PDF in the studies shown above.
In the right panel, the PDF parameters are fixed to their values at the global minimum.
The contours are slightly smaller than those shown in the left panel, which indicates weak correlations between SMEFT and PDF parameters.

\begin{table}[htpb]
  \centering
  \begin{tabular}{c|ccc}
  \hline
  TeV$^{-2}$  & $C_{1}$, $C_{tG}$ free & fix $C_{1}$ & fix $C_{tG}$ \\
    \hline
   $C_{1}/\Lambda^2$ & $-0.0015^{+0.0033}_{-0.0014}$  & 0& $-0.0015^{+0.0033}_{-0.0014}$  \\
  \hline
    $C_{tG}/\Lambda^2$ & $-0.120^{+0.248}_{-0.309}$  & $-0.117^{+0.247}_{-0.309}$ & 0\\
  \hline
  \end{tabular}
  \caption{Constraints on $C_{1}/\Lambda^2$ and $C_{tG}/\Lambda^2$ at 90\% CL from fits in which the SMEFT coefficients are fixed
	or allowed to freely vary.} 
  \label{tab:np_jet}
\end{table}

\begin{figure}[htbp]
  \centering
  \subcaptionbox{}[7.7cm]
    {\includegraphics[width=7.7cm]{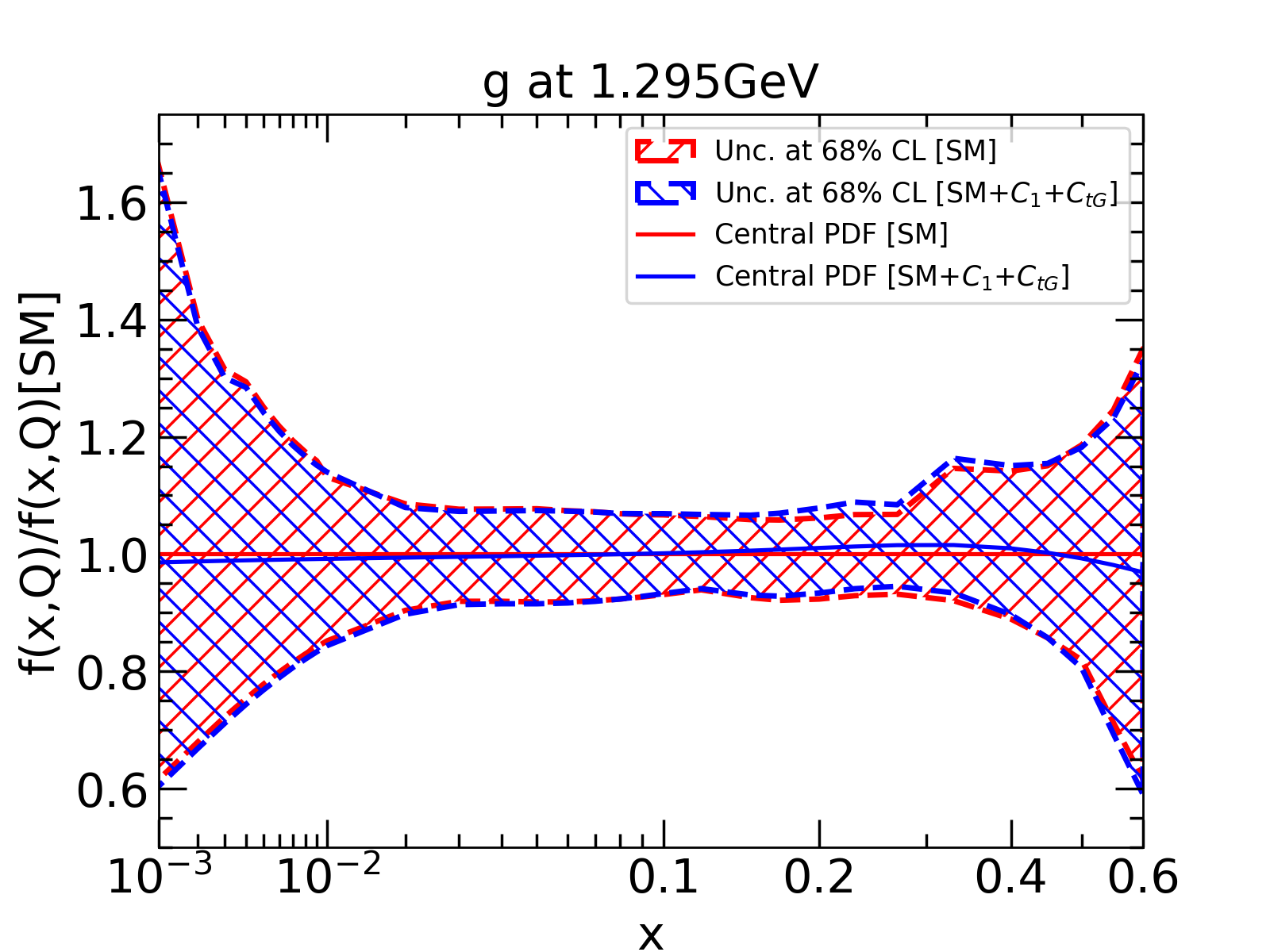}}
%  \subcaptionbox{}[7.7cm]
%    {\includegraphics[width=7.7cm]{figure/g_jet5_cctg_err.pdf}}
  \subcaptionbox{}[7.7cm]
    {\includegraphics[width=7.7cm]{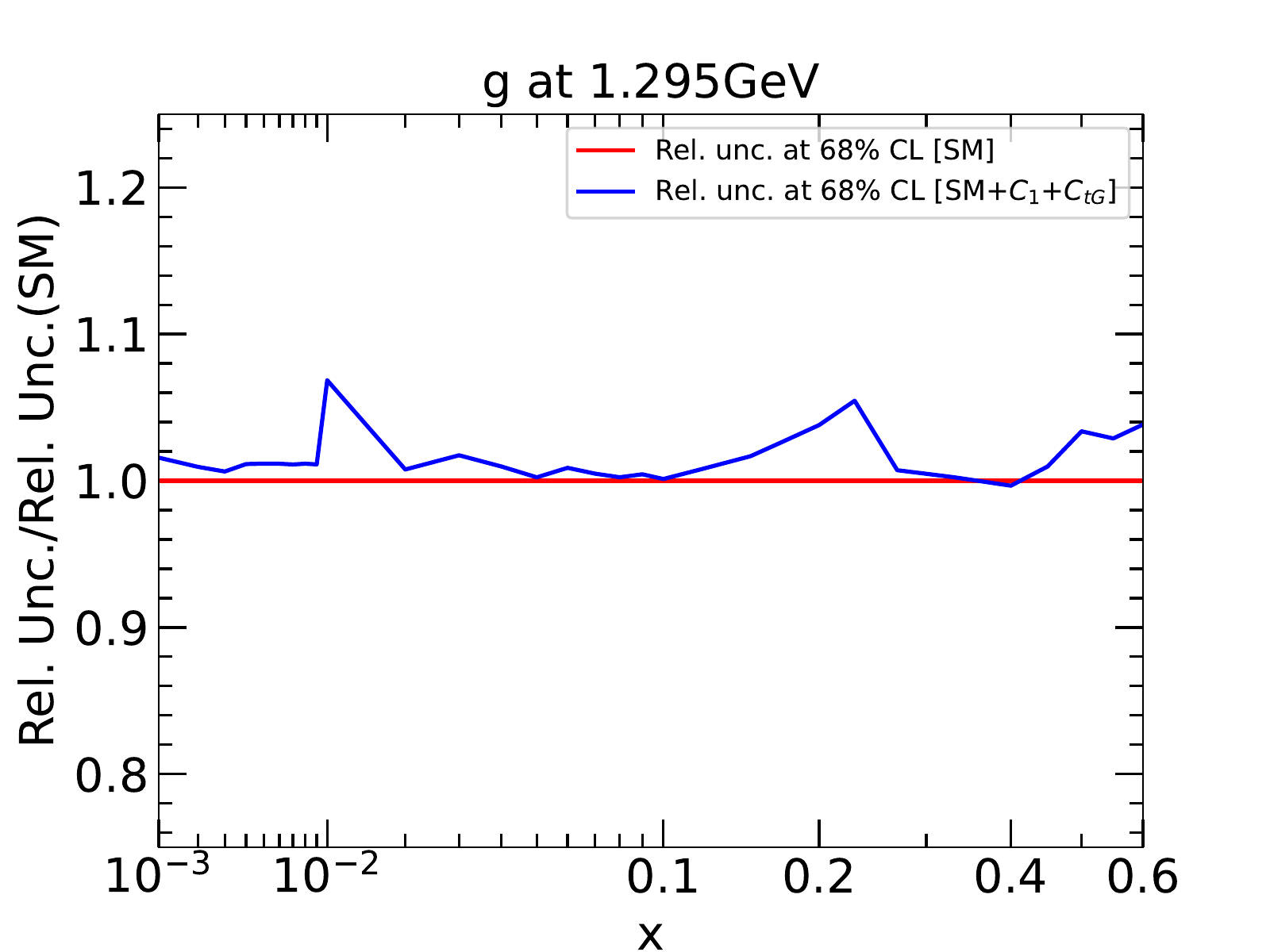}}
  \caption{The gluon PDF determined by fitting with and without SMEFT contributions from both $O_{1}$ and $O_{tG}$ at $Q_0 = 1.295$ GeV is shown in the left panel. The blue and red solid lines represent the central values determined by fitting with and without SMEFT contributions, respectively. The PDF uncertainties at the 68\% CL are shown via hatched areas in the corresponding colors. The relative uncertainties at 68\% CL are shown in the right panel, normalized as in previous plots.}
  \label{Fig:cj_ctq8_g}
\end{figure}

In Fig.~\ref{Fig:cj_ctq8_g}, we compare gluon PDFs at $Q_0 = 1.295$ GeV determined by fitting with and without
SMEFT contributions from $O_{1}$ and $O_{tG}$.
The impact on the gluon PDFs are mostly at large $x$.
In the left panel, a slight upward shifts on both the central value and uncertainty region can be found in the
region of $x\sim 0.3$ from SMEFT.
A slight downward shift, smaller than 5\%, on the central value can similarly be seen near $x\sim 0.6$.
In the right panel, the relative uncertainties at 68\% CL are shown, normalized for comparison as before.
We find that relative uncertainties are slightly enhanced when fitting SMEFT for
$x\sim 0.01$, $x\sim 0.2$, and $x\gtrsim 0.4$; while the $x$ dependence revealed in this case is somewhat different,
these enhancements do not exceed in size those seen when fitting $C_1$ or $C_{tG}$ separately

\section{Discussion}\label{sec:disc}

Following the detailed presentation of the various combined PDF+SMEFT fits shown above,
in the present section we briefly discuss a number topics which are particularly central to
these joint fits and their interpretation.
These issues include further discussion of correlations between the Wilson coefficients and PDFs
(Sec.~\ref{sec:corrs}) as well as the question of any dependence on the assumed statistical procedure
(Sec.~\ref{sec:tol}), for which we present several comparisons.

\subsection{Correlations between PDFs and SMEFT}
\label{sec:corrs}
\begin{figure}[htbp]
  \centering  \raisebox{0.35cm}{\includegraphics[width=0.49\textwidth,clip]{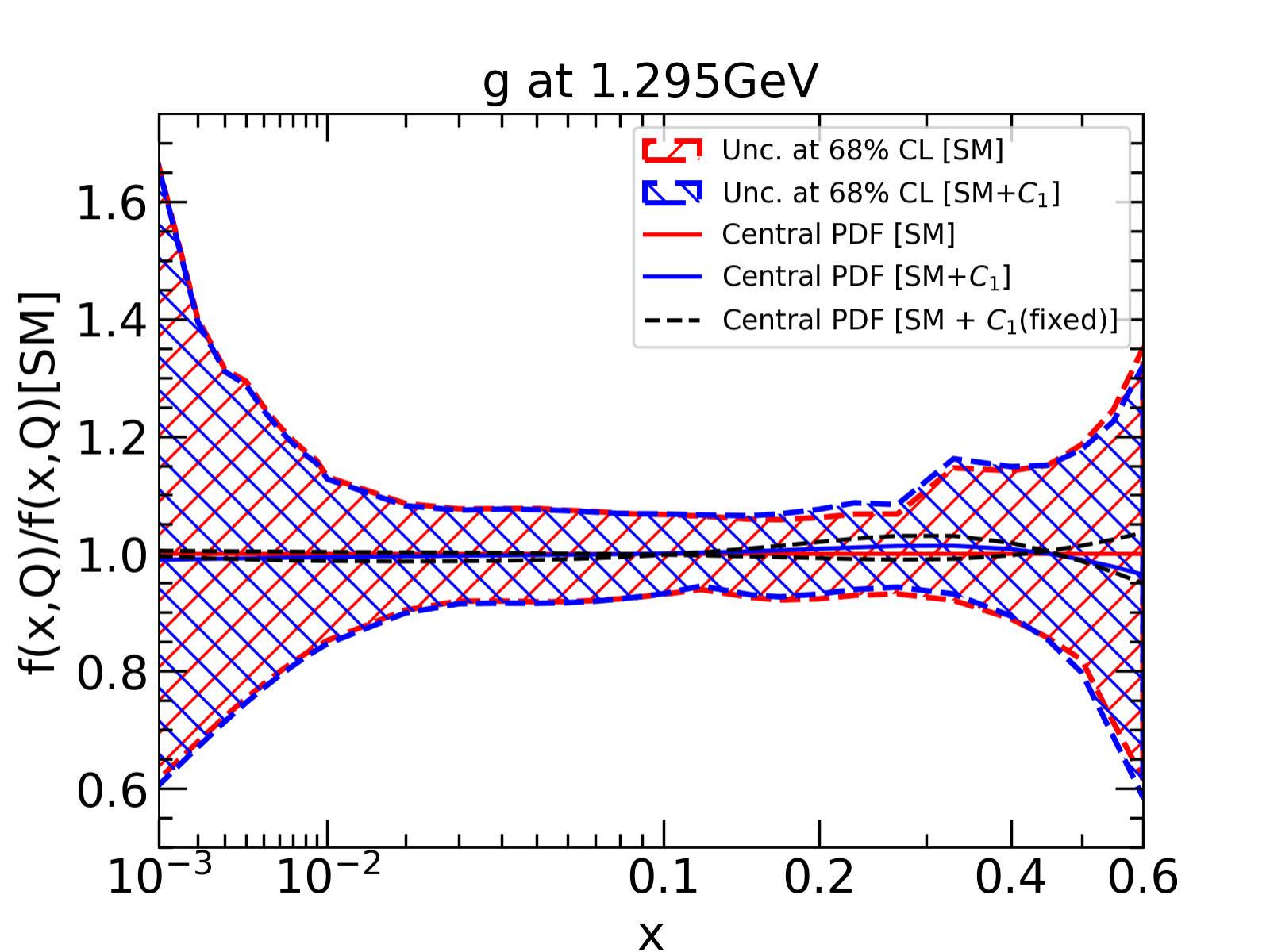}}
  \includegraphics[width=0.49\textwidth,clip]{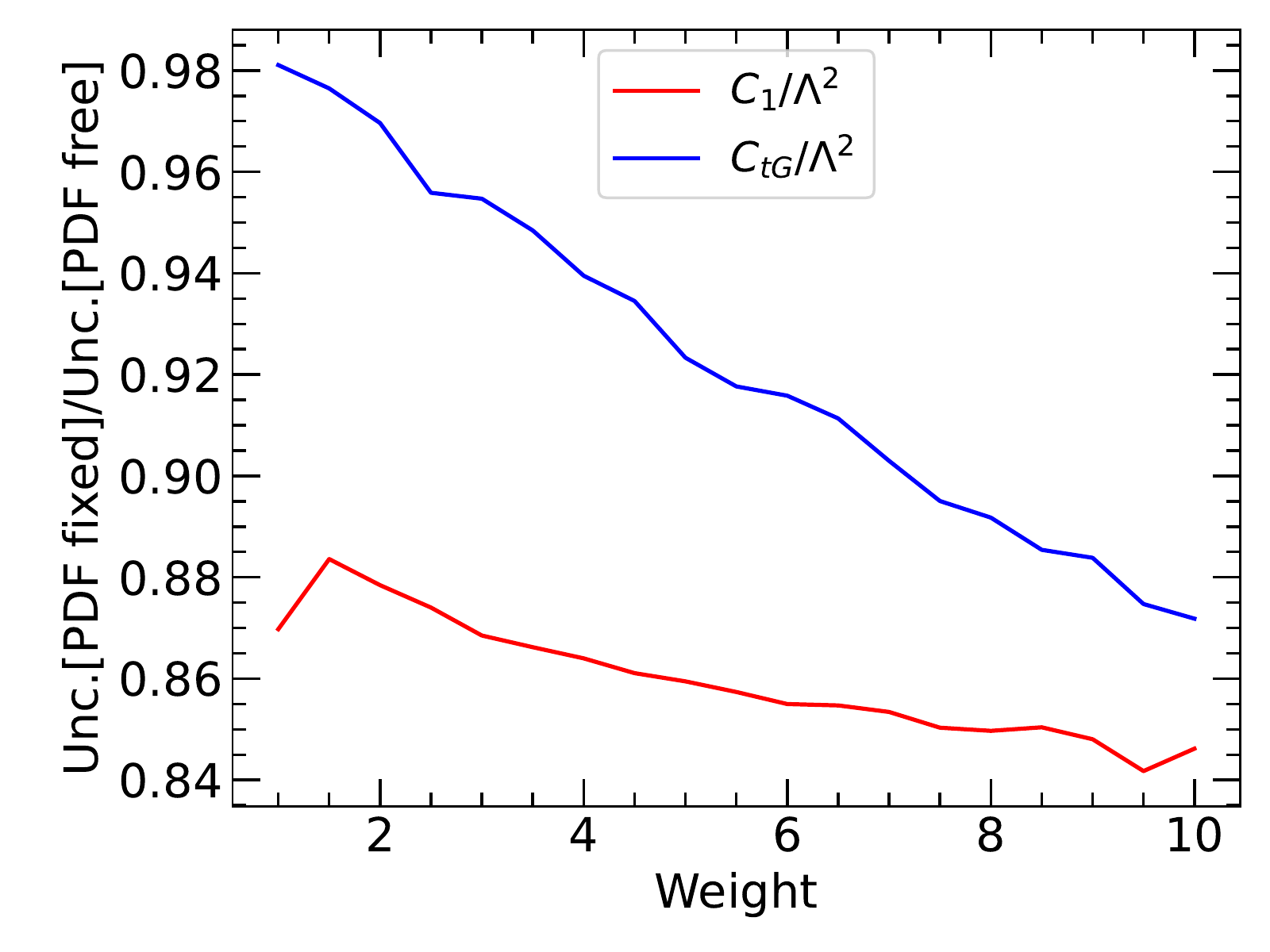}
  \caption{At left, we plot the gluon PDF ratio analogous to Fig.~\ref{Fig:jet_g} (left), now including two additional fits (dashed-black) in which the $C_1$ SMEFT coefficient is fixed to the extremal values allowed within a 90\% CL. In the right panel, we plot the ratio of the uncertainties determined with the PDF parameters fixed at the
  global minimum to the uncertainties determined with the PDF parameters free for $C_{1}/\Lambda^2$ and $C_{tG}/\Lambda^2$.}
  \label{Fig:corr_cj_ctg}
\end{figure}

Through the fitted PDFs and LM scans examined in the previous sections, we have seen evidence of mild correlations
between the extracted SMEFT coefficients and PDFs in joint fits of these quantities.
We observe PDF-SMEFT correlations as shifts in the PDF uncertainties once global fits are expanded to include
freely-fitted SMEFT Wilson coefficients, such as the $C_1$ operator associated with contact interactions probed by
jet production. Such correlations become more evident under scenarios in which the
SMEFT coefficients deviate more significantly from the pure, $C\!=\!0$, SM context. This can be seen in the
left panel of Fig.~\ref{Fig:corr_cj_ctg}, in which we again plot the fitted gluon PDF, normalized to the SM CT18 NNLO
baseline as in Fig.~\ref{Fig:jet_g}, but now including two additional fits in which $C_1$ is fixed at the extrema of
its 90\% uncertainty interval, resulting in the two additional dashed-black curves. While correlations remain
relatively modest, it is noteworthy that the deviations of the fitted gluon from the purely SM PDF fits under these
larger $C\!\neq\!0$ scenarios can rise to an appreciable fraction of the SM gluon PDF uncertainty, especially for
$x\! \gtrsim\! 0.1$ and above.
It is reasonable to expect potentially significant correlations between, {\it e.g.}, the gluon PDF and SMEFT coefficients,
especially when these quantities are fitted to individual data sets, as both top-quark pair and jet production
are generally thought to be sensitive to both.
In a realistic global analysis, however, the gluon PDF is constrained by a diverse collection of experiments
with unique pulls on the PDFs and their underlying $x$ dependence;
these fitted experiments include a variety of data sets other than $t\bar{t}$ or jet production, such as
DIS measured at high precision.
Moreover, even among the jet measurements we include, there are distinct center-of-mass energies and
various distributions in $y$, $p_T$, and $m_{t\bar{t}}$, each of which may differently probe the
gluon PDF.
These considerations have the effect of diluting the correlations between the PDF pulls of individual
data sets and the preferences of the full fit for specific SMEFT coefficients.
On the other hand, this point underscores the importance of extracting PDFs through a
global analysis with data sets spanning a wide range of energies in various channels.

The statements above are made on the basis of our analysis of contemporary hadronic
data; in principle, however, the mild correlations we find may grow in strength with
greater experimental precision at the HL-LHC or other future experiments.
We therefore explore this potential for enhanced correlations between PDFs and Wilson
coefficients at future runs of the LHC.
To maximize the likelihood of obtaining strong correlations, we take an extreme case of only keeping
the most sensitive data among all the top-quark pair and jet production sets explored in this work.
This corresponds to the total cross section measurements for top-quark pair
production as well as the CMS 13 TeV measurement of inclusive jet production. 
For the former, we then perform LM scans on the Wilson coefficient $C_{tG}$ in global fits
in which these $t\bar t$ data are overweighted by a multiplicative weight factor placed on their associated
$\chi^2$; this overweighting is statistically equivalent to an overall reduction in the uncorrelated
uncertainty of the $t\bar t$ cross sections, thereby mimicking future improvements in both experimental
precision and theoretical accuracy.
The final uncertainties on $C_{tG}$ are determined
with the criterion $\Delta\chi^2 = 2.706$ for simplicity, under separate scenarios in which the PDFs
are either frozen at their global minima or allowed to float freely.
The ratio of the uncertainties on the SMEFT coefficients for the fits with fixed or free PDFs can
be interpreted as an indication of the degree to which $C_{tG}$ might be correlated with the PDFs, which we trace
as a function of total the precision of the $t\bar{t}$ data ({\it i.e.}, the ``Weight'' on the data).
We carry out identical scans on the Wilson coefficient $C_{1}$, in this case, placing the additional
weight on the individual $\chi^2$ of the inclusive jet data, rather than the $t\bar{t}$.

In Fig.~\ref{Fig:corr_cj_ctg} (right), we show the ratios described above
as functions of the chosen Weight for both $C_{tG}$ and $C_1$.
Specifically, the ratio starts at 0.98 (0.87) for $\mathrm{Weight}\!=\! 1$, and decreases to approximately 0.87 (0.84)
near $\mathrm{Weight}\!=\! 10$ for $C_{tG}$ ($C_{1}$).
We thus find that the uncertainty ratio for $C_{1}$ is always smaller than the corresponding ratio for $C_{tG}$, indicating stronger
correlations with the fitted PDFs and a greater underestimate in the uncertainty for this SMEFT coefficient
when PDFs are not simultaneously fitted; this is true for all Weights considered on these leading experiments. At the same
time, it is noteworthy that the $\mathrm{Weight}\!=\! 10$ overweighting of the $t\bar{t}$ data leads to a more
rapid relative increase in the size of the PDF-SMEFT correlations than the corresponding shift found for $C_{1}$
and the jet data.
We note that taking $\mathrm{Weight}\!=\! 10$ may be interpreted in terms of a corresponding reduction in the total {\it uncorrelated} uncertainties for the fitted jet and $t\bar{t}$ experiments. Assuming the Weight to be an overall prefactor on the contribution to $\chi^2$ from
a given experimental data set, $\mathrm{Weight}\!=\!10$ then corresponds to a reduction
in the total uncorrelated (statistical and uncorrelated systematic) uncertainty by a factor
of $\sqrt{10}\!\approx\!3$. For comparison, the optimistic-scenario PDF projections of Ref~\cite{AbdulKhalek:2018rok}
assumed improvements by a factor of 2-3 in systematic uncertainties at HL-LHC as well as $\mathcal{L}\!=\!3\,\mathrm{ab}^{-1}$
data sets at ATLAS --- more than an order-of-magnitude increase in aggregated statistics.
Thus, the $\approx\!10\%$ under-estimate in $C_{tG}/\Lambda^2$ shown
in Fig.~18 (right) for $\mathrm{Weight}\!\sim\! 7$ reflects the enhanced correlations which might reasonably be expected at
HL-LHC under optimistic performance scenarios.
We conclude that the correlations between the PDFs and SMEFT coefficients become stronger with increasing
precision as expected --- a general observation that must inform future studies. These
projections are based on extrapolations starting from these particular $t\bar{t}$ and inclusive jet data;
future experiments with higher initial precision may steepen the trajectories shown in Fig.~\ref{Fig:corr_cj_ctg}
as uncertainties shrink.
These potential correlations can be further enhanced when using a realistic tolerance criterion,
which can only be studied with actual data rather than estimated via this simplified reweighting procedure. 

\subsection{Impact of different tolerance criteria}
\label{sec:tol}
\begin{figure}[htbp]
  \centering
  \subcaptionbox{}[7cm]
    {\includegraphics[width=7cm]{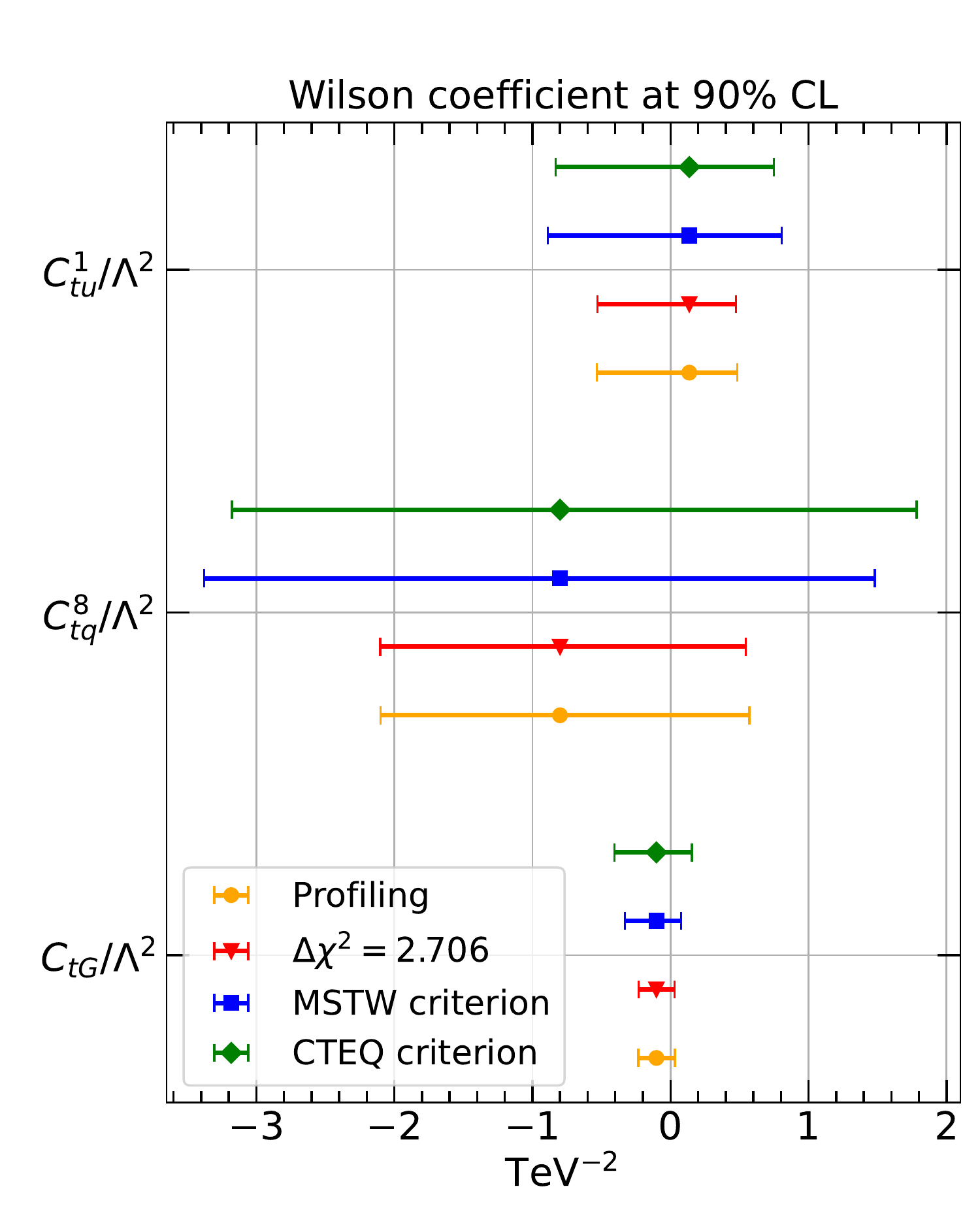}}
  \subcaptionbox{}[7cm]
    {\includegraphics[width=7cm]{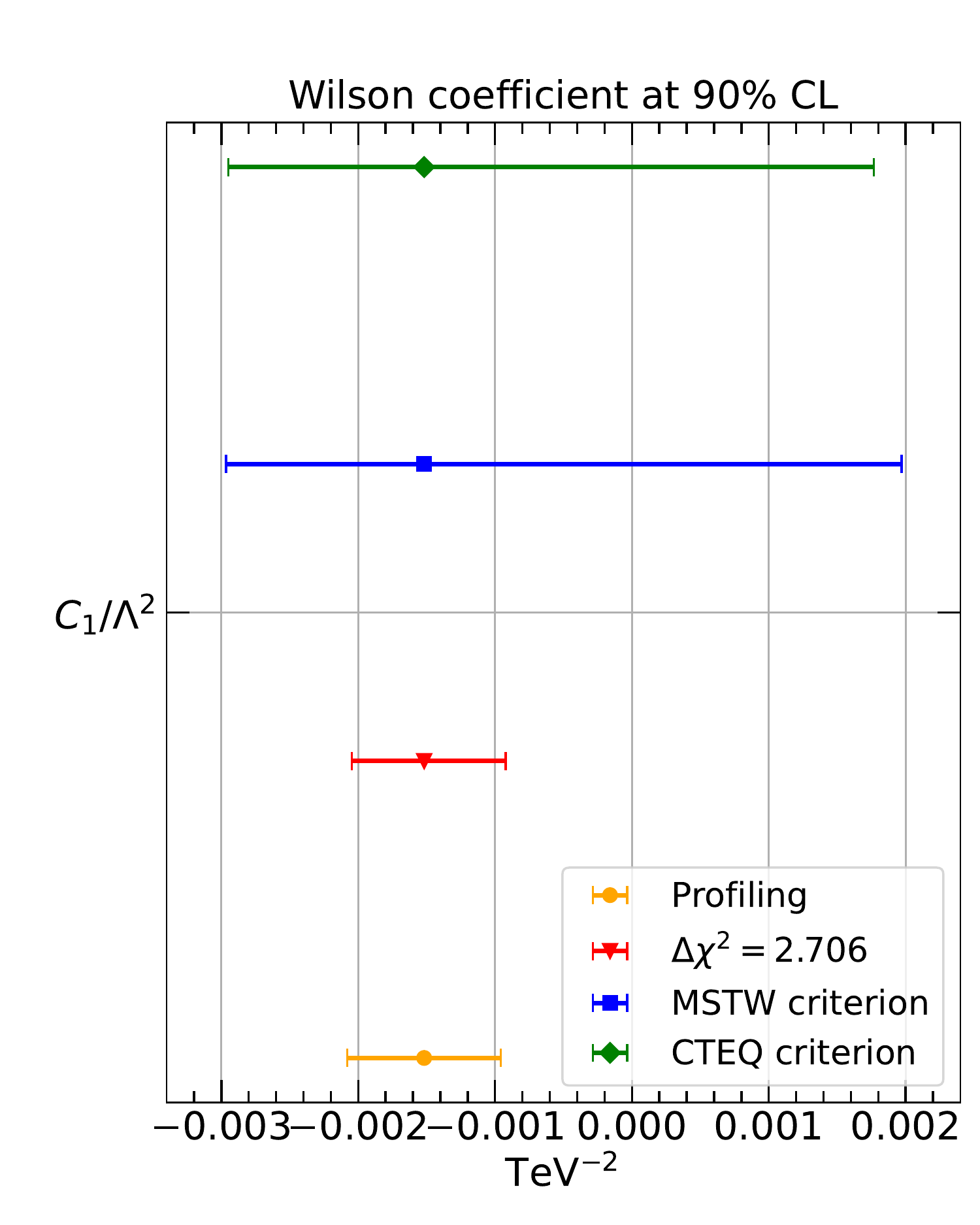}}
  \caption{Constraints on the Wilson coefficients with different tolerance criteria. The marks and error bars respectively indicate the central values and uncertainties at 90\% CL. The results with different tolerance criteria are shown with relevant colors.}
  \label{Fig:tolerance}
\end{figure}

In this study, uncertainties on the PDFs and Wilson coefficients were determined according to
the same tolerance criterion (1) as in the CT18 global analyses, namely,
with $\Delta\chi^2 + P = 100$ at the 90\% CL.
With this criterion, both the change in the global $\chi^2$ and disagreements
among individual data sets were considered at the same time.
In contrast, the MSTW~\cite{0901.0002} family of analyses employ a dynamical tolerance
criterion (2) in determinations of both PDF and parametric QCD uncertainties.
Though broadly similar, the use of dynamical tolerance somewhat differs from the CT18 criterion:
namely, variation in the global $\chi^2$ is not included in the dynamical tolerance, which instead
differently normalizes the $\chi^2$ values of individual data sets at the global minimum.
We emphasize that it is important to introduce the tolerance factors for a global analysis
with many different data sets in order to account for possible tensions among the fitted experiments. 
In experimental analyses using fewer data sets, the usual parameter-fitting criterion (3) is always used with uncertainties at the 90\% CL determined by requiring $\Delta\chi^2 = 2.706$.
We note that there is also the so-called PDF profiling method (4) to fit input parameters
together with PDFs through a series of nuisance parameters~\cite{1810.03639,1906.10127} using
Hessian PDFs from the global analyses.
This is approximately equivalent to performing a global fit with reduced weights for the data
sets used in the original PDF sets, when combining with the criterion of $\Delta\chi^2 = 2.706$.
We compare the extracted Wilson coefficients using the four (1-4) criteria noted above in Fig.~\ref{Fig:tolerance}.
In the last scenario, the $\chi^2$ of data sets other than the top-quark pair (jet) production
have been divided by a factor of 10, the average tolerance at 68\% CL, when included into
the global $\chi^2$ in the fit of Wilson coefficients associated with top-quark pair (jet) production.
In Fig.~\ref{Fig:tolerance}, we plot the central
values and uncertainties at 90\% CL for each of the various Wilson coefficients fitted in this study.
It can be seen that the CT18 and MSTW criteria show comparable results on the uncertainty range,
as similarly observed in Ref.~\cite{2203.05506} for PDF uncertainties.
The uncertainties from the CT18 criterion can be either slightly larger or smaller than those
governed by the MSTW criterion, depending on the Wilson coefficients considered.
The uncertainties determined with the usual parameter-fitting criterion are smaller by about a factor
of 2 in the case of the Wilson coefficients associated with top-quark pair production.
For the extraction of the contact-interaction coefficient, $C_1$, the dependence of the uncertainty on
the tolerance criterion is even larger, where the uncertainty from the parameter-fitting criterion is
smaller by a factor of 5.
The SM ($C_1=0$) is excluded already if using the uncertainty estimated from the parameter-fitting
criterion indicating the failure of such a criterion in the global analyses with large number of
data sets.
Results using the criterion with reduced weights applied to the data sets with minimal sensitivity show
almost no difference with respect to the ones without reweighting. 

Finally, we compare our results for the Wilson coefficients to those of previous studies.
The ATLAS collaboration reports a 95\% CL bound on $C_{tG}/\Lambda^2$ of [-0.52, 0.15] TeV$^{-2}$
using the transverse momentum distribution of the top quark measured at LHC 13 TeV in
the hadronic decay channel with an integrated luminosity of 139~fb$^{-1}$~\cite{2202.12134}.
Our nominal 90\% CL bound is $[-0.40, 0.16]$ TeV$^{-2}$, which is comparable with the ATLAS result.
In addition, the ATLAS collaboration reports a 95\% CL bound on $C_{tq}^{8}/\Lambda^2$ of [-0.64, 0.12] TeV$^{-2}$.
Our nominal 90\% CL bound is $[-3.2, 1.8]$ TeV$^{-2}$, which is much weaker than the ATLAS result.
There are two main reasons for this.
First, we use the default CT18 tolerance rather than the parameter-fitting criterion.
When using the criterion of $\Delta\chi^2 = 1.96^2$, we obtain a bound at 95\% CL of [-2.28, 0.72] TeV$^{-2}$.
Second, the ATLAS constraints are based on the transverse momentum distribution of the top quark.
That in general leads to stronger constraints on the Wilson coefficient $C_{tq}^{8}/\Lambda^2$ than that found when using the $m_{t\bar{t}}$ distribution of the top-quark pair.
For instance, in our study and the SMEFiT study of Ref.~\cite{2105.00006}, both of which use the $m_{t\bar{t}}$ distribution, the bounds on $C_{tq}^{8}/\Lambda^2$ are much weaker than the bounds on $C_{tG}/\Lambda^2$.
Meanwhile in the ATLAS measurement, the bounds on $C_{tq}^{8}/\Lambda^2$ and $C_{tG}/\Lambda^2$ are comparable.
In another analysis from CMS of inclusive jet production at 13 TeV (corresponding to the same data included in this work),
a 95\% CL result of $C_{1}/\Lambda^2 \in [-0.0013, -0.0001]$ TeV$^{-2}$ was reported based on a joint fit of PDFs
and contact interactions.
In comparison, our nominal result of $[-0.0029, 0.0018]$ TeV$^{-2}$ at the 90\% CL is compatible considering
the different criterion used.

\section{Summary}\label{sec:conc}
SMEFT model-independently parametrizes BSM physics as might typically be formulated via phenomenological Lagrangians
in the ultraviolet; this in turn provides a systematically-improvable framework for connecting BSM far above the electroweak
scale to empirical consequences at lower energies --- at the LHC or other facilities.
Problematically, SMEFT-based BSM searches often involve the same collider data as those fitted in studies of
proton PDFs, which are also core inputs to the SM theory predictions for BSM search baselines.
To understand the extent to which this might introduce statistical bias into extractions of SMEFT coefficients, we perform a
joint PDF+SMEFT fit based on an extension of the CT18 global analysis, and obtain a self-consistent determination of the possible
BSM effects.
The global analyses in this work are boosted with supervised machine learning techniques in the form
of multi-layer perceptron neural networks to ensure efficient scans of the full PDF+SMEFT parameter space.
To be specific, we compute $\chi^2$ profiles for all parameters, including the PDFs, strong coupling, $\alpha_s$,
top-quark mass, $m_{t\bar{t}}$, and the SMEFT Wilson coefficients, finding these can be learned efficiently and with high
fidelity by the neural network. 
In this study, we focused on several SMEFT operators that are relevant for top-quark pair and jet production at hadron colliders.
Regarding top-quark production, for the Wilson coefficients of the four-quark color-singlet and octet and gluonic operators, we obtain
$C_{tu}^{1}/\Lambda^2 = C_{td}^{1}/\Lambda^2 = 0.14^{+0.61}_{-0.97}$
TeV$^{-2}$, $C_{tq}^{8}/\Lambda^2 = -0.80^{+2.58}_{-2.38}$ TeV$^{-2}$ and $C_{tG}/\Lambda^2 = -0.10^{+0.26}_{-0.30}$ TeV$^{-2}$,
respectively, at the 90\% CL using the default CT tolerance.
For jet production, we get $C_1/\Lambda^2 = -0.0015^{+0.0033}_{-0.0014}$ TeV$^{-2}$ at 90\% CL for the four-quark contact interactions.
We find mild correlations between the extracted Wilson coefficients and PDFs, particularly, the gluon PDF at very high $x$, as well as other QCD
parameters like the strong coupling and top-quark mass. While we investigated the effects of the combined PDF+SMEFT analyses
on other PDF flavors, we generally found the impact in these cases to be much smaller than that observed for the gluon;
simultaneous fits of additional SMEFT operators probed by other data sets may alter this picture, which we reserve for
forthcoming studies.
Though presently mild, we also find that these correlations between SMEFT coefficients and PDFs may grow significantly with higher precision in $t\bar{t}$ and jet production,
as might be achievable at HL-LHC.

We have also examined the dependence of the Wilson coefficient uncertainties on the
statistical criteria used in joint fits.
We showed that, in the context of global analyses with a variety of experimental
data, the CT18 and MSTW tolerance criteria result in similar uncertainties while the
parameter-fitting and profiling criteria give much smaller uncertainties.
This work serves as a new basis for joint analyses of SM and BSM in the setting of the
CTEQ-TEA framework.
In addition to being generalizable with additional machine-learning developments, this approach may also be regularly updated with new data from the LHC and other
experiments, and a systematic study on SMEFT operators relevant for DY production and DIS
processes is underway.

\begin{acknowledgments}
This work was sponsored by the National Natural Science Foundation
of China under the Grant No.12275173 and No.11835005.
We would like
to thank Marco Guzzi, Keping Xie and other members of CTEQ-TEA collaboration for helpful discussions
and proofreading of the manuscript, and Katerina Lipka and Klaus Rabbertz for useful communications.
JG thanks the sponsorship from Yangyang Development Fund.
The work of TJH at Argonne National Laboratory was supported by the
U.S.~Department of Energy, Office of Science, under Contract
No.~DE-AC02-06CH11357.
\end{acknowledgments}

\bibliography{SMEFT_PDF}
\bibliographystyle{jhep}

\end{document}